%% file: JMP_v3.tex
\documentclass[12pt,twoside]{article}
\usepackage{chadstyle}  
\usepackage[margin=1in]{geometry}
\usepackage{apacite}
\usepackage{microtype}
\usepackage{amsmath}
\usepackage{amsfonts}
\usepackage{amssymb}
\usepackage{bbm}
\usepackage{dsfont}
\usepackage{tikz}
\usetikzlibrary{arrows.meta,decorations.pathreplacing}
\usepackage{xcolor}
\usepackage{bm}
\usepackage{booktabs}
\usepackage{placeins}
\usepackage{float}
\usepackage{bibunits}
\usepackage{titletoc}
\usepackage{pdflscape}
\usepackage{needspace}
\usepackage{bookmark}
\bookmarksetup{numbered,depth=2}
\hypersetup{
  pdftitle={Poverty Targeting with Imperfect Information},
  pdfauthor={Juan C. Yamin},
  pdfsubject={Statistical decision methods for poverty targeting under imperfect information},
  pdfkeywords={poverty targeting, empirical Bayes, statistical decision theory}
}

\usepackage{pgfplots}
\pgfplotsset{compat=1.18}
\definecolor{PlugInGreen}{RGB}{27,158,119}
\definecolor{JSPurple}{RGB}{117,112,179}
\definecolor{CentroidGray}{RGB}{140,140,140}

\definecolor{ao}{rgb}{0.0, 0.5, 0.0}
\definecolor{EBBlue}{HTML}{0072B2}      
\definecolor{EBLightBlue}{HTML}{56B4E9}
\definecolor{PlugOrange}{HTML}{D55E00}

\numberwithin{corollary}{section}

\newtheorem{remark}{Remark}[section]  
\newenvironment{titledRemark}[1]
  {\begin{remark}\normalfont\bfseries #1.\normalfont\mdseries}  
  {\end{remark}}
  
  \newcommand{\R}{\mathbb{R}}
\newcommand{\E}{\mathbb{E}}

\newcommand{\1}{\mathds{1}}

\newcommand{\PIlabel}{\mathrm{PI}}
\newcommand{\EBlabel}{\mathrm{EB}}
\newcommand{\pluglabel}{\mathrm{plug}}
\newcommand{\cslabel}{\mathrm{cs}}
\newcommand{\Gain}{\mathrm{Gain}}

\newtheorem{assumption}{Assumption}
\numberwithin{assumption}{section}

\defaultbibliographystyle{apacite}
\defaultbibliography{PovTarget}
\newcommand{\AppendixQED}[1]{%
  \unskip\nobreak\hspace{0.35em}\hyperref[#1]{QED}.}
\renewcommand{\Proof}[2]{%
  \par\Needspace{3\baselineskip}\noindent
  {\color{ChadGreen}\bfseries Proof of Proposition~\ref{#1}.}\ #2%
  \AppendixQED{#1}\par\smallskip}
\renewcommand{\ProofLemma}[2]{%
  \par\Needspace{3\baselineskip}\noindent
  {\color{ChadGreen}\bfseries Proof of Lemma~\ref{#1}.}\ #2%
  \AppendixQED{#1}\par\smallskip}
\renewcommand{\ProofCorollary}[2]{%
  \par\Needspace{3\baselineskip}\noindent
  {\color{ChadGreen}\bfseries Proof of Corollary~\ref{#1}.}\ #2%
  \AppendixQED{#1}\par\smallskip}
\newcommand{\AppendixProofHeading}[1]{%
  \par\Needspace{3\baselineskip}\noindent
  {\color{ChadGreen}\bfseries #1.}\ }
\input{\outdir/numbers/paper_text_numbers.tex}

\input{\outdir/numbers/paper_gain_percent_numbers.tex}
\providecommand{\poorReachedRatioEBOLSRounded}{1.8}

\begin{document}
\begin{bibunit}
\hypersetup{pageanchor=false}

\begin{spacing}{0.9}
\begin{titlepage}

\title{Poverty Targeting with Imperfect Information} 

\author{\large Juan C. Yamin\thanks{Brown University, Department of Economics, Robinson Hall, 64 Waterman Street, Providence, RI 02912. Email: \href{mailto:juan_yamin_silva@brown.edu}{\texttt{juan\_yamin\_silva@brown.edu}}. I am grateful for the generous advice and support of Toru Kitagawa, Soonwoo Kwon, and Jonathan Roth. I thank Andrew Chesher, Tim Christensen, Andrew Foster, Peter Hull, Patrick Kline, Eric D. Mbakop, Paul Niehaus, Kirill Ponomarev, Susanne Schennach, Jann Spiess, Stefan Wager, and participants at the 2025 World Congress of the Econometric Society and Brown's Econometric Coffee and Development Tea for their helpful comments. Caitlin Brown, Paul Corral, and Heath Henderson kindly shared their data for the simulation exercises.}\\ { Brown University }
   }

\date{September 22, 2026}
\maketitle
\thispagestyle{empty}

\vspace{-0.3in}

\begin{abstract}
How should antipoverty programs allocate transfers when household income is known only through noisy predictions? I formulate this as a statistical decision problem in which a policymaker chooses nonnegative transfers within a fixed budget to minimize squared deviations of post-transfer income from the poverty line. I show that the standard plug-in rule, which treats predictions as exact, is inadmissible. I then develop a nonparametric empirical Bayes allocation rule that replaces the estimated gaps with posterior mean poverty gaps. Its Bayes regret is bounded by the mean squared difference between its posterior mean gaps and the oracle's, so the budget and nonnegativity constraints do not slow its convergence to the oracle. The approach extends, with weaker guarantees, to a planner who cares only about how far households remain below the poverty line after transfers and to programs that pay a fixed set of benefit amounts. In simulations using household surveys from nine African countries, the allocations under the empirical Bayes rule reach about \poorReachedRatioEBOLSRounded\ times as many poor people as those under plug-in OLS with the same budget and achieve the same average poverty gap reduction with \equivBudgetSavingEBOLSPct\ less spending.
\end{abstract}

\bigskip 

{\footnotesize
\textbf{Keywords:} Poverty targeting, Empirical Bayes, Statistical decision theory.

\textbf{JEL:} I32, H53, C44, D81}

\end{titlepage}
\end{spacing}
\hypersetup{pageanchor=true}
\pagenumbering{arabic}
\section{Introduction}

Targeted cash transfers are a central antipoverty instrument in developing
countries, yet the governments that run them rarely observe the incomes of
the households the transfers are meant to reach \cite{banerjee2022social}. Targeting
these transfers well requires knowing who is poor and how far below the
poverty line each household falls. But because widespread informality and limited
administrative data make income or consumption difficult to measure reliably
for every potential beneficiary, governments predict it from information
they can observe. A widely used method, proxy means testing (PMT), predicts
consumption from household characteristics such as housing quality,
education, and asset ownership. Under geographic targeting, transfers
are assigned according to estimated poverty in the area where a household lives
\cite{baker1994poverty,elbers2007poverty}.\footnote{PMT underlies Colombia's
Sisb\'en registry, Pakistan's BISP, and Mexico's Progresa/Oportunidades.
Geographic targeting is used by the Productive Safety Net in Ethiopia,
Challenging the Frontiers of Poverty Reduction in Bangladesh, and Chile
Solidario \cite{barrientos2013social}.} In both cases, the policymaker chooses
recipients and transfer amounts from imperfect estimates of who is poor and
by how much.

Targeting performance depends both on how accurately household income is
predicted and on how the predictions are used to decide who receives a
transfer and how much. Predictions from a basic PMT, for example, miss most
of the poor and misclassify many nonpoor households, with average inclusion
and exclusion error rates of about 37\% and 72\% across nine African
countries \cite{brown2018poor}. Most efforts to reduce these errors have
focused on improving the predictions. More flexible machine learning methods
can predict more accurately, but the gains in targeting are often modest and
vary across settings and performance measures
\cite{baez2019adaptive,areias2022machine,corral2025poverty}. The allocation
step, by contrast, is usually taken as given, with transfers assigned by a
plug-in rule that treats the estimates as if they were exact.

This paper asks how a policymaker should assign transfers when household
income is known only through noisy predictions. I formulate the allocation
step as a statistical decision problem. The policymaker runs an unconditional
cash transfer program, observes an income estimate and its standard error for
each household, and chooses how much each household receives. Transfers
cannot be negative, so the program cannot tax, and total spending cannot
exceed a fixed budget. I evaluate rules by the expected average squared
distance between post-transfer income and the poverty line. This criterion
weights larger remaining poverty gaps more heavily and penalizes
transfers that raise income above the line, allowing for administrative,
social, or political costs of mistargeting
\cite{devereux2017targeting,della2022selective,cameron2014can}. I use it as a
tractable benchmark for targeting quality and also study a one-sided
objective that counts only the squared poverty gaps remaining after
transfers.

I show that the plug-in rule is inadmissible under the quadratic criterion. With
Gaussian prediction errors and conditional mean incomes in a bounded range,
another feasible rule has expected loss no greater than the plug-in rule's
for every income configuration within that range and strictly lower for some.
Because the plug-in rule treats each estimate as exact,
differences in estimated poverty gaps among recipient households pass
directly into differences in transfers, regardless of how precise each
estimate is. I show that this behavior is incompatible with a Bayes rule
under any proper prior on the bounded parameter space, and a complete-class
argument then yields inadmissibility.

To move beyond this negative result, I develop a nonparametric empirical
Bayes targeting rule that pools information across households while
accounting for the precision of each household's income estimate. The rule
approximates an oracle Bayes allocation, which would be optimal if the
distribution of conditional mean incomes across households were known. Under
the quadratic criterion, the oracle computes each household's posterior mean
poverty gap from its income estimate and standard error and then selects the
feasible allocation closest to the vector of posterior mean gaps. The
allocation step is the same as under the plug-in rule, with posterior mean
gaps in place of raw estimates.\footnote{The same approach applies to other policy
problems in which a fixed budget is divided across many units using noisy
estimates, such as health care funding distributed across areas according to
estimated local demand for care \cite{smith2001capitation}, school funding
distributed across districts according to estimated counts of poor children
\cite{zaslavsky2002interactions}, and medical supplies allocated across regions
from demand forecasts \cite{chen2021pandemic}.} Households with the same income estimate can
therefore receive different transfers when their standard errors differ. In
practice, the distribution of conditional mean incomes is unknown, so I
estimate it by nonparametric maximum likelihood.

My main theoretical result is that the empirical Bayes rule approaches
the oracle's performance as the number of households grows. I evaluate
the rule by its Bayes regret, the expected loss incurred under it
beyond the loss under the oracle Bayes allocation, which measures the cost of estimating the
distribution of conditional mean incomes rather than knowing it. This regret
is bounded by the expected average squared difference between the
estimated and oracle posterior mean gaps, even though the constraints make
the allocation a nonsmooth function of those gaps. For \(n\) households,
combining this inequality with the heteroskedastic nonparametric maximum
likelihood bound of \citeA{jiang2020general} gives a regret bound of order
\(n^{-p/(2(p+1))}\) up to logarithmic factors when the distribution of
conditional mean incomes has a finite \(p\)-th moment with \(p\ge 2\). The
same argument converts sharper bounds on posterior mean estimation error into
sharper targeting regret bounds, including a bound of order \(1/n\) up to
logarithmic factors when that distribution has compact support
\cite{JiangZhang2009,soloff2021multivariate}.

In simulations built from household survey data for nine African countries,
the empirical Bayes rule reaches about \poorReachedRatioEBOLSRounded\ times as
many poor people as plug-in ordinary least squares (OLS) under the same budget ceiling. Both rules use
the same out-of-sample consumption predictions from an OLS proxy means test,
so the comparison isolates the allocation step. The empirical Bayes rule also
achieves the same average poverty gap reduction as plug-in OLS with
\equivBudgetSavingEBOLSPct\ less transfer spending. It recovers \gainEBLtwoPct\ of
the reduction in quadratic loss that a policymaker who observed consumption
would attain relative to no transfers, compared with \gainOLSLtwoPct\ for plug-in OLS and \gainMLLtwoPct\ for
a plug-in rule built on XGBoost, a flexible machine learning predictor.
Plug-in OLS concentrates the budget on the households with the largest
predicted gaps, which often partly reflect noise, whereas empirical Bayes
trims those transfers, extends assistance to more poor households, and spends
less on transfers beyond recipients' poverty gaps.

The empirical Bayes approach also applies to the one-sided objective,
which does not penalize transfers that raise households' incomes above the poverty line. Up to normalization, this
objective is the Foster--Greer--Thorbecke index with parameter \(\alpha=2\),
evaluated at conditional mean income in the theoretical analysis. The
perfect-information allocation remains optimal, and the plug-in rule is still
inadmissible when the total budget exceeds the poverty line but can fall
short of the aggregate poverty gap. The oracle Bayes rule, however, is no
longer a projection of the posterior mean gaps. It equalizes the posterior expected
remaining gaps across recipients, so the empirical Bayes rule must estimate
how these gaps vary with the transfer amount. Under bounded support and
strict budget scarcity, I establish a weaker regret bound, of order
\(1/\log n\). In the simulations, the two empirical Bayes rules perform
nearly identically in the seven countries where the quadratic rule exhausts
the budget. Material differences in the two rules' performance arise in Ethiopia and Mali, where the
one-sided rule spends funds that the quadratic rule leaves unallocated and
attains a higher gain under the one-sided criterion.

Under the quadratic loss, the empirical Bayes approach also extends to
programs that restrict transfers to a fixed menu of benefit amounts. The
oracle again selects the feasible allocation closest to the vector of
posterior mean poverty gaps, and the empirical Bayes rule substitutes
estimated posterior means in the same assignment problem. The discrete
feasible set is nonconvex, and assignments can jump between payment levels.
With a fixed menu and a budget proportional to the population, I establish a
regret upper bound of order \(n^{-p/(4(p+1))}\) up to logarithmic factors
under the same finite \(p\)-th moment condition, slower than the
\(n^{-p/(2(p+1))}\) rate of the continuous problem.

In the poverty targeting literature, most work predicts income and then
applies a plug-in allocation rule \cite{banerjee2022social}.
\citeA{follett2023hybrid} bring community information into beneficiary
selection, and \citeA{sahoo2025cost} derive poverty-gap-minimizing transfers
from conditional consumption quantiles and implement them with estimated
quantiles. This paper takes income predictions and their standard errors as
given and studies how sampling uncertainty in the predictions should change
the transfers.

Statistical decision theory has entered policy design mainly
through treatment choice, where the policymaker decides whom to treat, or at
what dose, on the basis of estimated treatment effects
\cite{manski2004statistical, hirano2009asymptotics, stoye2009minimax,
tetenov2012statistical, kitagawa2018should, athey2021policy, ai2024data,
kitagawa2026treatment}, including under budget constraints
\cite{bhattacharya2012inferring, sun2026empirical} and with equality-minded
welfare criteria \cite{kitagawa2021equality}. In the targeting problem, by
contrast, the effect of a transfer on income is known, the uncertainty is
about income itself, and the loss measures squared deviations of post-transfer income from the poverty line. The inadmissibility result relates to the statistical literature on
estimation under restricted parameter spaces \cite{charras1992bayes,
marchand2001improving, hartigan2004uniform, fourdrinier2018shrinkage}, where
the unknown parameter is known to lie in a constrained set and estimators are
compared on that set. In the targeting problem, the constraint
falls on the action space instead, since the budget and nonnegativity
constraints define the set of feasible transfers.

The rule itself builds on the empirical Bayes literature on decisions with
noisy estimates, where these methods have served denoising, ranking, and
selection problems \cite{gu2023invidious, kline2022systemic,
montiel2021empirical, chetty2014measuring, chetty2018impacts,
kwon2026optimal}. \citeA{moon2025optimal} and
\citeA{kline2024discrimination} apply them to downstream decisions.
\citeA{chen2023empirical} studies binary selection rules based on posterior
means, including selection of a fixed number of units, and
\citeA{chen2025compound} choose selection rules by maximizing a Stein-type
unbiased estimate of their welfare. The single-grant case of the fixed-menu
extension is a budget-capped relative of these problems. The regret results draw on the Bayes regret literature for
empirical Bayes estimators \cite{kiefer1956consistency, zhang2009generalized,
JiangZhang2009, Saha, jiang2020general, soloff2021multivariate,
chen2023empirical}.

The paper is organized as follows. Section~\ref{sec:pov_target} introduces
the poverty targeting model, starting from a perfect-information benchmark
and extending it to noisy income estimates. Section~\ref{sec:inadmissible}
establishes the inadmissibility of the plug-in rule. Section~\ref{sec:Bayes}
presents the empirical Bayes framework, characterizes the oracle and feasible
rules, and derives regret bounds. Section~\ref{sec:sim_emp} reports the
simulation evidence. Section~\ref{sec:conclu} concludes.
Appendix~\ref{app:loss_functions} develops the one-sided extension and
Appendix~\ref{app:fixed_grants} the fixed-menu extension.

\section{Poverty Targeting Model}
\label{sec:pov_target}

When household incomes are observed, the optimal allocation raises the
poorest households' incomes to a common post-transfer level at or below the
poverty line and gives nothing to the others. When income
predictions are noisy, however, the policymaker is uncertain about both who should receive
the transfers and how far each household falls below the line. Targeting
therefore becomes a statistical decision problem.

\subsection{Perfect Information}
\label{sec:perfect}

I consider a policymaker who allocates a fixed antipoverty budget across
households through nonnegative transfers to reduce poverty, giving
priority to the households furthest below the poverty line. In this benchmark,
the policymaker observes household incomes without error.\footnote{Measuring household resources is
challenging in practice. Survey-based income and consumption measures are
sensitive to design choices, and poverty status can vary within the year
\cite{beegle2012methods,merfeld2023poverty}. I abstract from these issues
by taking a scalar income measure, defined by a fixed reference period and
accounting protocol, as the latent policy object and studying how noisy
signals of it should be used for targeting.}

\subsubsection{Setup}

The policymaker observes household incomes $y=(y_1,\ldots,y_n)^\top$,
where $y_i\ge 0$ is the realized income of household $i$. Throughout,
income stands for the scalar resource measure that the program targets,
such as per capita income or consumption. The budget $B>0$ is taken as given, its financing is not
modeled, and it may be insufficient to eliminate poverty. Transfers
$\tau=(\tau_1,\ldots,\tau_n)^\top$ satisfy $\tau_i\ge 0$ for all $i$, so
the policymaker cannot tax, and $\sum_{i=1}^n\tau_i\le B$. Let $z>0$
denote a common poverty line. In the benchmark, I treat each household as one
unit and abstract from differences in household size, so household $i$'s
post-transfer income is $y_i+\tau_i$.

\paragraph{Notation.}
Vectors are columns, $\mathbf{1}$ is the $n$-vector of ones, and
$\|\cdot\|$ is the Euclidean norm. For a scalar $x$, $(x)_+:=\max\{x,0\}$,
and for a vector, the positive part is taken coordinatewise. Hats denote
objects estimated from the data. I use uppercase letters for observed
random vectors, such as the noisy signals \(\hat Y\), and lowercase letters
for their realizations. Latent household quantities are also written in
lowercase.

\subsubsection{Targeting Objective}

The policymaker ranks feasible transfer schedules $\tau$ given the pre-transfer income profile $y$, the poverty line $z$, and the budget $B$. The budget constraint incorporates fiscal scarcity, since a dollar transferred to one household is a dollar unavailable for others. The main analysis uses the quadratic loss to account for remaining poverty gaps and nonfiscal costs of mistargeting. For household $i$,
\begin{equation}
\label{eq:L2_here}
\ell_i(\tau_i,y_i) := (z-y_i-\tau_i)^2 = (z-y_i-\tau_i)_+^2 + (y_i+\tau_i-z)_+^2 .
\end{equation}

The first term is the squared poverty gap that remains, and on its own, it is the one-sided loss $\ell_i^{+}(\tau_i,y_i):=(z-y_i-\tau_i)_+^2$. Divided by $z^2$, it is household $i$'s contribution to the Foster--Greer--Thorbecke (FGT) index with parameter $\alpha=2$ \cite{foster1984class}, a standard measure of poverty severity in empirical work \cite{ravallion2016economics}. The index is convex in the transfer, places greater weight on larger shortfalls, and satisfies focus, monotonicity, and the transfer principle.\footnote{In the FGT class, focus follows from $(\cdot)_+$, monotonicity holds for any positive exponent, and the transfer principle for any exponent above one; see \citeA{Sen} and \citeA{foster1984class}. In the perfect-information problem, the optimal allocation is the same for every exponent above one, so the choice of the squared gap is without loss for that benchmark.}

The second term is a penalty on post-transfer income above the poverty line, a reduced-form representation of the nonfiscal targeting costs reviewed by \citeA{devereux2017targeting}. These costs are distinct from the opportunity cost of funds, which the budget constraint already captures. The loss is a program objective for comparing transfer schedules and not a social welfare function. Pre-transfer incomes are fixed and transfers are nonnegative, so no feasible allocation lowers any household's income below its pre-transfer level. The second term therefore only discourages transfers that push a household above the poverty line or further above it.\footnote{\interlinepenalty=10000 A household's being above the poverty line before the program is not itself a targeting error. The loss it contributes when it receives nothing, $(y_i-z)^2$, is the same under every transfer schedule. Subtracting this baseline leaves comparisons between allocations and between decision rules unchanged. What matters for targeting is the additional loss caused by a transfer. Larger payments incur larger penalties, especially when the recipient is already far above the poverty line.}

In Indonesia, \citeA{cameron2014can} find that leakage of a cash transfer to ineligible households is associated with lower participation in community groups and higher crime, while undercoverage of eligible households shows no significant association with either outcome. This asymmetry matches the structure of \eqref{eq:L2_here}, which prices nonfiscal costs on transfers above the poverty line but not on gaps left unclosed. Beneficiary selection more broadly affects satisfaction, perceived legitimacy, and social cohesion \cite{alatas2012targeting,premand2018efficiency,della2022selective}. This evidence motivates pricing nonfiscal costs of mistargeting but does not identify their functional form. The quadratic loss is a tractable benchmark that delivers the paper's main ideas in their simplest setting.

Under perfect information, the allocation that minimizes the quadratic loss gives no household more than its poverty gap, and it also minimizes the one-sided loss. Decisions based on noisy estimates can instead exclude households that would receive transfers under perfect information or include nonpoor households. Both errors leave more deprivation unaddressed for a given budget, but only the quadratic loss penalizes a transfer to a nonpoor household beyond the budget it absorbs, so the two losses can rank transfer rules differently. In the main text, I focus on the quadratic loss, for which I derive an empirical Bayes rule whose Bayes regret converges to zero at a polynomial rate. In Appendix~\ref{app:loss_functions}, I consider an extension to the one-sided loss and show that Bayes regret still converges to zero, but at a slower rate.

\subsubsection{Optimal Allocation}
\label{sec:full_info_optimal_alloc}

Let $\mathcal A:=\{\tau\in\mathbb R_+^n:\mathbf{1}^\top\tau\le B\}$ denote the set of feasible transfer schedules, and let $P_{\mathcal A}(u):=\arg\min_{\tau\in\mathcal A}\|u-\tau\|^2$ denote the Euclidean projection onto $\mathcal A$. Under perfect information, the policymaker chooses transfers to minimize the average program loss,
\begin{equation}
\label{eq:theo_problem}
\tau^{\PIlabel}(y)
:=
\arg\min_{\tau\in\mathcal A}\ \frac{1}{n}\sum_{i=1}^n \ell_i(\tau_i,y_i).
\end{equation}
In many applications, the budget is not large enough to eliminate all pre-transfer poverty gaps, that is, $B<\sum_{i=1}^n (z-y_i)_+$, so the policymaker must prioritize who receives transfers and in what amounts.

Proposition~\ref{prop:full_info} gives three equivalent representations of the perfect-information benchmark. The leveling-up allocation follows \citeA{bourguignon1990poverty,bourguignon1997discontinuous}, and the Karush--Kuhn--Tucker (KKT) conditions give its threshold form. The projection representation is the one used in the rest of the paper, where the poverty gaps are replaced by estimated or posterior mean gaps. Appendix~\ref{app:proofs} collects the proofs of the main-text propositions and the supporting lemmas.

\begin{proposition}
\proptitle{\hyperlink{proof:full_info}{Perfect-Information Allocation}}
\label{prop:full_info} \hypertarget{prop:full_info}{}
Problem \eqref{eq:theo_problem} has a unique solution,
\[
\tau^{\PIlabel}(y)=P_{\mathcal A}(z\mathbf 1-y).
\]
Its household-level transfers are
\[
\tau_i^{\PIlabel}(y)=\bigl(z-y_i-\lambda^{\PIlabel}\bigr)_+,\qquad i=1,\ldots,n,
\]
where $\lambda^{\PIlabel}\ge 0$. If $\sum_{i=1}^n (z-y_i)_+\le B$, then $\lambda^{\PIlabel}=0$. Otherwise, $\lambda^{\PIlabel}>0$ is the unique solution of $\sum_{i=1}^n \bigl(z-y_i-\lambda^{\PIlabel}\bigr)_+=B$.
This allocation raises all households with incomes below $z-\lambda^{\PIlabel}$ to this common income level and gives nothing to the others.
\end{proposition}

The optimal rule applies a common cutoff $\lambda^{\PIlabel}$ to households' poverty gaps.\footnote{Cutoffs are expressed in income units. For the
average-loss objective with total budget $B$, the budget multiplier equals
$2\lambda^{\PIlabel}/n$.} Households with gaps at or below the cutoff receive nothing, while those with larger gaps receive the excess. All recipients therefore reach the same post-transfer income, $z-\lambda^{\PIlabel}$. This allocation levels the bottom of the income distribution upward. It raises the poorest household toward the income of the next poorest, then raises both together, and continues until the budget is exhausted or the poverty line is reached.

The budget determines how high this common income level can rise. If it suffices to close every poverty gap, $\lambda^{\PIlabel}=0$, and every poor household reaches the poverty line. Otherwise, the cutoff is positive and measures the gap left to each recipient. Increasing the budget lowers the cutoff and raises the common income level until the poverty line is reached.

The projection representation expresses the same allocation as the feasible schedule closest to $z\mathbf{1}-y$ in Euclidean distance. Setting transfers equal to this vector would bring every household exactly to the poverty line, but the schedule is infeasible whenever some household is above the line or closing every gap costs more than $B$. Projection onto $\mathcal A$ sets to zero the entries of $z\mathbf{1}-y$ below the cutoff and subtracts the cutoff from the rest.

Figure~\ref{fig:projection} shows this geometry for two households. Under the incomes $y$, the households are similarly poor, and the budget is large enough to close the difference between them and then raise both together. Each receives a transfer, and both are left with the same remaining gap $\lambda^{\PIlabel}$. Under $y'$, household 1 is so much poorer that even the whole budget would leave it below household 2. It therefore receives everything, the projection lands on the vertex $(B,0)$, and household 2 receives nothing despite being poor.

\subsection{Noisy Measurements}
\label{sec:noisy}

The allocation of Proposition~\ref{prop:full_info} requires the income vector $y$, which the policymaker typically does not observe. What is available instead is a prediction of each household's income from observable characteristics $W_i$, such as assets, housing quality, or location, together with its standard error. The predictions come from a model estimated on separate survey data, as in PMT or geographic targeting. I take them as given and study how the policymaker should use them to allocate the budget.

The policymaker observes the prediction vector $\hat Y=(\hat Y_1,\ldots,\hat Y_n)^\top$ and its standard errors $\sigma=(\sigma_1,\ldots,\sigma_n)^\top$. Each $\hat Y_i$ is a noisy estimate of the conditional mean income $\mu_i=\E[y_i\mid W_i]$, the part of household $i$'s income that its characteristics predict. I assume that the estimates are normally distributed around $\mu_i$, with known variances bounded away from zero and infinity.\footnote{The model abstracts from systematic prediction bias, which can arise, for example, when the predictions and the income measure refer to different seasons. For allocations obtained by projecting predicted gaps, bias changes transfers by at most its Euclidean norm. A bias common to all households leaves transfers unchanged if the budget binds both with and without it. The simulations in Section~\ref{sec:sim_emp} do not impose the signal model and still show gains from empirical Bayes over the plug-in rule.}

\vspace{0.25cm}
\begin{assumption}[Normality with Bounded Variance]
\label{assump:normality}
\hypertarget{assump:normality}{}
Conditional on $(\mu,\sigma)$, the income estimates $\hat Y$ are independent across households and satisfy
\begin{equation*}
\hat Y_i \mid \mu_i,\sigma_i \overset{\mathrm{ind.}}{\sim} \mathcal{N}(\mu_i,\sigma_i^2),
\qquad i=1,\ldots,n,
\end{equation*}
where $\mu_i=\E[y_i\mid W_i]\in\mathbb{R}_+$ and $\sigma_i^2\in\mathbb{R}_{++}$ is known and treated as fixed. Furthermore, for all $i=1,\ldots,n$, $0<\sigma_{\min}^2\le \sigma_i^2\le \sigma_{\max}^2$.
\end{assumption}
\vspace{0.25cm}

Remarks~\ref{remark:geo} and~\ref{remark:pmt} discuss why normality is a reasonable approximation under geographic targeting and PMT.

\begin{titledRemark}{Geographic Targeting} \label{remark:geo}
Geographic targeting allocates the budget across areas, such as villages or municipalities, rather than across households. In the benchmark case, every household in area $j$ receives the same transfer $\tau_j$. Let $N_j$ denote the number of households in area $j$ and $\mu_j$ their mean income. The policymaker chooses $\tau_1,\ldots,\tau_J$ to minimize $\sum_j N_j(z-\mu_j-\tau_j)^2$ subject to $\tau_j\ge 0$ and $\sum_j N_j\tau_j\le B$. The weights appear in both the loss and the budget because a transfer to a larger area reaches more households and costs more. With areas in place of households and these weights, the problem has the same structure as the household-level one. In particular, the optimal allocation keeps the threshold form of Proposition~\ref{prop:full_info}, $\tau_j=(z-\mu_j-\lambda)_+$, where $\lambda=0$ if $\sum_j N_j(z-\mu_j)_+\le B$ and otherwise $\lambda>0$ is such that $\sum_j N_j\tau_j=B$.

The policymaker does not observe $\mu_j$ but estimates it from a survey of the area. If $\hat Y_j$ is the survey mean, the central limit theorem makes it approximately normal around $\mu_j$ for large area samples under standard sampling conditions, and independent sampling across areas makes the estimates independent. Small-area estimates that combine survey and census data, as in \citeA{elbers2003micro}, rely on a shared fitted model, so their errors can be correlated across areas.
\end{titledRemark}

\begin{titledRemark}{Proxy Means Testing} \label{remark:pmt}
Under PMT, the policymaker estimates an income regression on an auxiliary survey $\{(\tilde y_j,\tilde W_j)\}_{j=1}^m$ and applies the estimated coefficients to the target households,
\begin{equation}
\label{eq:ols}
\tilde y_j=\tilde W_j'\beta+\tilde\epsilon_j,\qquad \hat Y_i=W_i'\hat\beta,
\end{equation}
where $\hat\beta$ is the OLS estimator. If the covariates take finitely many values and the regression is saturated, $\hat Y_i$ is the survey mean income of households with the same observable type as $i$. Targeting then proceeds by type, the normal approximation of Remark~\ref{remark:geo} applies to each type mean, and the means of different types are independent because they use disjoint survey observations.

In general, the approximation comes from the asymptotic normality of $\hat\beta$. If the linear model is correctly specified for the population from which both samples are drawn, then $\mu_i=W_i'\beta$, and under standard moment and rank conditions, $\hat Y_i$ is approximately normal around $\mu_i$ with variance $\sigma_i^2=W_i'\operatorname{Var}(\hat\beta\mid \tilde W_1,\ldots,\tilde W_m)W_i$. This variance measures the uncertainty in the estimated conditional mean. Because every prediction shares the estimation error in $\hat\beta$, prediction errors are correlated across households, and the correlations do not vanish as the auxiliary survey grows, since variances and covariances are both of order $1/m$. Independence in Assumption~\ref{assump:normality} is therefore a simplification rather than a consequence of OLS asymptotics.
\end{titledRemark}

The unknown state is the vector of conditional mean incomes $\mu=(\mu_1,\ldots,\mu_n)^\top$, with parameter space $\mathcal M:=\mathbb R_+^n$. After observing $\hat y$, the policymaker chooses a transfer schedule $\tau\in\mathcal A$. The program objective, however, is defined at realized incomes, which differ from the state by the residuals $\varepsilon_i:=y_i-\mu_i$. Assume that these residuals have finite second moments and mean zero conditional on the policymaker's information. Remark~\ref{remark:losses} shows that, under quadratic loss, the expected loss at realized incomes then differs from that at conditional mean incomes by a term that does not depend on the transfer rule, so the two criteria rank rules identically. I therefore evaluate transfers at state $\mu$ using the loss
\[
L(\tau,\mu):=\frac{1}{n}\sum_{i=1}^n\bigl(z-\mu_i-\tau_i\bigr)^2.
\]

A statistical decision rule is a measurable mapping $\delta:\mathbb R^n\to\mathcal A$ that assigns a feasible transfer schedule $\delta(\hat y)$ to each observed prediction vector $\hat y$. Unlike in a standard estimation problem, the action space differs from the parameter space, since the policymaker chooses transfers in $\mathcal A$ while the unknown state is a vector of incomes in $\mathcal M$. Let $\mathcal D$ denote the class of nonrandomized rules, which is essentially complete because the loss and $\mathcal A$ are convex \cite{ferguson2014mathematical}. The frequentist risk of $\delta\in\mathcal D$ at state $\mu$ is
\[
R(\delta,\mu):=\E_{\mu}\bigl[L\bigl(\delta(\hat Y),\mu\bigr)\bigr],
\]
where the expectation is over the predictions $\hat Y$ under state $\mu$.

\begin{titledRemark}{Realized versus Predictable Income}
\label{remark:losses}
Fix $(\mu,\sigma)$, and let $\E_{\mu,\varepsilon}$ denote the expectation over prediction noise and residual income variation at that state. Let $\mathcal I$ denote the pre-transfer information generated by the target-household covariates and the independent auxiliary survey used to estimate the prediction rule. The objects $\hat Y$, $\sigma$, $\mu$, and hence $\delta(\hat Y)$, are $\mathcal I$-measurable. At each state, the assumption on the residuals gives $\E_{\mu,\varepsilon}[\varepsilon_i\mid\mathcal I]=0$ and $\E_{\mu,\varepsilon}[\varepsilon_i^2]<\infty$, where $\varepsilon_i=y_i-\mu_i$.

Since the cross term vanishes after conditioning on $\mathcal I$,
\[
\E_{\mu,\varepsilon}\!\left[(z-y_i-\delta_i(\hat Y))^2\right]
=
\E_\mu\!\left[(z-\mu_i-\delta_i(\hat Y))^2\right]
+
\E_{\mu,\varepsilon}[\varepsilon_i^2].
\]
Averaging over households yields
\begin{equation}
\label{eq:risk_equiv}
\E_{\mu,\varepsilon}\!\left[L\bigl(\delta(\hat Y),y\bigr)\right]
=
R(\delta,\mu)
+
\frac{1}{n}\sum_{i=1}^n \E_{\mu,\varepsilon}[\varepsilon_i^2].
\end{equation}
The additional term in \eqref{eq:risk_equiv} does not depend on the transfer rule. Thus, expected loss at realized incomes and $R(\delta,\mu)$ rank transfer rules identically at every state.\footnote{For nonquadratic losses, averaging over residual income variation generally produces a correction that depends on the transfer, and expected loss at realized and at conditional mean incomes can then rank rules differently.}
\end{titledRemark}

\section{Inadmissibility of Plug-In Rule}
\label{sec:inadmissible}

The plug-in rule allocates transfers as if the estimated poverty gaps were exact. This section shows that, on a compact set of conditional mean incomes, this rule is inadmissible under quadratic loss. Some other feasible rule has risk no greater at every state in that set and strictly lower risk at some state.

\subsection{Poverty Gaps and the Plug-In Rule}
\label{subsec:equivalence}
\label{sec:plugin}

Expressing the problem in poverty gaps gives a heteroskedastic Gaussian location model. Define the observed poverty gap vector $X:=z\mathbf 1-\hat Y$ and the predictable poverty gap vector $\theta:=z\mathbf 1-\mu$.\footnote{The predictable gap $\theta_i$ differs from the realized gap $z-y_i$ by the residual $\varepsilon_i$. By Remark~\ref{remark:losses}, this difference does not affect risk comparisons.} Under Assumption~\ref{assump:normality},
\begin{equation}
\label{eq:gaussian_gap_model}
X_i\mid\theta,\sigma \overset{\mathrm{ind.}}{\sim} N(\theta_i,\sigma_i^2),\qquad i=1,\ldots,n.
\end{equation}
Since conditional mean incomes are nonnegative, the parameter space becomes $\Theta:=(-\infty,z]^n$. For a household with predictable income below the poverty line, $\theta_i$ is its shortfall from the line, and for one above the line, $-\theta_i$ is its excess over it.

Transfers remain in $\mathcal A$, and the loss becomes $L(\tau,\mu)=\frac{1}{n}\sum_{i=1}^n(\theta_i-\tau_i)^2$, the average squared difference between transfers and gaps. Since $\hat Y=z\mathbf 1-X$, observing the gap vector is equivalent to observing the income predictions. Every rule $\delta\in\mathcal D$ therefore has a counterpart $x\mapsto\delta(z\mathbf 1-x)$ in the gap formulation that assigns exactly the same transfers. The two rules have identical risk at the respective states $\mu$ and $\theta=z\mathbf 1-\mu$. The remainder of this section works in the gap formulation.

A plug-in rule replaces unknown parameters by their estimates when choosing an action \cite{hirano2019statistical}. If the predictable gap vector $\theta$ were known, the optimal allocation would be $P_{\mathcal A}(\theta)$ by Proposition~\ref{prop:full_info}. The plug-in rule applies the same projection to the observed gap vector $X$, treating it as if it were the true gap vector, $\tau^{\pluglabel}(X):=P_{\mathcal A}(X)$. Like the perfect-information allocation, it has a threshold form, $\tau_i^{\pluglabel}(X)=(X_i-\lambda(X))_+$ for $i=1,\ldots,n$, where $\lambda(X)=0$ if $\mathbf 1^\top X_+\le B$ and otherwise $\lambda(X)>0$ solves $\sum_{i=1}^n(X_i-\lambda(X))_+=B$. Households whose predicted income exceeds the poverty line have $X_i<0$ and receive nothing. The cutoff now depends on the data, and it is positive exactly when the observed poverty gaps cost more than the budget.

When the observed gaps cost more than the budget, the plug-in rule spends all of it, and its threshold form determines how the budget is divided. Each recipient receives a transfer equal to its observed gap minus the common cutoff $\lambda(X)$, so for any two recipients $i$ and $j$,
\[
\tau_i^{\pluglabel}(X)-\tau_j^{\pluglabel}(X)=X_i-X_j.
\]
Differences in observed gaps pass through to transfers one for one: a recipient whose gap is estimated to be one dollar larger than another's receives one dollar more. The rule does not use the standard errors $\sigma$, so this holds whether the gaps are estimated precisely or not.

\subsection{Bayes Rules}
\label{sec:bayes_risk_rules}

The inadmissibility argument compares the plug-in rule with Bayes rules, which minimize expected risk under a prior over the gap vector. Consider a proper prior $\pi$ on $\Theta$ with finite second moment, $\int_\Theta\|\theta\|^2\,d\pi(\theta)<\infty$, with the noise levels $\sigma$ known and held fixed. The Bayes risk of a rule $\delta\in\mathcal D$ is the expectation of its risk when $\theta$ is drawn from $\pi$, $r(\pi,\delta):=\int_\Theta R(\delta,\theta)\,d\pi(\theta)$. A rule that minimizes Bayes risk is a Bayes rule \cite{berger2013statistical}. Given an observation $x$, the prior and the Gaussian model \eqref{eq:gaussian_gap_model} determine the posterior distribution of the gap vector. Under quadratic loss, its mean, $m_\pi(x):=\E_\pi[\theta\mid X=x]$, determines the optimal allocation.

\begin{lemma}
\proptitle{\hyperlink{proof:bayes_rules}{Bayes rules}}
\label{lem:bayes_rules} \hypertarget{lem:bayes_rules}{}
Under Assumption~\ref{assump:normality}, for a proper prior $\pi$ on $\Theta$ with finite second moment, the rule $\delta_\pi(x):=P_{\mathcal A}(m_\pi(x))$ is a Bayes rule in $\mathcal D$ and uniquely minimizes the posterior expected loss at every $x\in\mathbb R^n$. Every Bayes rule in $\mathcal D$ agrees with $\delta_\pi$ for Lebesgue-almost every $x$.
\end{lemma}

The lemma keeps the projection step of the plug-in rule and changes its input, from the observed gaps $x$ to the posterior mean gaps $m_\pi(x)$. The reason is that, under quadratic loss, the posterior expected loss equals the loss evaluated at these means plus the average posterior variance of the gaps across the $n$ households. The variance term does not depend on the transfers, so it cannot affect which feasible allocation minimizes loss. By Proposition~\ref{prop:full_info}, the allocation again has the cutoff form $\delta_{\pi,i}(x)=(m_{\pi,i}(x)-\lambda_\pi(x))_+$, with $\lambda_\pi(x)\ge 0$ determined by the same budget condition applied to $m_\pi(x)$. Consequently, for any two recipients $i$ and $j$, differences in posterior mean gaps pass through to transfers one for one, $\delta_{\pi,i}(x)-\delta_{\pi,j}(x)=m_{\pi,i}(x)-m_{\pi,j}(x)$.

\subsection{Dominance Results}
\label{sec:dominance}

So far, the only restriction on a household's predictable income is that it be nonnegative. For the inadmissibility result, I also impose a finite upper bound, which can be chosen arbitrarily large. With an upper bound $z+M$, where $M>0$, predictable incomes in $[0,z+M]$ correspond to gaps in $[-M,z]$, giving the compact parameter space in Assumption~\ref{assump:bounded}.

\vspace{0.25cm}
\begin{assumption}[Bounded parameter space]
\label{assump:bounded}
\hypertarget{assump:bounded}{}
Let \(M\in(0,\infty)\) be fixed, and define the compact parameter space $\Theta_M:=[-M,z]^n \subset \Theta$.
\end{assumption}
\vspace{0.25cm}

The next proposition shows that, on this parameter space, the plug-in rule is dominated by another feasible rule.

\begin{proposition}
\proptitle{\hyperlink{proof:plugin_inadmissible}{Inadmissibility of the plug-in rule}}
\label{prop:plugin_inadmissible} \hypertarget{prop:plugin_inadmissible}{}
Assume $n\ge 2$. Under Assumptions~\ref{assump:normality} and \ref{assump:bounded}, the plug-in rule $\tau^{\pluglabel}(x)=P_{\mathcal A}(x)$ is inadmissible in the class of measurable nonrandomized decision rules $\mathcal D$. Equivalently, there exists a rule $\tilde\delta\in\mathcal D$ such that
\[
R(\tilde\delta,\theta)\le R(\tau^{\pluglabel},\theta)
\qquad\text{for all }\theta\in\Theta_M,
\]
with strict inequality for at least one $\theta\in\Theta_M$.
\end{proposition}

The proposition identifies an avoidable cost of the plug-in rule. With the same predictions and budget, another feasible rule lowers risk at some states without raising it anywhere on $\Theta_M$. The improvement comes from using the available information differently.

The proof asks whether any prior over the gap vector makes the plug-in allocation a Bayes rule. On the compact set $\Theta_M$, a complete class theorem makes this a necessary condition for admissibility \cite{wald1950statistical,berger2013statistical}, and Lemma~\ref{lem:compact_complete_class} verifies its conditions for the present problem. By Lemma~\ref{lem:bayes_rules}, a prior $\pi$ can do so only if projecting its posterior mean gaps reproduces the plug-in allocation. Among recipients, the plug-in rule preserves differences in observed gaps, while a Bayes rule preserves differences in posterior mean gaps. On an open region where the budget binds and households $i$ and $j$ both receive transfers, which exists by Lemma~\ref{lem:active_set_affine_regions}, the two allocations can agree only if
\[
m_{\pi,i}(x)-m_{\pi,j}(x)=x_i-x_j.
\]

Under the Gaussian model, posterior means are real analytic functions of the observations, so an identity that holds on an open region holds on the whole sample space \cite{krantz2002primer}. But a prior supported on $\Theta_M$ keeps each posterior mean gap between $-M$ and $z$, so the left side is bounded by $M+z$ in absolute value, while $x_i-x_j$ is not. Lemma~\ref{lem:plugin_not_bayes} concludes that no proper prior on $\Theta_M$ reproduces the plug-in allocation, and the rule is inadmissible.\footnote{The lower bound $-M$ makes the contradiction immediate by bounding posterior gap differences, but it is not essential to the non-Bayes conclusion. With $p_\pi$ the marginal density of $X$, the displayed identity and Tweedie's formula \cite{efron2011tweedie} give $\sigma_i^2\partial_i\log p_\pi=\sigma_j^2\partial_j\log p_\pi$ for every $x$, which forces the positive density $p_\pi$ to be constant along a fixed direction and contradicts its integrability.}

\begin{titledRemark}{Relation to Other Inadmissibility Results}
\label{remark:inadmissibility_lit}
The James--Stein theorem provides an explicit estimator that improves on the observation vector under squared error loss in the homoskedastic Gaussian location model when $n\ge 3$ \cite{stein1956inadmissibility,james1961estimation}, and \citeA{efron1973stein} give this shrinkage an empirical Bayes interpretation. In the targeting problem, estimates must be converted into transfers that satisfy the budget and nonnegativity constraints. Lower mean squared error in estimating poverty gaps does not automatically imply lower expected loss for the resulting allocation. Projection changes transfer amounts and can change which households receive them, so a dominance argument must be made for the allocation rules themselves. Proposition~\ref{prop:plugin_inadmissible} establishes that an improving feasible rule exists without constructing it.

Related work studies Bayes optimality and admissibility in restricted parameter spaces \cite{charras1991bayes,fourdrinier2018shrinkage}. In particular, \citeA{charras1991bayes} give conditions under which estimators that take values on the boundary of a convex parameter space are not Bayes and hence are inadmissible. That argument does not apply here, because the parameter space $\Theta_M$ and the feasible set $\mathcal A$ impose different restrictions. Even when posterior mean gaps lie in the interior of $\Theta_M$, their projection can lie on the boundary of $\mathcal A$, so exhausting the budget or assigning zero transfers is compatible with Bayes optimality.
\end{titledRemark}

\section{Empirical Bayes Rules}
\label{sec:Bayes}

Section~\ref{sec:inadmissible} evaluated transfer rules at each vector of
conditional mean incomes and established the plug-in rule's inadmissibility
without constructing an improving rule. I now develop a feasible alternative
with its own performance guarantee using an empirical Bayes approach. I model
the coordinates of \(\mu\) as draws from a common but unknown distribution
\(G\) and estimate it from the income predictions. The benchmark is the
oracle Bayes rule, the Bayes rule of Lemma~\ref{lem:bayes_rules} when \(G\)
is known. The empirical Bayes rule projects estimates of the oracle's
posterior mean gaps instead. Its Bayes regret, its expected loss beyond the
oracle's under this model, is bounded by the mean squared difference between
the estimated and the oracle posterior mean gaps.

\subsection{Prior Distribution}

Throughout this section, I condition on the noise levels
\(\sigma=(\sigma_1,\ldots,\sigma_n)^\top\) and treat them as known, and I
assume that, given \(\sigma\), the conditional mean incomes are independent
draws from a common distribution \(G\).
\begin{assumption}[Cross-sectional prior]
\label{assump:prior}
Conditional on \(\sigma=(\sigma_1,\ldots,\sigma_n)^\top\), the conditional mean incomes satisfy
\[
\mu_1,\ldots,\mu_n \mid \sigma \stackrel{\mathrm{iid}}{\sim} G,
\qquad
\int u^2\,dG(u)<\infty.
\]
\end{assumption}

The second-moment condition and the boundedness of \(\mathcal A\) ensure
that every rule in \(\mathcal D\) has finite Bayes risk. The assumption also
imposes precision independence, since the distribution of \(\mu_i\) does not
vary with \(\sigma_i\). This restriction is common in heteroskedastic
empirical Bayes models, and Remark~\ref{remark:independence_regret}
discusses how to relax it with the conditional location-scale empirical
Bayes (CLOSE) approach of \citeA{chen2023empirical}.

\subsection{Oracle Bayes Rule}

The oracle Bayes rule, available only when \(G\) is known, is the
benchmark for targeting under uncertainty and the object that the
empirical Bayes rule approximates. For each household \(i\), define the
posterior mean poverty gap by
\(g_{G,i}(\hat y_i,\sigma_i):=\E_G[z-\mu_i\mid \hat y_i,\sigma_i]\), and let
\(g_G(\hat y,\sigma):=(g_{G,1}(\hat y_1,\sigma_1),\ldots,g_{G,n}(\hat y_n,\sigma_n))^\top\).
Under Assumption~\ref{assump:prior}, \(g_G(\hat y,\sigma)\) is the posterior
mean of the gap vector, so by Lemma~\ref{lem:bayes_rules}, the oracle applies
the perfect-information allocation of Proposition~\ref{prop:full_info} to
these posterior mean gaps.

\begin{proposition}
\proptitle{\hyperlink{proof:oracle_bayes_rule}{Oracle Bayes Rule}}
\label{prop:oracle_bayes_rule}
\hypertarget{prop:oracle_bayes_rule}{}
Under Assumptions~\ref{assump:normality} and~\ref{assump:prior}, the oracle Bayes rule is
\[
\tau_G^\star(\hat y)=P_{\mathcal A}\bigl(g_G(\hat y,\sigma)\bigr).
\]
Equivalently, \(\tau_{G,i}^\star(\hat y)=\bigl(g_{G,i}(\hat y_i,\sigma_i)-\lambda_G(\hat y)\bigr)_+\) for \(i=1,\ldots,n\), where \(\lambda_G(\hat y)=0\) if \(\mathbf 1^\top g_G(\hat y,\sigma)_+\le B\) and otherwise \(\lambda_G(\hat y)>0\) solves \(\mathbf 1^\top\tau_G^\star(\hat y)=B\). The rule uniquely minimizes the posterior expected loss at every \(\hat y\), and every Bayes rule in \(\mathcal D\) agrees with it for Lebesgue-almost every \(\hat y\).
\end{proposition}

Proposition~\ref{prop:oracle_bayes_rule} differs from
Proposition~\ref{prop:full_info} only in what is projected. The
full-information rule projects the vector of poverty gaps onto the
feasible transfer set, and the oracle Bayes rule projects the vector of
their posterior means. This is why shrinkage matters for targeting under
uncertainty. The oracle replaces each observed gap with its posterior mean
under \(G\) before allocating, so the estimated poverty gaps of households
with noisy or unusually extreme signals are pulled toward values that are
more plausible given the cross section. Relative to the plug-in rule, the
oracle is therefore less willing to concentrate the budget on a few
households that look especially poor in the noisy data alone, and it
typically reaches more households with smaller transfers.

\subsection{Estimating the Prior Distribution}
\label{sec:npmle}

To make the oracle rule feasible, I estimate \(G\) by the nonparametric
maximum likelihood estimator (NPMLE) of \citeA{kiefer1956consistency},
which maximizes the marginal likelihood of the observed signals under the
Gaussian mixture model induced by Assumptions~\ref{assump:normality}
and~\ref{assump:prior} without imposing a parametric form on \(G\). The
shape of \(G\) determines how noisy signals are translated into posterior
mean poverty gaps and hence into transfers. A parametric family would
restrict that shape in advance, whereas the NPMLE learns it from the cross
section, so the feasible rule can aim at the oracle for the actual \(G\)
rather than at the best rule within a prespecified family
\cite{JiangZhang2009}.

Let \(\mathcal G\) denote the class of all probability distributions on \(\mathbb R\).
For \(H\in\mathcal G\), define
\[
f_{H,\sigma_i}(\hat y_i)
=
\int \frac{1}{\sigma_i}\varphi\!\left(\frac{\hat y_i-\mu}{\sigma_i}\right)\,dH(\mu),
\]
where \(\varphi\) is the standard normal density. The NPMLE is any maximizer of the marginal
log-likelihood,
\[
\hat G \in \arg\max_{H\in\mathcal G}\,\frac{1}{n}\sum_{i=1}^n \log f_{H,\sigma_i}(\hat y_i).
\]

Two features of the NPMLE matter for computation. First, the optimization
ranges over an infinite-dimensional class, but the log-likelihood is
concave in \(H\), so the problem can be solved by convex programming over
probability weights on a fixed grid of income values
\cite{koenker2014convex}. Second, a maximizer can be chosen to be discrete
with at most \(n\) support points \cite{koenker2017rebayes}, whether or not
\(G\) itself is discrete.

\subsection{Feasible Empirical Bayes Rule}

Given \(\hat G\), the feasible rule is obtained by substituting \(\hat G\) for \(G\) in
the oracle construction. For each household \(i\), define the feasible posterior mean poverty
gap by $\hat g_i(\hat y_i,\sigma_i):=\E_{\hat G}[z-\mu_i\mid \hat y_i,\sigma_i]$, and let $\hat g(\hat y,\sigma):=(\hat g_1(\hat y_1,\sigma_1),\ldots,\hat g_n(\hat y_n,\sigma_n))^\top$. The feasible empirical Bayes rule is then
\[
\hat\tau^{\EBlabel}(\hat y)=P_{\mathcal A}\bigl(\hat g(\hat y,\sigma)\bigr).
\]
As in Proposition~\ref{prop:oracle_bayes_rule}, this rule admits an equivalent thresholding
representation, with \(\E_{\hat G}\) in place of \(\E_G\) and a data-dependent cutoff
\(\hat\lambda^{\EBlabel}(\hat y)\). Both rules apply shrinkage and then projection. Under the oracle, household
\(i\)'s posterior mean gap depends only on its own signal and noise level,
and other households affect its transfer only through the common budget
cutoff. Under the feasible rule, \(\hat G\) is estimated from all \(n\)
signals, so other households also enter through the shrinkage step.

\subsection{Statistical Properties}
\label{sec:bayes_regret_results}

The Bayes regret of the feasible rule, its expected loss in excess of the
oracle's, is bounded by the mean squared error of the estimated
posterior mean gaps relative to the oracle's. Under two assumptions taken
from \citeA{jiang2020general}, his finite-sample oracle inequality bounds
that error, so the feasible rule inherits the convergence rate of NPMLE
posterior means.

\subsubsection{Assumptions}

The maximizer of the marginal likelihood over all distributions is not
available in closed form, so the first assumption allows \(\hat G\) to be
an approximate maximizer of the kind that numerical implementations of
the NPMLE return.

\begin{assumption}[Approximate NPMLE]
\label{assump:approx}
Let \(f_{G,\sigma_i}\) be the marginal density defined in Section~\ref{sec:npmle}. The
computed estimator \(\hat G\) satisfies
\[
\prod_{i=1}^{n} f_{\hat G,\sigma_i}(\hat y_i)
\ge q_n \sup_{\tilde G\in\mathcal G}\prod_{i=1}^{n} f_{\tilde G,\sigma_i}(\hat y_i),
\qquad
q_n=\min\{e\sqrt{2\pi}/n^2,1\}.
\]
\end{assumption}

The assumption requires the likelihood at \(\hat G\) to be at least a
fraction \(q_n\) of the maximum, which is the tolerance in equation (2.1)
of \citeA{jiang2020general}, under which his Theorem 1 holds. Since
\(q_n\) is of order \(n^{-2}\), the average log-likelihood at \(\hat G\)
may fall below the maximum by a term of order \((\log n)/n\).

The second assumption controls the tails of \(G\). A finite absolute
moment of order \(p\) limits how much mass \(G\) places in its tails, and
a larger \(p\) limits it more tightly.

\begin{assumption}[Moment condition]
\label{assump:moment}
Let
\(
\mathcal M_p(G):=\left(\int |u|^p\,dG(u)\right)^{1/p}
\)
denote the \(p\)-th root of the \(p\)-th absolute moment of the prior \(G\). Fix \(p\ge 2\) such that \(\mathcal M_p(G)<\infty\).
\end{assumption}

Assumption~\ref{assump:prior} already guarantees the condition for
\(p=2\), so the assumption adds a requirement only when the bound is
applied with \(p>2\). Even then, only a finite moment of some order is
required, so \(G\) may have polynomial tails. A log-normal distribution,
for example, has every finite moment and satisfies the assumption for
every finite \(p\).

\subsubsection{Bayes Regret}

Following the empirical Bayes literature, I evaluate the feasible rule by comparing it with
the oracle Bayes rule under the true prior distribution \(G\). Conditional on \(\sigma\), let \(\E_{G,\hat Y}\) denote the
expectation under the joint distribution induced by
\(\mu_1,\ldots,\mu_n \stackrel{\mathrm{iid}}{\sim} G\) and
\(\hat Y_i\mid(\mu_i,\sigma_i)\sim N(\mu_i,\sigma_i^2)\). The Bayes regret of
\(\hat\tau^{\EBlabel}\) relative to \(\tau_G^\star\) is
\[
\operatorname{BR}(\hat\tau^{\EBlabel},\tau_G^\star\mid \sigma)
:=
\E_{G,\hat Y}\!\left[
L\!\left(\hat\tau^{\EBlabel}(\hat Y),\mu\right)
-
L\!\left(\tau_G^\star(\hat Y),\mu\right)
\right].
\]
Bayes regret measures the statistical cost of estimating \(G\). When it is
close to zero, the feasible rule performs, on average, almost as well as
the oracle that knows the prior. Proposition~\ref{prop:regret} bounds it
for all \(n\) large enough.

\begin{proposition}
\proptitle{\hyperlink{proof:regret}{Finite-Sample Bayes Regret Bound}}
\label{prop:regret}
\hypertarget{prop:regret}{}
Let Assumptions~\ref{assump:normality}, \ref{assump:prior},
\ref{assump:approx}, and \ref{assump:moment} hold, and condition on \(\sigma\). Let
\(p\ge 2\) be as in Assumption~\ref{assump:moment}. Define
\[
\varepsilon(n,G,p)
:=
\max\left\{
\sqrt{2\log n},
\left(n^{1/p}\sqrt{\log n}\,\mathcal M_p(G)\right)^{\frac{p}{2+2p}}
\right\}
\sqrt{\frac{\log n}{n}},
\]
and let $\Delta_n:=M_0\,\varepsilon(n,G,p)\,(\log n)^{3/2}$, where \(M_0\) is a constant depending only on \(p\) and the noise bounds \(\sigma_{\min}\) and \(\sigma_{\max}\). Then, for all \(n\)
large enough such that \(\log n>1/p\),
\[
\operatorname{BR}(\hat\tau^{\EBlabel},\tau_G^\star\mid \sigma)
\le
2\bar\sigma\,\Delta_n+\Delta_n^2,
\]
where $\bar{\sigma}^2 = 1/n \sum_{i=1}^n \sigma_i^2$. 
\end{proposition}

For fixed \(G\) and \(p\), \(\Delta_n\) converges to zero, so the cost of
estimating the prior vanishes as the cross section grows. The leading term
is \(2\bar\sigma\,\Delta_n\), since \(\Delta_n^2\) is of smaller order.
Up to logarithmic factors, the bound is of order \(n^{-p/(2(p+1))}\), so
\(p=2\) gives \(n^{-1/3}\) and the exponent approaches \(1/2\) as \(p\)
grows. A smaller \(p\) permits heavier tails at the cost of a slower
rate, and a larger \(p\) imposes tighter control of the tails and yields
a faster one.

The proof has a geometric step and a statistical step. The former bounds
the cost of allocating with estimated rather than oracle posterior mean
gaps. Both rules project their posterior mean gaps onto the same feasible
set, and this projection has kinks where the recipient set changes or the
budget switches between slack and binding. The allocation is therefore a
nonsmooth function of the estimated posterior mean gaps. A small error in
these gaps changes recipient status only for households near the cutoff,
and for these households, little is at stake. Their transfers are small
because a recipient's transfer falls continuously to zero as its posterior
mean gap approaches the cutoff. Each unit they receive because of the
error also costs little, because their marginal reduction in loss is
almost that of a recipient. Similarly, near the switch between a binding
and a slack budget, the marginal reduction in loss from additional spending
is close to zero, and a small error in how much to spend costs little. The
loss from these mistakes is the product of the amount misallocated and its
cost per unit, and both are at most of the order of the error.
Lemma~\ref{lem:proj_ineq} makes this precise, using only that the loss is
quadratic and the feasible set is convex. The regret is at most
\(\tfrac{1}{n}\E_{G,\hat Y}\|\hat g(\hat Y,\sigma)-g_G(\hat Y,\sigma)\|^2\),
the mean squared error of the estimated posterior mean gaps relative to
the oracle's.

The statistical step is Theorem~1 of
\citeA{jiang2020general}, which bounds the excess root mean squared error
of the NPMLE posterior means relative to the oracle's by \(\Delta_n\).
The oracle's root mean squared error is at most \(\bar\sigma\), because
its posterior means have no greater mean squared error than the raw
signals. Jiang's bound therefore controls how accurately the feasible
rule learns the oracle's posterior mean gaps, and the projection
inequality shows that this is enough to allocate almost as well as the
oracle.

The moment condition is chosen for generality, and the rate it delivers
need not be the sharpest possible.\footnote{Lower bounds for posterior mean estimation, such as those of
\citeA{polyanskiy2021sharp}, do not by themselves establish targeting lower
bounds, because projection can leave transfers unchanged despite estimation
error. Lemma~\ref{lem:proj_lower} bounds the Bayes regret below by
\(n^{-1}\E\|\delta(\hat Y)-\tau_G^\star(\hat Y)\|^2\).
A matching minimax lower bound on this allocation error, over the same
model class, would establish optimality of the rate in
Proposition~\ref{prop:regret}.} The argument is
modular, since any bound on the mean squared error of the estimated
posterior means yields a Bayes regret bound of the same order through
Lemma~\ref{lem:proj_ineq}. Under compact support, for example, the bounds
of \citeA{JiangZhang2009} and \citeA{soloff2021multivariate} give an
error of order \(1/n\) up to logarithmic factors, so the Bayes regret has
an upper bound of the same order.

\begin{titledRemark}{Relaxing Independence Assumptions}
\label{remark:independence_regret}
The regret bound in Proposition~\ref{prop:regret} is derived under two independence
restrictions, precision independence and cross-household independence. Neither
restriction enters the projection argument.

The first restriction fails when the conditional distribution of \(\mu_i\) given
\(\sigma_i\) depends on \(\sigma_i\). In that case, one can instead use the CLOSE
approach of \citeA{chen2023empirical}. CLOSE models the conditional
distribution of \(\mu_i\) given \(\sigma_i\) as a location-scale family whose location
and scale vary with \(\sigma_i\), while the common shape is estimated nonparametrically
from residualized data. By the modularity of the argument, any bound on the
discrepancy between the CLOSE posterior means and their oracle counterparts yields a
regret bound for the feasible rule.

The second restriction fails when the signals share estimation error, as in the PMT
setting of Remark~\ref{remark:pmt}. In the saturated case, however, collapsing to
types restores independence, and the approach applies with the weights of
Remark~\ref{remark:geo} and one signal per type. The precision of a type mean
reflects how common the type is, and CLOSE handles the case in which mean income
varies with type frequency. In general, the criterion in Section~\ref{sec:npmle}
is a composite marginal likelihood rather than the joint likelihood, studied by
\citeA{han2026correlated} under correlated Gaussian noise and by \citeA{chen2026normal}
under approximate normality and estimated variances. Neither result establishes
vanishing regret for PMT with a fixed number of coefficients. All predictions share the
estimation error of those coefficients, and the effective sample size stays bounded as
the target population grows. Extending Proposition~\ref{prop:regret} to this case
remains an open question. The simulations of Section~\ref{sec:sim_emp} do not impose
normality or independence, and there the empirical Bayes rule still shows gains over
the plug-in rule.
\end{titledRemark}

Appendix~\ref{sec:L_plus_regret} studies regret rates under the one-sided poverty gap loss
\(\ell_+\). Relative to the squared-loss case, the key difference is that performance depends on
posterior tail behavior near the poverty line rather than on posterior means alone. The resulting problem is therefore more difficult because it requires the
procedure to recover the parts of the latent income distribution that govern those tail
probabilities, and the associated regret rates are slower.

\section{Empirical Simulation Evidence}
\label{sec:sim_emp}

This section evaluates the allocation rules in an empirical simulation built
from the household consumption data for nine African countries assembled by
\citeA{brown2018poor}. In each replication, the policymaker observes only
consumption predicted from PMT models estimated on a small
training sample, as in the setting studied by \citeA{corral2025poverty}, and
allocates a fixed budget to the held-out population. I compare the plug-in OLS
rule, a plug-in XGBoost rule, and the empirical Bayes rule of
Section~\ref{sec:Bayes}. The empirical Bayes rule reaches
\poorReachedRatioEBOLS\ times as many poor people as plug-in OLS under the same
budget ceiling and achieves the same poverty gap reduction with
\equivBudgetSavingEBOLSPct\ less transfer spending. Under
quadratic loss, it outperforms plug-in OLS in all \nCountries\ countries and
XGBoost in \countriesEBBeatsMLLtwo\ of the \nCountries. I also evaluate the rules
under the one-sided poverty gap loss developed in
Appendix~\ref{app:loss_functions}.

\subsection{Setup}

The simulation uses the household survey data assembled by
\citeA{brown2018poor} for nine African countries: Burkina Faso, Ethiopia,
Ghana, Malawi, Mali, Niger, Nigeria, Tanzania, and Uganda. For each country, I
set the poverty line at the person-weighted 40th percentile of per capita
consumption and let $y_{ic}$ denote the consumption of household $i$ in country
$c$ as a share of that line, so the normalized poverty line is $z_c=1$ for every
country.\footnote{Consumption is winsorized at the person-weighted 95th
percentile before normalization. Both percentiles are computed on all
households with valid consumption and survey weights, before the sample is
restricted to households with complete PMT covariates, so the target-population
poverty rates are close to but not exactly 40 percent.} The losses, budgets, and
allocation rules use person weights $\omega_{ic}$, obtained by multiplying
each household's survey weight by its household size and rescaling to average
one within each replication's target population, so the simulated population
is representative of people.

The prediction exercise is repeated \nReps\ times for each country. In each
replication, I draw a training sample of \nTrain\ households, stratified by
urban status and region, and estimate two PMT models for
normalized consumption on the same ``basic PMT'' covariate set from
\citeA{brown2018poor}.\footnote{OLS is estimated without survey weights.
XGBoost is tuned within each training sample by fivefold cross-validation
over a prespecified grid of tree depth and minimum child weight, with early
stopping selecting the number of boosting rounds.} The target
population consists of the remaining households with complete PMT covariates,
for which both models predict consumption. Appendix
Table~\ref{tab:country_environments} reports country-level descriptives and
prediction quality.

I compare three feasible allocation rules. The plug-in OLS rule treats the
OLS prediction as true consumption. The plug-in XGBoost rule does the same
with the XGBoost prediction. The empirical Bayes rule uses the OLS prediction
and its heteroskedasticity-robust standard error, computed with the HC1
degrees-of-freedom correction. This standard error measures sampling
uncertainty in the estimated conditional mean. The rule estimates the prior
distribution of conditional mean consumption by nonparametric maximum
likelihood using the person weights $\omega_{ic}$ on a grid restricted to
nonnegative values, computes each
household's posterior mean poverty gap, and projects the vector of posterior
mean gaps onto the feasible transfer set.\footnote{The grid has 600 equally
spaced points from zero to the largest OLS prediction. Appendix
Table~\ref{tab:robustness} evaluates the standard GLmix support, which starts
at the smallest prediction. The prior-estimation solver reports an optimal
solution in all \nReps\ replications in each country, and all three headline
rules are evaluated on these same replications.} The projection imposes the budget as
an inequality, so the rule leaves part of the budget unspent when the positive
posterior mean gaps sum to less than the budget.

The design is an empirical simulation rather than a fully model-generated Monte
Carlo. Household consumption is not drawn from a known prior, and the prediction
error is not generated by adding synthetic normal noise to survey consumption.
Instead, the empirical distribution of consumption, the relationship between the PMT
covariates and consumption, and the resulting prediction errors all arise from
the observed data and from repeated finite-sample estimation of the PMT model.
Thus, the simulation does not impose the assumptions used in the theoretical
analysis, including normality of the signals, conditional independence
across households, and correct specification of the linear conditional mean. The empirical Bayes rule is nevertheless implemented as if the Gaussian
signal model were exact, so the exercise asks whether it remains useful in a
realistic environment where the conditions that motivate it do not hold
exactly.

Let $B_c$ denote the total transfer budget for country $c$'s target population.
Each country has the same transfer budget per represented person,
$b_c:=B_c/\sum_i\omega_{ic}=0.005$, or 0.5 percent of its poverty line. The
budget therefore has the same fiscal cost as paying a transfer equal to 10
percent of the poverty line to 5 percent of the population. A common value
keeps program size comparable across countries. On this per-person basis, the
benchmark is generous relative to the program scenarios documented in Appendix
Table~\ref{tab:program_budgets}, exceeding 10 of the 11 reported budgets. Even
so, the budget amounts to only \budgetShareGapMin--\budgetShareGapMax\ of the
aggregate poverty gap, depending on the country. The results are broadly the
same when the budget varies over a sixteen-fold range or matches each
documented program scenario, as Appendix Table~\ref{tab:robustness} shows.

The main performance measure is gain under quadratic loss. For
any feasible allocation $\hat\tau_c$, define
\[
    \Gain_{L_2,c}(\hat\tau_c)
    =
    \frac{
        L_{2,c}(0)-L_{2,c}(\hat\tau_c)
    }{
        L_{2,c}(0)-L_{2,c}(\tau_c^{\PIlabel})
    } ,
\]
where $L_{2,c}(\tau)$ is the person-weighted quadratic loss of allocation
$\tau$ evaluated at realized consumption in country $c$, $L_{2,c}(0)$ is the
loss without transfers, and $\tau_c^{\PIlabel}$ is the perfect-information
allocation of Proposition~\ref{prop:full_info}. I report this ratio as a
percentage of the perfect-information improvement over no transfers. A gain of
0\% means no improvement over no transfers, 100\% means that the allocation
attains the same improvement as the perfect-information allocation, and a negative gain means that
it increases quadratic loss relative to no transfers. Differences in gains
are reported in percentage points. I also report
the analogous measure, $\Gain_{L_+}$, under the one-sided poverty gap loss.

\subsection{Main Results}

Relative to the plug-in rules, the empirical Bayes rule based on OLS predictions (EB-OLS) reaches more poor people at a given
budget and requires less transfer spending to achieve a given reduction in
poverty. With the same budget, EB-OLS gives a positive transfer to
\poorReachedEBPct\ of initially poor people,
\poorReachedRatioEBOLS\ times the \poorReachedOLSPct\ reached by plug-in OLS,
while XGBoost reaches \poorReachedMLPct. The same advantage appears as lower
spending when poverty reduction is held fixed. In Appendix
Figure~\ref{fig:budget_required}, I lower the budget available to EB-OLS until
its mean realized poverty gap reduction across countries and replications
equals that of a plug-in rule at the headline budget. Comparing the transfers
actually paid out, EB-OLS matches the poverty gap reduction of plug-in OLS
with \equivBudgetSavingEBOLSPct\ less spending and that of XGBoost with
\equivBudgetSavingEBXGBPct\ less. For a program that already runs a PMT
regression, these savings come at no additional financial cost. The survey
data, the regression, and the predictions remain the same. Only the
allocation step changes, with transfers assigned using posterior mean poverty
gaps rather than raw predictions.

\begin{figure}[!t]
    \centering
    \setlength{\abovecaptionskip}{0pt}
    \setlength{\belowcaptionskip}{6pt}
    \caption{Targeting performance under quadratic loss}
    \label{fig:gains_L2}
    \includegraphics[width=6.25in]{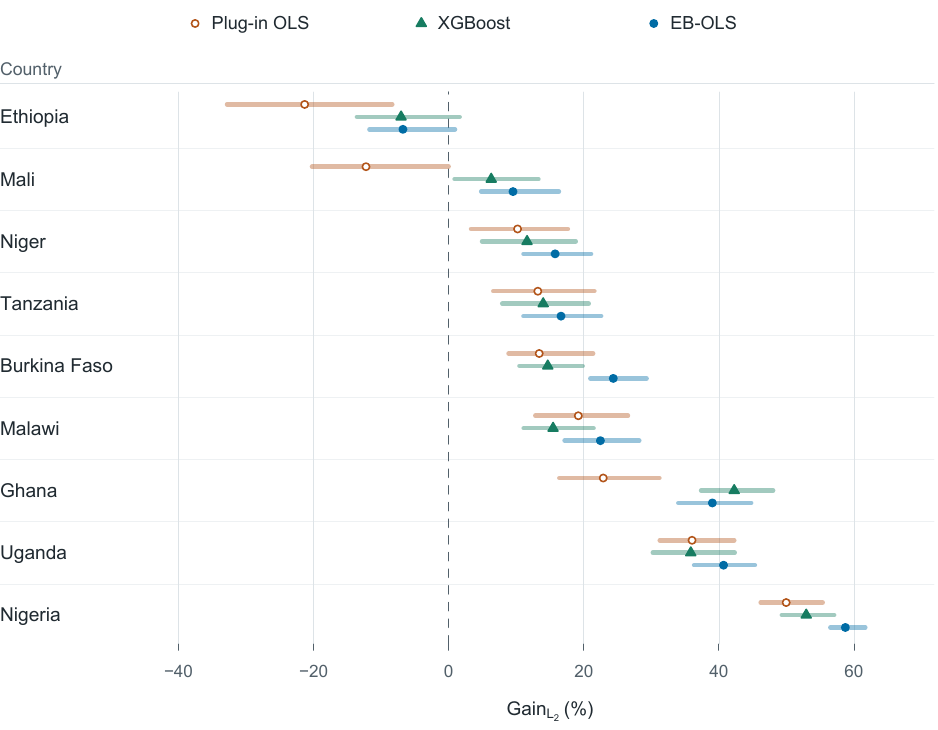}

    \vspace{8pt}
    \begin{minipage}{6.25in}
    \footnotesize
    \setstretch{1}
    \setlength{\baselineskip}{12pt}
    \emph{Notes.} Symbols report country-specific mean gains under quadratic
    loss over \nReps\ simulation replications at the common budget $b=0.005$.
    Gains are reported as percentages of the perfect-information improvement over no transfers.
    A gain of 0\% matches the no-transfer benchmark, and 100\% matches the
    perfect-information benchmark, which is outside the displayed range.
    Pale intervals span the 25th to 75th percentiles across
    replications. Negative gains indicate higher quadratic loss
    than with no transfers. Countries are ordered by mean plug-in OLS gain,
    from lowest to highest.
    \end{minipage}
\end{figure}

EB-OLS also delivers the highest average gain under quadratic loss. On
average, it recovers \gainEBLtwoPct\ of the reduction in quadratic loss attainable
under perfect information relative to no transfers, compared with
\gainOLSLtwoPct\ for plug-in OLS and \gainMLLtwoPct\ for XGBoost. Figure~\ref{fig:gains_L2} reports the comparison country by country.
EB-OLS yields higher gains than plug-in OLS in all \nCountries\ countries and
XGBoost in \countriesEBBeatsMLLtwo\ of the \nCountries. When the same
allocations are instead evaluated under the one-sided loss, their
cross-country averages retain the same ordering, with $\Gain_{L_+}$ equal to
\gainEBLtwoUnderLplusPct\ for EB-OLS, \gainMLLplusPct\ for XGBoost, and
\gainOLSLplusPct\ for plug-in OLS.

\Needspace{3\baselineskip}
The comparisons of EB-OLS with plug-in OLS and with XGBoost isolate different
changes. Plug-in OLS and EB-OLS start from the same predictions, but EB-OLS uses
their uncertainty to replace the predicted poverty gaps with posterior mean
poverty gaps. This replacement
raises the gain in every country, with increases ranging from
\gainDiffEBOLSMinLtwoPp\ to \gainDiffEBOLSMaxLtwoPp\ percentage points. XGBoost instead changes the
prediction model while retaining the plug-in allocation. EB-OLS attains a
higher average gain than XGBoost in every country except Ghana, although its
margin in Ethiopia is negligible.\footnote{Across 4,500 country--replication
pairs, EB-OLS outperforms plug-in OLS in 85.6\% of cases, with an average
quadratic-gain advantage of \gainEBOLSWinPp\ percentage points when it wins
and a disadvantage of \gainEBOLSLossPp\ percentage points when it loses.
Against XGBoost, it wins in 65.4\% of cases; the corresponding advantage and
disadvantage are \gainEBMLWinPp\ and \gainEBMLLossPp\ percentage points.}

The absolute level of the gain also varies widely across countries, and it
is associated with two features of each country. The first is how much the
predictions reveal about who is poor. Although \povertyRateTarget\ of people
are poor, OLS predicts only \olsFixedLinePredPoorMin--\olsFixedLinePredPoorMax\
of people to be below the line and correctly identifies only
\olsPoorRecallMin--\olsPoorRecallMax\ of poor people as poor. Shrinkage uses
an imperfect prediction more cautiously, and a flexible model extracts
different patterns from the same covariates, but neither can recover
information that the covariates do not contain, so all three feasible rules
tend to yield lower gains where the predictions reveal less. The second feature
is the share of people near the poverty line, where small prediction errors
turn intended gap closure into overshoot or leakage. In countries with a share
of people within 10 percent of the line at or above the median, the mean gain
of EB-OLS is \gainEBHighNearLinePct, compared with \gainEBLowNearLinePct\ for the other countries,
and the plug-in rules show the same pattern.

Ethiopia and Nigeria mark the two ends of this range. In Ethiopia, the
out-of-sample OLS regression has $R^2=\ethiopiaRtwoOLS$ and predicts only
\ethiopiaFixedLinePredPoorOLS\ of people to be poor. All three rules perform worse
than making no transfers under the quadratic loss. The households with the
largest predicted gaps are often not the poorest, so much of the budget
overshoots or leaks to the nonpoor, and the quadratic criterion counts both
against the rule. EB-OLS increases quadratic loss much less than plug-in OLS and about as much
as XGBoost, because it shrinks the largest predicted gaps and spends only
where posterior mean gaps remain positive, which leaves about a third of the
budget unspent. Its gain is \ethiopiaGainEBLtwoPct, compared with
\ethiopiaGainOLSLtwoPct\ for plug-in OLS and \ethiopiaGainMLLtwoPct\ for XGBoost.
By contrast, in Nigeria, the out-of-sample regression has
$R^2=\nigeriaRtwoOLS$, and plug-in OLS already attains a gain of
\nigeriaGainOLSLtwoPct. EB-OLS raises it to \nigeriaGainEBLtwoPct, above the
\nigeriaGainMLLtwoPct\ attained by XGBoost. Shrinkage therefore improves on
plug-in OLS whether the predictions are weak or strong, but it can only use
the information the covariates contain more cautiously, not replace what they
lack.\footnote{The empirical Bayes rule is not built on XGBoost because the
signal model requires approximately unbiased predictions with standard
errors for their sampling variation. OLS supplies both when the linear model is correctly specified, as
Remark~\ref{remark:pmt} shows, whereas XGBoost predictions carry
regularization bias that bootstrap standard errors would not measure.}

Appendix Table~\ref{tab:robustness} varies the budget and the design. Over
the sixteen-fold budget range, the advantage of EB-OLS over plug-in OLS grows
with the budget, while its advantage over XGBoost narrows. EB-OLS remains the
best-performing rule under every prespecified change to the design. Its
advantage over plug-in OLS is largest with 250 training households---where the
predictions are noisiest and plug-in OLS does no better than making no
transfers---and shrinks as the training sample grows to 1,000. Clustering
the OLS covariance matrix by enumeration area leaves the results unchanged,
and the CLOSE-NPMLE rule of \citeA{chen2023empirical}, which lets the prior
depend on the standard error, performs somewhat below EB-OLS and above both
plug-in rules. At the documented program budgets, EB-OLS has the highest gain
in \nProgramScenariosEBBest\ of the \nProgramScenarios\
scenarios.\footnote{The exception is Malawi at the smallest budget evaluated,
where plug-in OLS transfers only to the households with the most extreme
predicted gaps, a group it targets very accurately, whereas shrinkage spreads
the same budget over more than twice as many households and admits more
nonpoor recipients. At the larger Malawi budget in the table, EB-OLS leads by
\malawiProgramGainAdvantagePp\ percentage points.}

\subsection{Allocation Mechanisms}
\label{subsec:policy_value}

\begin{figure}[!t]
    \centering
    \setlength{\abovecaptionskip}{0pt}
    \setlength{\belowcaptionskip}{6pt}
    \caption{How the empirical Bayes rule reallocates transfers under prediction uncertainty}
    \label{fig:uncertainty_mechanism}
    \includegraphics[width=6.25in]{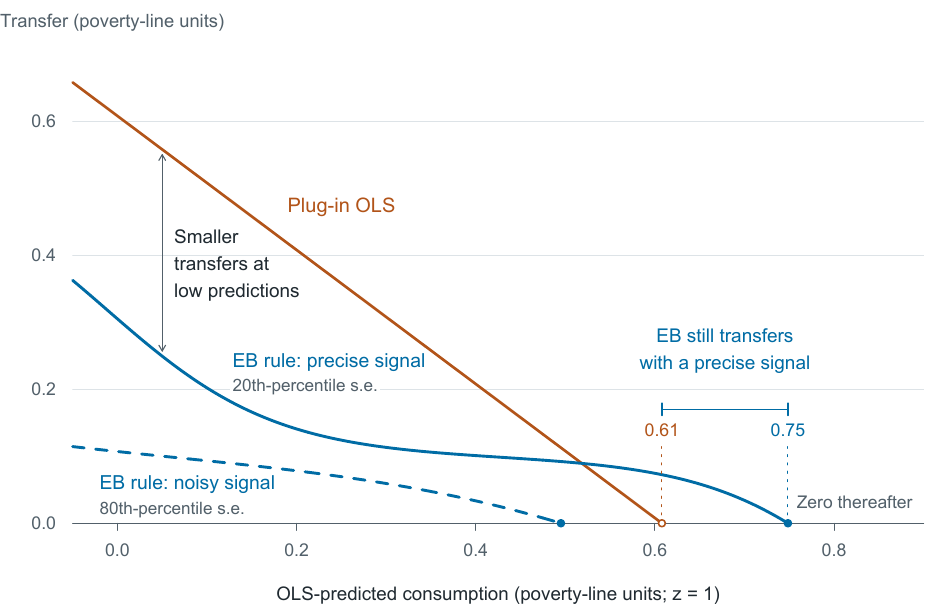}

    \vspace{8pt}
    \begin{minipage}{6.25in}
    \footnotesize
    \setstretch{1}
    \setlength{\baselineskip}{12pt}
    \emph{Notes.} Mali, replication 221, selected mechanically from the allocation
    schedules. Predictions and transfers are expressed in poverty-line units.
    The EB curves hold the prediction standard error at its person-weighted 20th
    percentile (precise) and 80th percentile (noisy), with all other components
    of the fitted EB allocation rule held fixed. Actual household allocations use each
    household's own standard error. Plug-in OLS and EB-OLS each exhaust the same
    budget, $b=0.005$. The marked schedules remain zero beyond their respective
    cutoffs.
    \end{minipage}
\end{figure}

Figure~\ref{fig:uncertainty_mechanism} plots the transfer assigned by plug-in
OLS and by EB-OLS against predicted consumption for \figureTwoCountry\ in one
replication. Under plug-in, the transfer equals the predicted poverty gap minus
a common cutoff. It falls linearly as predicted consumption rises and reaches
zero at the cutoff. Under EB-OLS, the transfer depends on the prediction and on
its standard error, and the figure draws the schedule for a precise and for a
noisy prediction. Both curves are nonlinear in predicted consumption because
the estimated posterior mean is. The noisy curve lies below the precise curve
for all low predictions, so a household whose consumption is estimated less
precisely receives a smaller transfer at the same low prediction. At the
lowest predictions, the EB curves also lie below the plug-in line because
EB-OLS trims the largest transfers. The resources trimmed there help finance
transfers to households with precise signals and predicted consumption above
the plug-in cutoff. The EB allocation is less concentrated, with
smaller transfers at the lowest predictions and positive transfers for a
broader set of households.

\begingroup
\widowpenalty=10000
Averaged across countries and replications, EB-OLS reaches more people with
smaller transfers. Table~\ref{tab:main_mechanism} reports transfer coverage
and budget use for the three rules, Appendix
Table~\ref{tab:country_main_mechanism} repeats it by country, and Appendix
Table~\ref{tab:detailed_mechanisms} reports transfer amounts among recipients.
Plug-in OLS treats \popTreatedOLS\ of people, with a 90th-percentile transfer
among recipients of \pNinetyTransferOLS\ poverty-line units. EB-OLS treats
\popTreatedEB\ of people and lowers the 90th-percentile transfer to
\pNinetyTransferEB. EB-OLS gives each recipient less and extends assistance
to substantially more poor people, instead of making large transfers to a
small group selected on noisy predictions.

\endgroup
\input{\outdir/tables/table1_policy_value_percent.tex}

Table~\ref{tab:main_mechanism} also shows how the rules use the budget. For
every 100 units of available budget, overshoot---spending on poor recipients
beyond their initial poverty gaps---falls from \overshootOLSperHundred\ under
plug-in OLS to \overshootEBperHundred\ under EB-OLS, while leakage to
initially nonpoor recipients changes little, from \leakageOLSperHundred\ to
\leakageEBperHundred. The principal change in budget use is therefore lower
overshoot rather than lower payments to the nonpoor. XGBoost also reduces
overshoot, to \overshootMLperHundred, but raises leakage to
\leakageMLperHundred. EB-OLS and XGBoost leave \unspentEBperHundred\ and
\unspentMLperHundred\ per 100 units unspent on average, mostly in Ethiopia and
Mali.\footnote{Appendix Table~\ref{tab:robustness} shows that requiring every
rule to spend the full budget leaves the gain of EB-OLS essentially unchanged,
so the unspent budget does not explain its advantage.}

The broader transfer coverage extends to the poorest people. Appendix
Table~\ref{tab:detailed_mechanisms} shows that the share of extremely poor
people---those with consumption below $0.5z$---who receive a transfer rises from
\covExtremePoorOLS\ under plug-in OLS to \covExtremePoorEB\ under EB-OLS.
Despite reaching more of them, EB-OLS closes \extrGapClosedShareEB\ of their
aggregate pre-transfer poverty gap---a share similar to the \extrGapClosedShareOLS\
under plug-in OLS and \extrGapClosedShareML\ under XGBoost. Under EB-OLS,
fewer extremely poor people receive nothing, but the average reduction in the
poverty gap among those reached is smaller. Appendix
Figure~\ref{fig:headcount_reductions} shows the consequence for headcounts.
EB-OLS and XGBoost lift more people out of extreme poverty, while plug-in OLS
lifts more people out of poverty. The small transfers of EB-OLS and XGBoost
raise the consumption of deeply poor households but usually leave them below
the line, and the large transfers of plug-in OLS more often carry their
recipients all the way across it.

The empirical Bayes rule for the one-sided loss, developed in
Appendix~\ref{app:loss_functions}, uses the same estimated posterior
distribution but does not penalize overshoot and therefore exhausts the
budget. Evaluated under the one-sided loss, it attains a cross-country mean
gain of \gainEBLplusPct, compared with \gainOLSLplusPct\ for plug-in OLS and
\gainMLLplusPct\ for XGBoost. It yields higher gains than plug-in OLS in
\countriesEBBeatsOLSLplus\ of the \nCountries\ countries and XGBoost in
\countriesEBBeatsMLLplus\ of the \nCountries. Appendix
Figure~\ref{fig:gains_Lplus} reports the comparison by country, and Appendix
Table~\ref{tab:eb_L2_vs_Lplus} evaluates both EB rules under both losses. In
the seven countries where the quadratic rule spends the full budget, the two
EB rules perform nearly identically, with gains within 0.1 percentage points of each other
under the one-sided loss and within 0.5 percentage points under the quadratic loss. Material
differences are confined to Ethiopia and Mali, where the quadratic rule leaves
part of the budget unspent and the one-sided rule allocates it, raising the
gain under the one-sided loss. The choice of objective therefore matters
empirically mainly when the penalty on overshoot leads the quadratic rule to
stop spending.

\section{Conclusion}
\label{sec:conclu}
Targeting programs invest in better data and better prediction models and
then convert the predictions into transfers with a rule that treats them as
if they were true. This paper shows that the conversion step discards
information the budget could use. The plug-in rule is inadmissible. The
empirical Bayes rule pools information across households, shrinking each
estimated poverty gap toward what the distribution of incomes in the
population makes plausible and taking the precision of each estimate into
account, before imposing the budget. Its regret relative to the oracle Bayes
allocation is bounded by the estimation error in the posterior mean gaps, so
the budget and nonnegativity constraints do not slow the rate at which it
approaches the oracle. The one-sided objective and fixed benefit menus show
that the same approach carries over to program designs closer to practice,
with weaker guarantees.

The practical implication is that antipoverty programs can improve targeting
with the information they already have. The empirical Bayes rule needs the
PMT predictions and the standard errors of the estimated
conditional means, which can be computed from the fitted regression. It
changes nothing about data collection, the prediction model, or the
eligibility survey. In the nine-country simulations, that change alone reaches about
\poorReachedRatioEBOLSRounded\ times as many poor people as plug-in OLS under
the same budget, or attains the same average poverty gap reduction with
\equivBudgetSavingEBOLSPct\ less transfer spending. Tanzania has the smallest
estimated saving in the simulations, 1.9 percent. If that rate carried over
to its Productive Social Safety Net at the program's 2015 scale, the
estimated annualized transfer outlay of about 140 million dollars would
imply savings of about 2.6 million dollars a year.
Better
targeting can come from the predictions programs already have, once the step that converts them into transfers is properly designed.

This paper takes the prediction model as given and optimizes the allocation
rule. Designing the prediction and the allocation together is a natural
direction for future work. In the simulations, OLS and XGBoost predict
consumption about equally well, yet their plug-in allocations perform very
differently, so overall prediction accuracy does not determine targeting
performance. Programs should evaluate a prediction method together with the
allocation rule built on it, under the targeting loss studied in this paper.
Extending the analysis to joint design would require accounting for local bias
in machine learning predictors and estimation error shared across households,
both of which lie outside the regret bound.

\clearpage
\nocite{boyd2004convex,berger2013statistical}
\renewcommand{\refname}{References}
\putbib

\clearpage
\appendix
\setlength{\abovedisplayskip}{6pt plus 2pt minus 2pt}
\setlength{\belowdisplayskip}{6pt plus 2pt minus 2pt}
\setlength{\abovedisplayshortskip}{4pt plus 2pt minus 1pt}
\setlength{\belowdisplayshortskip}{4pt plus 2pt minus 1pt}
\setlength{\jot}{3pt}
\renewcommand{\thefigure}{\Alph{section}.\arabic{figure}}
\renewcommand{\thetable}{\Alph{section}.\arabic{table}}
\renewcommand{\theHfigure}{\thesection.\arabic{figure}}
\renewcommand{\theHtable}{\thesection.\arabic{table}}
\setcounter{figure}{0}
\section{Proofs and Supporting Results for the Main Text}
\label{app:proofs}
This appendix collects the proofs of the main-text results
and the supporting lemmas.

\subsection{Proofs of Main-Text Results}
\label{app:proposition_proofs}

\hypertarget{proof:full_info}{} 
\Proof{prop:full_info}{
The objective is
\[
\frac{1}{n}\sum_{i=1}^n (z-y_i-\tau_i)^2=\frac{1}{n}\|\tau-(z\mathbf 1-y)\|^2 .
\]
Because $\mathcal A$ is nonempty, closed, and convex, the projection of $z\mathbf 1-y$ onto $\mathcal A$ exists and is unique, so the unique minimizer is $P_{\mathcal A}(z\mathbf 1-y)$.

To obtain the threshold representation, multiply the objective by $n$. Since $B>0$, the allocation $\tau_i=B/(2n)$ is strictly feasible, so the KKT conditions are necessary and sufficient. With nonnegative multipliers $\nu$ for the budget constraint and $\xi_i$ for nonnegativity, stationarity gives $2(\tau_i-z+y_i)+\nu-\xi_i=0$. Together with $\tau_i\ge 0$, $\xi_i\ge 0$, and $\xi_i\tau_i=0$, this implies $\tau_i=\bigl(z-y_i-\nu/2\bigr)_+$. Write $\lambda^{\PIlabel}:=\nu/2$, so that $\tau_i=\bigl(z-y_i-\lambda^{\PIlabel}\bigr)_+$. Budget feasibility and complementary slackness require
\[
T(\lambda^{\PIlabel})\le B,\qquad \lambda^{\PIlabel}\bigl(T(\lambda^{\PIlabel})-B\bigr)=0,
\qquad\text{where}\quad
T(a):=\sum_{i=1}^n \bigl(z-y_i-a\bigr)_+ .
\]
If $T(0)\le B$, setting $\lambda^{\PIlabel}=0$ satisfies these conditions. Otherwise, $\lambda^{\PIlabel}>0$ and $T(\lambda^{\PIlabel})=B$. The function $T$ is continuous, strictly decreasing wherever positive, and eventually zero, so this equation has a unique positive solution.

Finally, the threshold formula implies $y_i+\tau_i^{\PIlabel}(y)=z-\lambda^{\PIlabel}$ for every recipient and zero transfers otherwise, which is the leveling representation.}

\hypertarget{proof:bayes_rules}{}
\ProofLemma{lem:bayes_rules}{
Fix $x$. The Gaussian likelihood is bounded and the marginal density of $X$ is strictly positive everywhere, so the posterior defined by Bayes' formula exists at every $x$ and, by the second-moment condition on $\pi$, has finite second moments. Expanding the square, $\E_\pi[L(\tau,\theta)\mid X=x]=n^{-1}\|\tau-m_\pi(x)\|^2+n^{-1}\E_\pi[\|\theta-m_\pi(x)\|^2\mid X=x]$. The first term is strictly convex in $\tau$, so $P_{\mathcal A}(m_\pi(x))$ is its unique minimizer over $\mathcal A$. The map $m_\pi$ is measurable and $P_{\mathcal A}$ is 1-Lipschitz \cite{combettes2018monotone}, hence continuous, so $\delta_\pi\in\mathcal D$. Since $\mathcal A$ is bounded and $\pi$ has a finite second moment, every rule in $\mathcal D$ has finite Bayes risk, which equals the integral of posterior expected loss against the marginal law of $X$. Hence $\delta_\pi$ minimizes Bayes risk, and any other minimizer attains the posterior minimum for marginal-almost every $x$ and therefore agrees with $\delta_\pi$ there. The marginal has a strictly positive Lebesgue density, so the exceptional set is Lebesgue-null.}

\hypertarget{proof:plugin_inadmissible}{} 
\Proof{prop:plugin_inadmissible}{Suppose, toward a contradiction, that \(\tau^{\pluglabel}\) is admissible in
\(\mathcal D\). By Lemma~\ref{lem:compact_complete_class}, there exists a proper
Borel prior \(\pi\) on \(\Theta_M\) and a Bayes rule \(\delta_\pi\) under \(\pi\)
such that
\(\tau^{\pluglabel}(X)=\delta_\pi(X) \qquad P_\theta\text{-a.s. for every }\theta\in\Theta_M.\)
Thus \(\tau^{\pluglabel}\) is Bayes, up to the usual almost-sure equivalence of
decision rules, with respect to a proper prior on \(\Theta_M\).

This contradicts Lemma~\ref{lem:plugin_not_bayes}, which shows that no proper
prior on \(\Theta_M\) can make $\tau^{\pluglabel}(x)=P_{\mathcal A}(x)$ Bayes, even up to \(P_\theta\)-almost-sure equivalence for every
\(\theta\in\Theta_M\), under squared-error loss. Therefore
\(\tau^{\pluglabel}\) is inadmissible in \(\mathcal D\).}

\hypertarget{proof:oracle_bayes_rule}{} 
\Proof{prop:oracle_bayes_rule}{
Let \(\pi\) be the prior that Assumption~\ref{assump:prior} induces on the
gap vector \(\theta\). It is proper with finite second moment, and
conditional on \(\sigma\), Assumptions~\ref{assump:normality}
and~\ref{assump:prior} imply that the posterior factorizes across
households, so \(m_\pi(z\mathbf 1-\hat y)=g_G(\hat y,\sigma)\) for every
\(\hat y\). Lemma~\ref{lem:bayes_rules} then shows that
\(P_{\mathcal A}(g_G(\hat y,\sigma))\) uniquely minimizes posterior expected
loss over \(\mathcal A\) at every \(\hat y\), minimizes Bayes risk over
\(\mathcal D\), and agrees with every other Bayes rule in \(\mathcal D\) for
Lebesgue-almost every \(\hat y\). The threshold form is
Proposition~\ref{prop:full_info} applied with the posterior mean incomes
\(z\mathbf 1-g_G(\hat y,\sigma)\) in place of \(y\), with \(\lambda_G(\hat y)\)
the cutoff determined by its budget condition.}

\hypertarget{proof:regret}{} 
\Proof{prop:regret}{
Throughout, condition on \(\sigma\), and write \(\E[\cdot]\) for expectation under the joint
distribution of \((\mu,\hat Y)\) induced by Assumptions~\ref{assump:normality}--\ref{assump:prior}.

Let \(\tilde\mu_{G,i}:=\E_G[\mu_i\mid \hat Y_i,\sigma_i]\) and
\(\tilde\mu_G:=(\tilde\mu_{G,1},\ldots,\tilde\mu_{G,n})^\top\). Since
\(g_G(\hat Y,\sigma)=z\mathbf 1-\tilde\mu_G\), expanding the squared loss, taking conditional
expectation given \((\hat Y,\sigma)\), and canceling the posterior-variance term gives
\[
\operatorname{BR}(\hat\tau^{\EBlabel},\tau_G^\star\mid \sigma)
=
\frac{1}{n}
\E\!\left[
\bigl\|z\mathbf 1-\tilde\mu_G-\hat\tau^{\EBlabel}\bigr\|^2
-
\bigl\|z\mathbf 1-\tilde\mu_G-\tau_G^\star\bigr\|^2
\right].
\]

Now set \(u:=g_G(\hat Y,\sigma)\) and \(v:=\hat g(\hat Y,\sigma)\). By definition,
\(\tau_G^\star=P_{\mathcal A}(u)\) and \(\hat\tau^{\EBlabel}=P_{\mathcal A}(v)\), so the integrand is
\(\|u-P_{\mathcal A}(v)\|^2-\|u-P_{\mathcal A}(u)\|^2\). Applying
Lemma~\ref{lem:proj_ineq} pointwise yields
\[
\operatorname{BR}(\hat\tau^{\EBlabel},\tau_G^\star\mid \sigma)
\le
\frac{1}{n}\,
\E\!\left[
\|\hat g(\hat Y,\sigma)-g_G(\hat Y,\sigma)\|^2
\right].
\]

It remains to relate the posterior mean poverty-gap discrepancy to a posterior-mean risk
difference in the heteroskedastic normal-means problem. Define
\(\tilde\mu_{\hat G,i}:=\E_{\hat G}[\mu_i\mid \hat Y_i,\sigma_i]\). Since
\(\hat g_i(\hat Y_i,\sigma_i)=z-\tilde\mu_{\hat G,i}\) and
\(g_{G,i}(\hat Y_i,\sigma_i)=z-\tilde\mu_{G,i}\), we have
\(\hat g_i(\hat Y_i,\sigma_i)-g_{G,i}(\hat Y_i,\sigma_i)=\tilde\mu_{G,i}-\tilde\mu_{\hat G,i}\). Also,
$\tilde\mu_{\hat G,i}-\mu_i = (\tilde\mu_{\hat G,i}-\tilde\mu_{G,i}) + (\tilde\mu_{G,i}-\mu_i)$.
The estimated prior \(\hat G\) is computed from the full sample, so
\(\tilde\mu_{\hat G,i}\) is generally not fixed after conditioning only on
\((\hat Y_i,\sigma_i)\). I therefore condition on the full observed data vector
\((\hat Y,\sigma)\). Under Assumptions~\ref{assump:normality}--\ref{assump:prior},
conditional on \(\sigma\), the pairs \((\mu_i,\hat Y_i)\) are independent across
\(i\). Hence
\(\E[\mu_i\mid \hat Y,\sigma] = \E[\mu_i\mid \hat Y_i,\sigma_i] = \tilde\mu_{G,i}.\)
Therefore $\E[\tilde\mu_{G,i}-\mu_i\mid \hat Y,\sigma]=0$. Since \(\tilde\mu_{\hat G,i}-\tilde\mu_{G,i}\) is fixed after conditioning on
\((\hat Y,\sigma)\), the cross-term vanishes:
\[
\begin{aligned}
\E\!\left[
(\tilde\mu_{\hat G,i}-\tilde\mu_{G,i})(\tilde\mu_{G,i}-\mu_i)
\right]
&=
\E\!\left[
(\tilde\mu_{\hat G,i}-\tilde\mu_{G,i})
\E[\tilde\mu_{G,i}-\mu_i\mid \hat Y,\sigma]
\right] =0.
\end{aligned}
\]
Expanding the square, summing over \(i\), and rearranging gives
\[
\frac{1}{n}
\E\!\left[
\|\hat g(\hat Y,\sigma)-g_G(\hat Y,\sigma)\|^2
\right]
=
\frac{1}{n}
\E\!\left[
\sum_{i=1}^n(\tilde\mu_{\hat G,i}-\mu_i)^2
\right]
-
\frac{1}{n}
\E\!\left[
\sum_{i=1}^n(\tilde\mu_{G,i}-\mu_i)^2
\right].
\]
In words, the expected posterior-mean poverty-gap discrepancy equals the excess average
squared-error risk of \(\tilde\mu_{\hat G}\) relative to the oracle posterior mean \(\tilde\mu_G\).

By \citeA{jiang2020general}, Theorem~1,
\[
\Bigg\{
\frac{1}{n}
\E\!\left[
\sum_{i=1}^n(\tilde\mu_{\hat G,i}-\mu_i)^2
\right]
\Bigg\}^{1/2}
-
\Bigg\{
\frac{1}{n}
\E\!\left[
\sum_{i=1}^n(\tilde\mu_{G,i}-\mu_i)^2
\right]
\Bigg\}^{1/2}
\le
\Delta_n.
\]
Writing
\(R^*:=\frac{1}{n}\E\!\left[\sum_{i=1}^n(\tilde\mu_{G,i}-\mu_i)^2\right]\),
this implies
\(\frac{1}{n}\E[\|\hat g-g_G\|^2]\le 2\Delta_n\sqrt{R^*}+\Delta_n^2\).

To bound \(R^*\), note that under squared loss the posterior mean minimizes conditional risk.
Taking the competitor \(\delta_i=\hat Y_i\) therefore gives
\(\E[(\tilde\mu_{G,i}-\mu_i)^2]\le \E[(\hat Y_i-\mu_i)^2]=\sigma_i^2\), where the last equality
uses Assumption~\ref{assump:normality}. Hence
\(R^*\le n^{-1}\sum_{i=1}^n\sigma_i^2=:\bar\sigma^2\), so \(\sqrt{R^*}\le \bar\sigma\).

Combining the preceding bounds yields $\operatorname{BR}(\hat\tau^{\EBlabel},\tau_G^\star\mid \sigma) \le 2\bar\sigma\,\Delta_n+\Delta_n^2$. This proves the proposition.
}

\subsection{Supporting Lemmas}
\label{app:supporting_lemmas}

\begin{lemma}
\proptitle{\hyperlink{proof:active_set_affine_regions}{Local form of the plug-in rule}}
\label{lem:active_set_affine_regions}
\hypertarget{lem:active_set_affine_regions}{}
For every nonempty $K\subseteq\{1,\ldots,n\}$ there is a nonempty open set
$\mathcal U_K\subseteq\mathbb R^n$ on which the plug-in rule exhausts the
budget, has active set $K$, and satisfies
\[
\tau_i^{\pluglabel}(x)=
\begin{cases}
x_i-\lambda_K(x), & i\in K,\\
0, & i\notin K,
\end{cases}
\qquad
\lambda_K(x):=\frac{\sum_{\ell\in K}x_\ell-B}{|K|},
\]
with $\lambda_K(x)>0$ on $\mathcal U_K$. The rule is therefore affine on
$\mathcal U_K$, and $\tau_i^{\pluglabel}(x)-\tau_j^{\pluglabel}(x)=x_i-x_j$
for all $i,j\in K$ and $x\in\mathcal U_K$.
\end{lemma}

\hypertarget{proof:active_set_affine_regions}{}
\AppendixProofHeading{Proof of Lemma~\ref{lem:active_set_affine_regions}}
Let $x^0$ have $x_i^0=2B/|K|$ for $i\in K$ and $x_j^0=0$ for
$j\notin K$. Then $\lambda_K(x^0)=B/|K|>0$, $x_i^0>\lambda_K(x^0)$ for
$i\in K$, and $x_j^0<\lambda_K(x^0)$ for $j\notin K$. Since $\lambda_K$ is
continuous, these strict inequalities hold on an open neighborhood
$\mathcal U_K$ of $x^0$. Fix $x\in\mathcal U_K$. The vector
$t:=(x-\lambda_K(x)\mathbf 1)_+$ has positive coordinates exactly on $K$
and satisfies $\mathbf 1^\top t=B$, and
$\mathbf 1^\top x_+\ge\sum_{i\in K}x_i=B+|K|\lambda_K(x)>B$. Thus $t$
satisfies the KKT characterization of the projection established in the
proof of Proposition~\ref{prop:full_info}, so $t=\tau^{\pluglabel}(x)$.
The affine form and the difference identity follow.
\AppendixQED{lem:active_set_affine_regions}

\begin{lemma}
\proptitle{\hyperlink{proof:compact_complete_class}{Admissible rules are Bayes}}
\label{lem:compact_complete_class}
\hypertarget{lem:compact_complete_class}{}
Under Assumptions~\ref{assump:normality} and~\ref{assump:bounded}, every
rule $\delta\in\mathcal D$ that is admissible on $\Theta_M$ agrees
$P_\theta$-almost surely, for every $\theta\in\Theta_M$, with a Bayes rule
under some proper prior on $\Theta_M$.
\end{lemma}

\hypertarget{proof:compact_complete_class}{}
\AppendixProofHeading{Proof of Lemma~\ref{lem:compact_complete_class}}
The parameter space $\Theta_M$ and the action space $\mathcal A$ are
compact, the loss is continuous and bounded on $\mathcal A\times\Theta_M$,
and the Gaussian density is jointly continuous in $(x,\theta)$. Every rule
in $\mathcal D$ has continuous risk, because the Gaussian location family
is continuous in total variation and the loss is uniformly continuous on
$\mathcal A\times\Theta_M$. Since $\mathcal A$ is convex and the loss is
convex in the action, replacing a randomized action by its conditional
mean preserves feasibility and weakly lowers risk, so a rule admissible in
$\mathcal D$ is admissible among randomized rules as well. The
complete-class theorem of \citeA[Theorem~12, Section~8.8]{berger2013statistical}
therefore implies that $\delta$ is Bayes under some proper prior $\pi$ on
$\Theta_M$.

A prior on $\Theta_M$ has finite second moments, so by
Lemma~\ref{lem:bayes_rules} $\delta$ agrees with
$\delta_\pi(x)=P_{\mathcal A}(m_\pi(x))$ Lebesgue-almost everywhere and
hence $P_\theta$-almost surely for every $\theta\in\Theta_M$.
\AppendixQED{lem:compact_complete_class}

\begin{lemma}
\proptitle{\hyperlink{proof:plugin_not_bayes}{The plug-in rule is not Bayes}}
\label{lem:plugin_not_bayes}
\hypertarget{lem:plugin_not_bayes}{}
Assume $n\ge2$. Under Assumptions~\ref{assump:normality}
and~\ref{assump:bounded}, the plug-in rule is not Bayes under any proper
prior on $\Theta_M$, even up to $P_\theta$-almost-sure equivalence for
every $\theta\in\Theta_M$.
\end{lemma}

\hypertarget{proof:plugin_not_bayes}{}
\AppendixProofHeading{Proof of Lemma~\ref{lem:plugin_not_bayes}}
Suppose otherwise, and let $\pi$ be the corresponding proper prior, with
posterior mean $m_\pi(x):=\E_\pi[\theta\mid X=x]$. By
Lemma~\ref{lem:bayes_rules} and the strict positivity of the
Gaussian densities, $\tau^{\pluglabel}(x)=P_{\mathcal A}(m_\pi(x))$ for
Lebesgue-almost every $x$. Each coordinate of $m_\pi$ is real analytic
on $\mathbb R^n$, being a ratio of Gaussian integrals over the compact set
$\Theta_M$ with strictly positive denominator, so both sides of this
equality are continuous and it holds everywhere on any open set on which
it holds almost everywhere.

Fix distinct indices $i\neq j$ and let $K=\{i,j\}$. By
Lemma~\ref{lem:active_set_affine_regions}, there is a nonempty open set
$\mathcal U_K$ on which the plug-in rule exhausts the budget and has
active set $K$. On $\mathcal U_K$ the Bayes projection has the same two
positive coordinates, and its common threshold cancels in their
difference, so
\[
m_{\pi,i}(x)-m_{\pi,j}(x)
=
\tau_i^{\pluglabel}(x)-\tau_j^{\pluglabel}(x)
=
x_i-x_j
\qquad\text{for all }x\in\mathcal U_K.
\]
The real analytic function $x\mapsto m_{\pi,i}(x)-m_{\pi,j}(x)-(x_i-x_j)$
vanishes on the nonempty open set $\mathcal U_K$, so by the identity
theorem it vanishes on all of $\mathbb R^n$. This is impossible, because
$|m_{\pi,i}(x)-m_{\pi,j}(x)|\le M+z$ for every $x$ while $|x_i-x_j|$
is unbounded.
\AppendixQED{lem:plugin_not_bayes}

\begin{lemma}
	\proptitle{\hyperlink{proof:proj_ineq}{Projection mismatch inequality}}
\label{lem:proj_ineq} \hypertarget{lem:proj_ineq}{}
	Let \(\mathcal A\subset\mathbb R^n\) be nonempty, closed, and convex, and let
\(P_{\mathcal A}:\mathbb R^n\to\mathcal A\) denote Euclidean projection. Then, for any
\(u,v\in\mathbb R^n\),
\begin{equation}
\label{eq:proj_mismatch}
\|u-P_{\mathcal A}(v)\|^2-\|u-P_{\mathcal A}(u)\|^2
\le
\|u-v\|^2.
\end{equation}
The constant \(1\) is sharp whenever \(\mathcal A\) contains two distinct points.
\end{lemma}
\hypertarget{proof:proj_ineq}{}
\AppendixProofHeading{Proof of Lemma~\ref{lem:proj_ineq}} Write \(p:=P_{\mathcal A}(u)\) and \(q:=P_{\mathcal A}(v)\).
By convexity, \(q+t(p-q)\in\mathcal A\) for \(t\in[0,1]\). Since \(q\) minimizes the squared
distance to \(v\) over \(\mathcal A\), the right derivative of \(\|v-q-t(p-q)\|^2\) at
\(t=0\) is nonnegative, which gives \(\langle v-q,\,p-q\rangle\le 0\). Hence
\[
\|v-q\|^2-\|v-p\|^2
=
2\langle v-q,\,p-q\rangle-\|p-q\|^2
\le
-\|p-q\|^2.
\]
Expanding the squared distances and applying this inequality,
\[
\begin{aligned}
\|u-q\|^2-\|u-p\|^2
&=
2\langle u-v,\,p-q\rangle+\|v-q\|^2-\|v-p\|^2\\
&\le
2\langle u-v,\,p-q\rangle-\|p-q\|^2\\
&=
\|u-v\|^2-\|(u-v)-(p-q)\|^2\\
&\le
\|u-v\|^2.
\end{aligned}
\]
If \(u,v\in\mathcal A\) are distinct, then \(p=u\) and \(q=v\), so
\eqref{eq:proj_mismatch} holds with equality and the constant cannot be reduced.
\AppendixQED{lem:proj_ineq}

\begin{lemma}
	\proptitle{\hyperlink{proof:proj_lower}{Projection mismatch lower bound}}
\label{lem:proj_lower} \hypertarget{lem:proj_lower}{}
	Let \(\mathcal A\subset\mathbb R^n\) be nonempty, closed, and convex, and let
\(P_{\mathcal A}:\mathbb R^n\to\mathcal A\) denote Euclidean projection. Then, for any
\(g\in\mathbb R^n\) and any \(\tau\in\mathcal A\),
\begin{equation}
\label{eq:proj_lower}
\|\tau-g\|^2-\|P_{\mathcal A}(g)-g\|^2
\ge
\|\tau-P_{\mathcal A}(g)\|^2 .
\end{equation}
\end{lemma}
\hypertarget{proof:proj_lower}{}
\AppendixProofHeading{Proof of Lemma~\ref{lem:proj_lower}} Write \(\tau^\star:=P_{\mathcal A}(g)\). Expanding,
\[
\|\tau-g\|^2=\|\tau-\tau^\star\|^2+2\langle \tau-\tau^\star,\ \tau^\star-g\rangle+\|\tau^\star-g\|^2 .
\]
Since \(\tau^\star\) minimizes the distance to \(g\) over \(\mathcal A\) and \((1-t)\tau^\star+t\tau\in\mathcal A\) for all \(t\in[0,1]\), expanding \(\|g-(1-t)\tau^\star-t\tau\|^2\ge\|g-\tau^\star\|^2\) and letting \(t\to 0\) gives \(\langle g-\tau^\star,\ \tau-\tau^\star\rangle\le 0\). The middle term above is therefore nonnegative, and \eqref{eq:proj_lower} follows. \AppendixQED{lem:proj_lower}

Taking \(g=g_G(\hat y,\sigma)\) and \(\tau=\delta(\hat y)\) for a rule \(\delta\in\mathcal D\), and using that posterior expected loss equals \(\|\tau-g\|^2/n\) plus a term that does not depend on \(\tau\), Lemma~\ref{lem:proj_lower} implies
\[
\operatorname{BR}(\delta,\tau_G^\star\mid\sigma)\ \ge\ \frac{1}{n}\,\E\bigl\|\delta(\hat Y)-\tau_G^\star(\hat Y)\bigr\|^2
\qquad\text{for every }\delta\in\mathcal D .
\]

\clearpage
\end{bibunit}

\phantomsection
\pdfbookmark[0]{Online Appendix}{online-appendix}
\thispagestyle{empty}
\vspace*{\fill}
\begin{center}
  {\LARGE\bfseries Online Appendix\par}
  \vspace{1.25em}
  {\Large Poverty Targeting with Imperfect Information\par}
  \vspace{0.9em}
  {\large Juan C. Yamin\par}
  \vspace{0.6em}
  {\normalsize September 22, 2026\par}
\end{center}
\vspace*{\fill}
\clearpage

\begin{bibunit}
\startcontents[onlineappendix]
\phantomsection
\pdfbookmark[1]{Contents of the Online Appendix}{online-appendix-contents}
\section*{Contents of the Online Appendix}
{\small\contentsmargin{2.8em}\printcontents[onlineappendix]{}{1}{\setcounter{tocdepth}{2}}}
\clearpage

\setcounter{table}{0}
\setcounter{figure}{0}
\renewcommand{\thetable}{\thesection.\arabic{table}}
\renewcommand{\thefigure}{\thesection.\arabic{figure}}
\renewcommand{\theHtable}{\thesection.\arabic{table}}
\renewcommand{\theHfigure}{\thesection.\arabic{figure}}

\section{One-Sided Poverty Gap Objective}
\label{app:loss_functions}

This appendix replaces the quadratic loss of the main text with the
one-sided poverty-gap objective
\[
L_{+}(\tau,y)
=
\frac{1}{n}\sum_{i=1}^{n}(z-y_i-\tau_i)_{+}^{2},
\]
which equals \(z^{2}\) times the Foster--Greer--Thorbecke index with
parameter \(\alpha=2\). Transfers have value only insofar as they close
poverty gaps, and inclusion errors carry no cost beyond the budget they
use. Under imperfect information, I evaluate the objective at conditional
mean income. Unlike in the quadratic loss case, this is not equivalent to
minimizing expected realized poverty, because the residual term no longer
separates from the transfer.

The perfect-information rule of Proposition~\ref{prop:full_info} remains
optimal, and the plug-in rule is again inadmissible when \(B>z\). The
oracle equalizes posterior expected remaining gaps across recipients, and
the empirical Bayes rule uses estimated counterparts of these
expectations. Posterior means alone no longer determine the allocation,
since the expected remaining gap depends on the posterior distribution of
the gap above the transfer received, not only on its mean. Under bounded
support and strict fiscal scarcity, I obtain an \(O(1/\log n)\)
Bayes-regret bound for the feasible rule, a weaker guarantee than the
polynomial bound in Proposition~\ref{prop:regret}.

\subsection{Perfect Information}
\label{sec:L_plus_perfect_info}

Any transfer can be capped at the household's poverty gap
\((z-y_i)_+\) without changing \(L_+\) or violating the budget. On this
restricted set, the one-sided and quadratic losses differ only by a
constant, so they have the same minimizer.

\begin{corollary}
\proptitle{\hyperlink{proof:full_info_L_plus}{Perfect-Information Policy under $L_+$}}
\label{cor:full_info_L_plus} \hypertarget{cor:full_info_L_plus}{}
The policy \(\tau^{\PIlabel}(y)\) of Proposition~\ref{prop:full_info} is
optimal under \(L_+\). It is the unique minimizer if
\(B\le\sum_{i=1}^n(z-y_i)_+\). If \(B>\sum_{i=1}^n(z-y_i)_+\), the
minimum loss is zero, and the minimizers are exactly the feasible
allocations satisfying \(\tau_i\ge(z-y_i)_+\) for every household.
\end{corollary}

Transfers beyond the poverty line leave \(L_+\) unchanged, which is why
uniqueness fails once the budget covers every gap. Since the
perfect-information policy is unchanged, so is the plug-in rule of
Section~\ref{sec:plugin}, which applies \(\tau^{\PIlabel}\) to estimated
incomes.

\subsection{Predictable versus Realized Poverty}
\label{sec:L_plus_y_mu}

Under imperfect information, the one-sided loss of a rule \(\delta\) can be
evaluated at realized income or at conditional mean income, and the two
evaluations, the realized-income objective \(\E[L_+(\delta(\hat Y),y)]\)
and the predictable-income objective \(\E[L_+(\delta(\hat Y),\mu)]\), can
rank rules differently. Under the quadratic loss, the residual adds its
variance to every household's expected loss whatever the transfer, which
is what makes the two evaluations agree in Remark~\ref{remark:losses}.
Under the one-sided loss, the amount it adds depends on where the transfer
leaves the household. A household brought exactly to the poverty line in
conditional mean income has zero loss under the predictable-income
objective but can have positive expected loss under the realized-income
objective, because negative residual draws leave it poor. Whether this
residual poverty enters the objective is a choice of targeting mandate.

\begin{lemma}
\proptitle{\hyperlink{proof:sandwich_Lplus}{Bounds between realized and predictable one-sided risk}}
\label{lem:sandwich_Lplus} \hypertarget{lem:sandwich_Lplus}{}
Suppose \(\E[\varepsilon_i\mid \mu,\hat Y]=0\) and \(\E[\varepsilon_i^2]<\infty\)
for every \(i\), as in Remark~\ref{remark:losses}, and let
\(\bar v_n:=\frac{1}{n}\sum_{i=1}^n\E[\varepsilon_i^2]\). Then, for every
decision rule \(\delta\in\mathcal D\),
\begin{equation}
\label{eq:sandwich_Lplus}
0 \le \E\big[L_+(\delta(\hat Y),y)\big]-\E\big[L_+(\delta(\hat Y),\mu)\big] \le \bar v_n,
\end{equation}
where the expectation is over the signals and the residuals at fixed \(\mu\).
\end{lemma}

For every rule, the realized-income objective exceeds the
predictable-income objective by at most the average residual variance
\(\bar v_n\). Switching objectives therefore changes the risk difference
between any two rules by at most \(\bar v_n\), so a rule that beats
another by more than \(\bar v_n\) under one objective also beats it under
the other. The realized-income objective depends on the probability and
depth of poverty generated by residual variation, and the maintained
model does not specify the conditional residual distribution needed to
evaluate it. The rest of this appendix therefore studies the
predictable-income objective.

\subsection{Inadmissibility of Plug-In Rule}
\label{sec:L_plus_inadmissibility}

Under \(L_+\), the plug-in rule can misallocate budget even when it
exhausts it, and a direct dominance argument shows this without the
Bayes-risk and complete-class arguments of Section~\ref{sec:inadmissible}.
In gap notation, \(L_+(\tau,\theta)=\frac{1}{n}\sum_{i=1}^n(\theta_i-\tau_i)_+^2\)
with \(\theta_i=z-\mu_i\), and since \(\mu_i\ge 0\), every predictable
poverty gap is at most \(z\).\footnote{The gap formulation of Section~\ref{subsec:equivalence} applies verbatim under \(L_+\), since \((z-\mu_i-\tau_i)_+^2=(\theta_i-\tau_i)_+^2\).}
A transfer above \(z\) is not penalized under \(L_+\), but the excess
spends budget without reducing poverty. The plug-in rule makes such
transfers with positive probability. It projects the noisy gap vector
\(X\) onto the budget set, and because the Gaussian signal has unbounded
support, some component of \(P_{\mathcal A}(X)\) exceeds \(z\) on a set
of positive probability whenever \(B>z\).

Capping every transfer at \(z\) and spreading the freed budget uniformly
across households gives the cap-and-spread rule
\[
\tau_i^{\cslabel}(x)
:=
\min\!\bigl\{\tau_i^{\pluglabel}(x),\,z\bigr\}
+ \frac{1}{n} \sum_{j=1}^{n}\bigl(\tau_j^{\pluglabel}(x)-z\bigr)_{+},
\qquad i=1,\ldots,n,
\]
which preserves total spending and uses no information beyond \(x\).
Capping leaves the loss of the affected households at zero, since a
transfer of \(z\) already closes any gap, and spreading cannot raise the
loss of anyone else, since \(L_+\) is nonincreasing in transfers. When
total poverty gaps exceed the budget, the plug-in allocation leaves some
gap uncovered. On realizations where the cap releases funds, that
household receives more, and its loss falls strictly.

\begin{proposition}
\proptitle{\hyperlink{proof:Lplus_simple_dominance}{A simple domination of the plug-in rule under $L_+$}}
\label{prop:Lplus_simple_dominance}
\hypertarget{prop:Lplus_simple_dominance}{}
Let $R_+(\delta,\theta):=\E_\theta[L_+(\delta(X),\theta)]$ denote the frequentist risk under the loss \(L_+\). Under Assumption~\ref{assump:normality} and $B>z$,
\begin{itemize}
    \item[(i)] For every $\theta\in(-\infty,z]^n$, $R_+(\tau^{\cslabel},\theta)\le R_+(\tau^{\pluglabel},\theta)$.

    \item[(ii)] For every $\theta\in(-\infty,z]^n$ satisfying $\sum_{i=1}^n (\theta_i)_+>B$, the inequality in \textup{(i)} is strict.
\end{itemize}
In particular, the plug-in rule is inadmissible on any parameter space $\Theta\subseteq(-\infty,z]^n$ that contains at least one $\theta$ with $\sum_{i=1}^n (\theta_i)_+>B$.
\end{proposition}

Nothing recommends \(\tau^{\cslabel}\) beyond this comparison. It shows
only that noisy signals lead the plug-in rule to spend budget on
transfers no household could need, and that under \(L_+\) those dollars
have no poverty-reduction value.

\subsection{Oracle and Empirical Bayes Rules under \texorpdfstring{\(L_+\)}{L+}}
\label{sec:L_plus_EB}

Under the quadratic loss, household \(i\) enters the oracle problem only
through its posterior mean gap \(g_{G,i}\), and the oracle of
Proposition~\ref{prop:oracle_bayes_rule} projects the vector of posterior
mean gaps onto the budget set. Under \(L_+\), the posterior expected loss
of household \(i\) at transfer \(t\) is
\(\E_G[(z-\mu_i-t)_+^2\mid \hat y_i,\sigma_i]\), and an additional dollar
lowers it only in the posterior states where the household is still below
the poverty line after receiving \(t\). Under the moment condition of
Proposition~\ref{prop:oracle_bayes_rule_Lplus},
\[
-\frac{\partial}{\partial t}
\E_G\!\left[(z-\mu_i-t)_+^2\mid \hat y_i,\sigma_i\right]
=
2\, g^+_{G,i}(t\mid \hat y_i,\sigma_i),
\qquad
g^+_{G,i}(t\mid \hat y_i,\sigma_i)
:=
\E_G\!\left[(z-\mu_i-t)_+\mid \hat y_i,\sigma_i\right],
\]
so \(2g^+_{G,i}(t\mid \hat y_i,\sigma_i)\) is the value of the next dollar
to household \(i\), and \(g^+_{G,i}(t\mid \hat y_i,\sigma_i)\) is its
posterior expected remaining gap after a transfer of \(t\). The expected
remaining gap is the posterior probability that the household remains poor
after the transfer times its expected shortfall in that event,
\[
g^+_{G,i}(t\mid \hat y_i,\sigma_i)
=
\Pr_G\!\left(z-\mu_i>t\mid \hat y_i,\sigma_i\right)\,
\E_G\!\left[z-\mu_i-t\mid z-\mu_i>t,\hat y_i,\sigma_i\right].
\]
Panel A of Figure~\ref{fig:Lplus_oracle} compares the marginal values for
household 1 under the two losses. Under quadratic loss,
\(2(g_{G,1}-t)\) reaches zero at the posterior mean gap and becomes negative
beyond it, because the loss also penalizes predictable income above the
poverty line after the transfer. Under one-sided loss,
\(2g^+_{G,1}(t)\) lies weakly above the quadratic line. With a nondegenerate
posterior, it remains positive at \(t=g_{G,1}\), because states with gaps
above the mean remain poor and an additional transfer still reduces
expected loss in those states. The size of this expected shortfall depends
on how the posterior distributes probability around its mean. In
particular, because \((z-\mu_i-t)_+\) is convex in \(\mu_i\), a
mean-preserving spread of the posterior weakly raises \(g^+_{G,i}(t)\) at
every \(t\). Households with the same posterior mean gap can therefore
value an additional dollar differently and, as the next result shows,
receive different transfers under \(L_+\).

\begin{figure}[H]
    \centering
    \setstretch{1}
    \setlength{\abovecaptionskip}{0pt}
    \setlength{\belowcaptionskip}{6pt}
    \caption{Marginal values and allocation under one-sided loss}
    \label{fig:Lplus_oracle}
    \input{figures/jmp_v3/figure_Lplus_oracle.tex}

    \vspace{8pt}
    \begin{minipage}{6.25in}
    \footnotesize
    \emph{Notes.} Curves show marginal reductions in expected
household loss; under $L_+$, these are $2g^+_{G,i}(t)$.
A common bounded uniform prior and Gaussian signals yield the illustrated
truncated-normal posteriors. Households 1 and 2 share the same posterior mean,
but household 1's noisier signal produces a more dispersed posterior.
Panel A uses household 1 from Panel B.
Recipients reach $2\lambda_{G,+}$, while household 3 receives nothing because
$2g^+_{G,3}(0)<2\lambda_{G,+}$.
    \end{minipage}
\end{figure}

Because the posterior expected loss is nonincreasing in the transfer, an
optimal allocation can always spend the whole budget, so the proposition
imposes \(\sum_i\tau_i=B\).

\begin{proposition}
\proptitle{\hyperlink{proof:oracle_bayes_rule_Lplus}{Oracle Bayes Rule under $L_+$}}
\label{prop:oracle_bayes_rule_Lplus}
\hypertarget{prop:oracle_bayes_rule_Lplus}{}
Under Assumptions~\ref{assump:normality} and~\ref{assump:prior}, fix
\((\hat y,\sigma)\), and assume that each household has finite posterior
one-sided second moment, $\E_G[(z-\mu_i)_+^2\mid \hat y_i,\sigma_i]<\infty$. Consider the equality-constrained oracle Bayes problem
\[
\min_{\tau\in\mathbb R_+^n}
\frac1n\sum_{i=1}^n
\E_G\!\left[(z-\mu_i-\tau_i)_+^2\mid \hat y_i,\sigma_i\right]
\quad\text{subject to}\quad
\sum_{i=1}^n \tau_i = B .
\]
Then, there exist an optimal allocation \(\tau^\star_{G,+}\) and a scalar \(\lambda_{G,+}\ge 0\) such that $\sum_{i=1}^n \tau^\star_{G,+,i}=B$ and, for each household \(i\), $\tau^\star_{G,+,i}=0$ only if $g^+_{G,i}(0\mid \hat y_i,\sigma_i)\le \lambda_{G,+}$, while any household with \(\tau^\star_{G,+,i}>0\) satisfies $g^+_{G,i}(\tau^\star_{G,+,i}\mid \hat y_i,\sigma_i) = \lambda_{G,+}$.

\end{proposition}

Under \(L_+\), the oracle equalizes the posterior expected remaining gaps
across recipients, as the perfect-information rule equalizes the actual
remaining gaps. Equivalently, recipients reach the common marginal value
\(2\lambda_{G,+}\). If two recipients had different marginal values, a
small reallocation toward the one with the higher value would reduce
the expected loss.

Panel B of Figure~\ref{fig:Lplus_oracle} illustrates this allocation for
three households, with the cutoff chosen so that the transfers exhaust the
budget. Households 1 and 2 share the same posterior mean gap, so the
quadratic oracle would give them equal transfers. Here, household 1's
noisier signal produces a more dispersed posterior, so its marginal value
reaches \(2\lambda_{G,+}\) at a larger transfer. Household 3 receives
nothing because its marginal value at a zero transfer is already below
\(2\lambda_{G,+}\), even though poverty remains possible.

The empirical Bayes rule replaces \(G\) with the Kiefer--Wolfowitz NPMLE
\(\hat G\). For each household \(i\) and transfer level \(t\ge 0\), let
\(\hat g_i^+(t\mid \hat y_i,\sigma_i):=\E_{\hat G}[(z-\mu_i-t)_+\mid \hat y_i,\sigma_i]\).
Since \(\hat G\) is discrete, \(\hat g_i^+\) is a finite posterior-weighted
sum of the lines \((z-u-t)_+\) over the support points \(u\) of \(\hat G\),
hence continuous and piecewise linear in \(t\). In the binding-budget case,
the empirical Bayes allocation \(\hat\tau^{\EBlabel}_+\) is defined by
replacing \(g^+_{G,i}\) with \(\hat g_i^+\) and \(\lambda_{G,+}\) with the
budget-clearing cutoff \(\hat\lambda_+\) in the oracle
characterization.\footnote{If several allocations are optimal, \(\hat\tau^{\EBlabel}_+\) denotes a fixed measurable choice among those that spend the full budget.} Its input is a remaining-gap function for each household
rather than a vector of posterior mean gaps.

\subsection{Bayes Regret under \texorpdfstring{\(L_+\)}{L+}}
\label{sec:L_plus_regret}

Under the assumptions below, the Bayes regret of the feasible rule is
\(O(1/\log n)\). Two ingredients deliver the bound. Strict scarcity keeps the
oracle's cutoff bounded away from zero with high probability,
and a positive cutoff makes the posterior expected loss grow at least
quadratically with the distance from the oracle allocation, so errors in
the estimated remaining-gap functions translate into regret. Those errors
are controlled on average across households by the Wasserstein accuracy of
the NPMLE, which is where the slow rate of Gaussian deconvolution enters.

\subsubsection{Regret Criterion and Remaining-Gap Error}

As in Section~\ref{sec:bayes_regret_results}, the feasible rule is
evaluated by its Bayes regret relative to the oracle rule under the true
prior,
\[
\operatorname{BR}_+\bigl(\hat\tau^{\EBlabel}_+,\tau^\star_{G,+}\mid\sigma\bigr)
:=
\E_{G,\hat Y}\!\left[
L_+\!\left(\hat\tau^{\EBlabel}_+(\hat Y,\sigma),\mu\right)
-
L_+\!\left(\tau^\star_{G,+}(\hat Y,\sigma),\mu\right)
\,\middle|\,\sigma\right],
\]
where the expectation is over \((\mu,\hat Y)\) at fixed \(\sigma\). The two
rules solve the same allocation problem and differ only in the prior they
use, so the regret is the cost of replacing \(G\) with \(\hat G\).

Throughout, the true prior \(G\) and the fitted prior \(\hat G\) are
assumed to be supported on a bounded interval of nonnegative incomes, and
\(\hat\tau^{\EBlabel}_+\) is the rule of Section~\ref{sec:L_plus_EB}
computed from \(\hat G\). Nonnegative incomes mean that no poverty gap
exceeds \(z\), so a transfer above \(z\) cannot reduce loss further. When
\(0<B<nz\), any allocation that spends \(B\) can be capped at \(z\) per
household and the amount removed redistributed within the caps without
raising loss. Lemma~\ref{lem:Lplus_cap} therefore lets both rules be
chosen with \(0\le\tau_i\le z\) and \(\sum_i\tau_i=B\) without changing
their optimal values. From here on, \(\tau^\star_{G,+}\) and
\(\hat\tau^{\EBlabel}_+\) denote these capped rules, and
\(\lambda_{G,+}\ge 0\) is the common posterior expected remaining gap at
the oracle allocation.

The statistical input to regret is the discrepancy between the estimated
and true remaining-gap functions, measured by its root mean square across
households, \(r_+(\hat y,\sigma)\), where
\[
r_+^2(\hat y,\sigma)
:=
\frac1n\sum_{i=1}^n
\Bigl[\,\sup_{t\in[0,z]}
\bigl|\hat g_i^+(t\mid\hat y_i,\sigma_i)-g^+_{G,i}(t\mid\hat y_i,\sigma_i)\bigr|
\,\Bigr]^2 .
\]
The supremum over \(t\) is needed because each household's transfer is
chosen after the data are observed, so the estimated function must be accurate
at a data-dependent point, and \([0,z]\) covers every transfer the capped
rules can choose. Averaging across households matches the normalization
of the loss, and since both functions take values in \([0,z]\), each
household contributes at most \(z^2/n\).

\subsubsection{Assumptions}

The regret bound needs two conditions beyond the maintained assumptions,
one on the estimated prior and one on the budget.

\begin{assumption}[Compact support and support-constrained approximate NPMLE]
\label{assump:Lplus_support}
\hypertarget{assump:Lplus_support}{}
For some \(M<\infty\), the true and the estimated mixing distributions satisfy
\(G([0,M])=1\) and \(\hat G([0,M])=1\). In addition, there is a constant
\(C_\ell<\infty\) such that
\begin{equation}
\label{eq:Lplus_likelihood}
\ell_n(G)-\ell_n(\hat G)\le C_\ell\,\frac{(\log n)^2}{n},
\qquad\text{where}\quad
\ell_n(H):=\frac1n\sum_{i=1}^n\log\int\phi_{\sigma_i}(\hat Y_i-u)\,dH(u).
\end{equation}
Here, \(\phi_{\sigma}\) denotes the \(N(0,\sigma^2)\) density.
\end{assumption}

The support condition has two roles. Nonnegative incomes bound every
poverty gap by \(z\), which the capped rules use. The upper bound \(M\)
keeps the posterior stable in the prior. A household's remaining-gap
function is a ratio of integrals against the prior, and by
Lemma~\ref{lem:Lplus_prior_stability}, its estimation error on a bounded
support is at most the Wasserstein distance \(W_2(G,\hat G)\) times a
factor that depends only on the household's own signal. Accuracy of
\(\hat G\) in Wasserstein distance is therefore what the allocation needs.

The likelihood condition \eqref{eq:Lplus_likelihood} is the
approximate-maximizer condition of \citeA{soloff2021multivariate}. It
requires \(\hat G\) to explain the data nearly as well as the true prior
does, and their deconvolution theory shows that this alone makes an
approximate NPMLE close to the true prior. By Lemma~\ref{lem:Lplus_w2},
their argument adapted to the bounded support gives
\(W_2^2(G,\hat G)\lesssim1/\log n\) with probability at least
\(1-2/n^8\). The condition is easy to meet. An exact maximizer over
distributions supported on \([0,M]\) satisfies it with zero gap, since
\(G\) itself is a candidate. By Lemma~\ref{lem:Lplus_grid}, a
maximizer over a grid of mesh \(\delta_n\) computed to a likelihood
tolerance \(\eta_n\) satisfies it whenever
\(\delta_n^2+\eta_n\lesssim(\log n)^2/n\).\footnote{The empirical
implementation estimates \(\hat G\) on 600 equally spaced points over
\([0,\max\{0,\max_{1\le i\le n}\hat Y_i\}+10^{-4}]\), with a GLmix relative
tolerance of \(10^{-8}\). The upper endpoint is data dependent rather than
a known bound on the support of \(G\), and the mesh and tolerance were not
chosen to satisfy the rate condition of Lemma~\ref{lem:Lplus_grid}, so
Proposition~\ref{prop:Lplus_regret_rate} does not formally cover the
simulation implementation.}

The deepest poverty gap the prior allows is
\(z-\inf\operatorname{supp}(G)\), the gap of a household at the lowest
income in the support of \(G\). Scarcity requires spending per household
to stay below this gap by a margin.

\begin{assumption}[Strict fiscal scarcity]
\label{assump:Lplus_scarcity}
\hypertarget{assump:Lplus_scarcity}{}
There is a constant \(\bar b<z-\inf\operatorname{supp}(G)\) such that
\(B/n\le\bar b\) for all sufficiently large \(n\).
\end{assumption}

A constant total budget satisfies the condition once the population is
large enough, provided \(G\) puts positive mass below \(z\), and a budget
proportional to the population satisfies it whenever the proportion is
below \(z-\inf\operatorname{supp}(G)\). The threshold is the deepest gap
the prior allows, so the condition can hold even when \(B\) exceeds the
total poverty gap of the population at hand. If spending per household
reached the threshold, the program could lift every household from the
deepest possible poverty to the poverty line, and targeting would stop
being a rationing problem.

The role of scarcity is to keep the oracle's cutoff, the common expected
remaining gap at the oracle allocation, away from zero. A small cutoff
means that every household is left with a small expected remaining gap,
so every household with a signal in a moderate range must already have
received a sizable transfer. Under scarcity, the budget cannot afford
sizable transfers to that many households, and such signals occur for a
stable share of the population, so a small cutoff is an event of
exponentially small probability. By Lemma~\ref{lem:Lplus_shadow}, there
are constants \(\underline\lambda>0\) and \(c_\lambda>0\) such that
\(P_{G,\hat Y}(\lambda_{G,+}<\underline\lambda\mid\sigma)\le e^{-c_\lambda n}\)
whenever \(B/n\le\bar b\). Together, the first assumption controls the
average error in the remaining-gap functions, and the second controls the
cutoff that converts that error into regret.

\subsubsection{Bayes Regret}

When the oracle's cutoff is positive, the additional posterior
expected loss from using estimated remaining-gap functions in place of
the true ones is at most a multiple of their average squared error.

\begin{lemma}
\proptitle{\hyperlink{proof:Lplus_global_stability}{Global stability of the \(L_+\) allocation}}
\label{lem:Lplus_global_stability}
\hypertarget{lem:Lplus_global_stability}{}
Let Assumptions~\ref{assump:normality}, \ref{assump:prior},
and~\ref{assump:Lplus_support} hold, and let \(0<B<nz\). For the capped
oracle and empirical Bayes rules defined above, any realization
\((\hat y,\sigma)\) with \(\lambda_{G,+}>0\) satisfies
\[
\E_G\!\left[L_+(\hat\tau^{\EBlabel}_+,\mu)-L_+(\tau^\star_{G,+},\mu)
\,\middle|\,\hat y,\sigma\right]
\le
\frac{4z}{\lambda_{G,+}}\,r_+^2(\hat y,\sigma).
\]
\end{lemma}

The bound comes from how the posterior expected loss rises when transfers
move away from the oracle allocation. At the oracle allocation, every
recipient has expected remaining gap \(\lambda_{G,+}\), and every
household that receives nothing has a gap no larger. A reallocation
raises some transfers and lowers others while keeping total spending at
\(B\). Because the remaining-gap functions are convex and reach zero at
\(z\), raising a transfer by \(d\) puts its expected remaining gap at
least \(\lambda_{G,+}d/z\) below the common level, and lowering a
transfer by \(d\) puts it at least that far above. The marginal
reduction in household loss is twice the expected remaining gap. Valued
at \(\lambda_{G,+}\), the additions and the
cuts cancel, since total spending is unchanged, so the loss from the
reallocation is the accumulation of these departures, which grows with
the square of the change. For every allocation \(\tau\) with
\(0\le\tau_i\le z\) and \(\sum_{i=1}^n\tau_i=B\),
\[
\E_G\!\left[L_+(\tau,\mu)\,\middle|\,\hat y,\sigma\right]
-
\E_G\!\left[L_+(\tau^\star_{G,+},\mu)\,\middle|\,\hat y,\sigma\right]
\ \ge\
\frac{\lambda_{G,+}}{z}\,
\frac{\|\tau-\tau^\star_{G,+}\|^2}{n}.
\]
The empirical Bayes allocation minimizes the estimated objective, so its
excess true loss is at most the estimation error at the coordinates it
moves, and the quadratic growth converts that into the lemma.

The support and likelihood condition of
Assumption~\ref{assump:Lplus_support} bounds the average squared error
itself.

\begin{lemma}
\proptitle{\hyperlink{proof:Lplus_curve_error}{Average remaining-gap error}}
\label{lem:Lplus_curve_error}
\hypertarget{lem:Lplus_curve_error}{}
Let Assumptions~\ref{assump:normality}, \ref{assump:prior},
and~\ref{assump:Lplus_support} hold. There are constants \(C_r<\infty\) and
\(n_r<\infty\), depending only on \(z\), \(M\), \(\sigma_{\min}\),
\(\sigma_{\max}\), and the constant in the approximate-likelihood condition,
such that for all \(n\ge n_r\),
\[
\E_{G,\hat Y}\!\left[r_+^2(\hat Y,\sigma)\,\middle|\,\sigma\right]
\le
\frac{C_r}{\log n}+\frac{C_r}{n^{8}} .
\]
\end{lemma}

A household's remaining-gap function moves at most in proportion to the
Wasserstein distance between the estimated and the true prior, and by
Lemma~\ref{lem:Lplus_w2}, that distance is of order \(1/\sqrt{\log n}\)
except on an event of polynomially small probability. Squaring and
averaging gives the first term, and the exceptional event gives the
second. This is where the rate falls behind the main text. The
quadratic-loss bound of Section~\ref{sec:bayes_regret_results} needs
only accurate posterior means, which empirical Bayes theory delivers at
polynomial rates without recovering \(G\), whereas the remaining-gap
functions depend on the whole posterior, and the proof recovers \(G\)
itself in Wasserstein distance, a Gaussian deconvolution problem whose
rates are logarithmic.

Under strict scarcity, the cutoff stays away from zero with high
probability, and the two lemmas combine into the regret bound.

\begin{proposition}
\proptitle{\hyperlink{proof:Lplus_regret_rate}{Finite-Sample Bayes Regret Bound under \(L_+\)}}
\label{prop:Lplus_regret_rate}
\hypertarget{prop:Lplus_regret_rate}{}
Let Assumptions~\ref{assump:normality}, \ref{assump:prior},
\ref{assump:Lplus_support}, and~\ref{assump:Lplus_scarcity} hold, and
consider the capped oracle and empirical Bayes rules defined above. There
are constants \(C<\infty\), \(c>0\), and \(n_0<\infty\), none depending
on \(n\) or on the particular vector \(\sigma\), such that for all
\(n\ge n_0\),
\[
\operatorname{BR}_+\bigl(\hat\tau^{\EBlabel}_+,\tau^\star_{G,+}\mid\sigma\bigr)
\le
\frac{C}{\log n}+Ce^{-cn}+\frac{C}{n^{8}} .
\]
In particular,
\(\operatorname{BR}_+(\hat\tau^{\EBlabel}_+,\tau^\star_{G,+}\mid\sigma)
=O(1/\log n)\) uniformly over noise-level vectors satisfying
Assumption~\ref{assump:normality}.
\end{proposition}

On the event that the cutoff is at least \(\underline\lambda\),
Lemma~\ref{lem:Lplus_global_stability} bounds conditional regret by
\((4z/\underline\lambda)\,r_+^2\), and Lemma~\ref{lem:Lplus_curve_error}
bounds its expectation. Off the event, conditional regret is at most
\(z^2\), since incomes and transfers are nonnegative. By
Lemma~\ref{lem:Lplus_shadow}, the event fails with exponentially small
probability. The three terms are the price of Gaussian deconvolution,
the probability of a small cutoff, and the failure probability of the
Wasserstein guarantee. Without Assumption~\ref{assump:Lplus_scarcity},
the comparison behind Lemma~\ref{lem:Lplus_global_stability} still bounds
conditional regret by \(2z\,r_+\), and the same argument gives regret of
order \((\log n)^{-1/2}\), so scarcity improves the rate rather than
enabling vanishing regret.

\clearpage

\setcounter{table}{0}
\setcounter{figure}{0}
\renewcommand{\thetable}{\thesection.\arabic{table}}
\renewcommand{\thefigure}{\thesection.\arabic{figure}}
\renewcommand{\theHtable}{\thesection.\arabic{table}}
\renewcommand{\theHfigure}{\thesection.\arabic{figure}}

\section{Fixed Benefit Menus}
\label{app:fixed_grants}

The main text lets the policymaker give every household its own amount,
and each of its allocation rules projects a vector of poverty gaps onto
the convex budget set. Many programs instead use a short list of
benefit levels set in advance, which simplifies administration and makes
costs easy to forecast. The targeting system decides who receives
nothing, who receives the basic amount, and who receives
more.\footnote{Georgia's Targeted Social Assistance pays a per-member
amount graded by the household's PMT score
\cite{unicef2019georgia}, and South Africa's Child Support Grant pays a
fixed amount per eligible child \cite{barrientos2013social}.} This
appendix extends the quadratic targeting problem to such a menu. The
policymaker observes the noisy income estimates and chooses which
level, if any, each household receives, subject to the budget. The
simplest case is a single common grant, for which only the beneficiary
list is chosen, a budget-constrained relative of the selection problems
of \citeA{chen2023empirical}.

Replacing the budget set by the finite set of menu allocations changes
the geometry but not the logic. Under perfect information, the optimal
allocation is the menu allocation nearest to the vector of poverty
gaps, the oracle is the menu allocation nearest to the vector of
posterior mean poverty gaps, and the empirical Bayes rule is the menu
allocation nearest to the estimated posterior means. What changes is
the cost of estimation error. Projection onto the convex budget set
does not increase distances, so in the main text, an error moves the
allocation by no more than the error itself, and the regret bound is
quadratic in the error. With a menu, a small error can switch a
household from one level to another and move the allocation by a full
menu step. The switch is cheap in loss, because it happens only when
the two allocations are nearly tied, but the regret bound becomes
linear in the error rather than quadratic. With a fixed menu and a
budget proportional to the population, it is of the order of the
square root of the bound in the main text, a weaker guarantee.

\subsection{Setup}
\label{sec:ds_setup}

The decision problem is that of the main text with the budget set
replaced by a menu. A menu of payment levels
\(\mathcal{C}=\{0,c_{1},\ldots,c_{J}\}\) with \(0<c_{1}<\cdots<c_{J}\)
is announced before the signals are observed, zero denotes no benefit,
and each household is assigned one level. The feasible allocations are
$\mathcal{A}_{\mathcal{C}} := \bigl\{\tau\in\mathcal{C}^{n} \,\bigm|\, \mathbf{1}^{\top}\tau\le B\bigr\}$,
and I assume \(c_{1}\le B\), so that some positive level is affordable.
The set \(\mathcal{A}_{\mathcal{C}}\) is a finite subset of the budget
set \(\mathcal{A}\) of the main text, so every menu allocation is
feasible in the continuous problem, but it is not convex.

Every rule in this section has the form of its continuous counterpart,
a nearest point of the feasible set to a vector of poverty gaps. For
\(v\in\mathbb{R}^{n}\), let \(P_{\mathcal{A}_{\mathcal{C}}}(v)\) denote a
menu allocation nearest to \(v\),
$P_{\mathcal{A}_{\mathcal{C}}}(v) \in \arg\min_{\tau\in\mathcal{A}_{\mathcal{C}}}\|\tau-v\|^{2}$.
Without the budget, the nearest allocation rounds each
coordinate of \(v\) to its nearest menu level. When that rounding is
unaffordable, the budget couples the households, and the nearest
allocation must be chosen jointly. Unlike the projection onto
\(\mathcal{A}\), this map can send two nearby vectors to allocations a
full menu step apart.

\subsection{Perfect Information}
\label{sec:ds_pi}

When the income vector \(y\) is observed, the objective is proportional
to the squared distance between \(\tau\) and the poverty gap vector
\(z\mathbf{1}-y\), as in Section~\ref{sec:perfect}, so an optimal
allocation is $\tau^{\PIlabel}_{\mathcal{C}}(y) = P_{\mathcal{A}_{\mathcal{C}}}(z\mathbf{1}-y)$.
Without the budget, the projection rounds each poverty gap to its
nearest menu level, and the midpoints between adjacent levels mark
where the assignment switches. With the budget, this rounding remains
optimal whenever it is affordable. When it is not, some households must
be moved down the menu, and the projection chooses which ones jointly,
but it never moves a poorer household below a richer one.

\begin{proposition}
\proptitle{\hyperlink{proof:ds_pi_menu}{Perfect-Information Menu Allocation}}
\label{prop:ds_pi_menu}
\hypertarget{prop:ds_pi_menu}{}
Relabel households so that \(y_{1}\le y_{2}\le\cdots\le y_{n}\). An
optimal allocation exists.
\begin{enumerate}
\item[(i)] If some allocation assigning each household a menu level
nearest to its poverty gap is feasible, the optimal allocations are
exactly the feasible allocations of this form. In this case, the optimal
allocation is unique when no poverty gap lies at a midpoint between
adjacent menu levels.
\item[(ii)] Every optimal allocation satisfies \(\tau_{i}\ge\tau_{k}\)
whenever \(y_{i}<y_{k}\), and an optimal allocation with
\(\tau_{1}\ge\tau_{2}\ge\cdots\ge\tau_{n}\) exists.
\end{enumerate}
\end{proposition}

If a poorer household received less than a richer one, swapping the
two payments would keep the cost and lower the loss, and swapping
payments between households with equal incomes changes neither. The
rationed problem therefore reduces to choosing how many households
receive benefits at each level, with the highest level going to the poorest
households, the next level to the next poorest, and nothing to the rest.
What remains is a tradeoff between depth---raising the level paid to a
poorer household---and breadth---extending benefits at a lower level to more
households. Its resolution depends on the sizes of the poverty gaps
and the spacing of the menu rather than on the income ranking
alone.\footnote{Interestingly, tightening the budget can bring a
household into the program. With menu \(\{0,2,5\}\) and poverty gaps
\(5\) and \(1.1\), a budget of \(5\) pays the poorer household \(5\),
while a budget of \(4\) pays \mbox{each household \(2\)}.} Figure~\ref{fig:menu_projection} illustrates how the fixed menu changes the projection for two households.

\begin{figure}[H]
    \centering
    \setstretch{1}
    \setlength{\abovecaptionskip}{0pt}
    \setlength{\belowcaptionskip}{6pt}
    \caption{Fixed-menu allocation as a projection}
    \label{fig:menu_projection}
    \input{figures/jmp_v3/figure_menu_projection.tex}

    \vspace{8pt}
    \begin{minipage}{6.25in}
    \footnotesize
    \setlength{\baselineskip}{12pt}
    \emph{Notes.} The squares form the feasible menu set
    $\mathcal A_{\mathcal C}$; the triangle outlines the continuous feasible set
    $\mathcal A$. Under quadratic loss, the poverty gap vector projects to the blue
    point under continuous transfers and to the orange square under the fixed menu.
    \end{minipage}
\end{figure}

\subsection{Oracle Bayes Rule}
\label{sec:ds_oracle}

The oracle knows \(G\) and observes the estimates \(\hat{y}\) with
their standard errors \(\sigma\). By the posterior loss decomposition in the proof of
Lemma~\ref{lem:bayes_rules}, the posterior expected loss
equals \(n^{-1}\|\tau-g_{G}(\hat{y},\sigma)\|^{2}\) plus a term that
does not depend on \(\tau\), so the oracle projects the posterior mean
poverty gap vector onto the menu.\footnote{In Section~\ref{sec:noisy}, the convexity of \(\mathcal{A}\)
lets nonrandomized rules dominate randomized ones at every state. Here,
randomization is unnecessary for the Bayes problem, since the posterior
risk of a randomized rule averages the posterior risks of the
allocations it mixes over.}

\begin{proposition}
\proptitle{\hyperlink{proof:ds_oracle_menu}{Oracle Bayes Menu Rule}}
\label{prop:ds_oracle_menu}
\hypertarget{prop:ds_oracle_menu}{}
Under Assumptions~\ref{assump:normality} and~\ref{assump:prior}, an
oracle Bayes rule over \(\mathcal{A}_{\mathcal{C}}\) is $\tau^{\star}_{G,\mathcal{C}}(\hat{y},\sigma) = P_{\mathcal{A}_{\mathcal{C}}}\bigl(g_{G}(\hat{y},\sigma)\bigr)$,
and a measurable rule with values in \(\mathcal{A}_{\mathcal{C}}\) is
Bayes if and only if it selects a menu allocation nearest to
\(g_{G}(\hat{y},\sigma)\) at almost every realization of the data.
\end{proposition}

As in the continuous problem, uncertainty changes only the vector that
is projected. Proposition~\ref{prop:ds_pi_menu} therefore applies with
posterior mean poverty gaps in place of poverty gaps. When assigning
each household the level nearest to its posterior mean gap is
affordable, the oracle does exactly that, so it first shrinks the
estimates and then assigns levels. Otherwise it rations, and
households with larger posterior mean gaps receive weakly larger
transfers. In either case, a household receives level \(c_{j}\) only if
its posterior mean gap is at least \(c_{j}/2\), since below that value,
giving it nothing would lower the posterior expected loss and free budget.
These comparisons use the magnitudes of the posterior mean gaps, not
only their ranking, so shrinkage can change how many households
receive benefits at each level even when it leaves the ranking
unchanged.\footnote{Under a common standard error, the posterior mean
is the same nondecreasing function of every household's estimate, so the
plug-in and oracle rules rank households in the same order and differ
only in the number assigned to each level. With household-specific
standard errors, rank reversals become possible.}

With a single positive benefit level, \(\mathcal{C}=\{0,c\}\), only the
beneficiary list remains to be chosen.

\begin{corollary}
\proptitle{\hyperlink{proof:ds_one_grant}{One Common Grant}}
\label{cor:ds_one_grant}
\hypertarget{cor:ds_one_grant}{}
Let \(\mathcal{C}=\{0,c\}\). At each realization of the data, an oracle
Bayes rule enrolls households in decreasing order of their posterior
mean poverty gaps, stopping when the next household's posterior mean
poverty gap falls below \(c/2\) or when another grant is unaffordable.
\end{corollary}

A grant lowers household \(i\)'s posterior expected loss by
\(2c\,g_{G,i}(\hat{y}_{i},\sigma_{i})-c^{2}\), which is positive exactly
when the posterior mean gap exceeds \(c/2\), the cutoff in the
corollary. Enrollment therefore proceeds down a ranked list and stops
for one of two reasons. Either the next household no longer benefits
from the grant, in which case budget is left over although poverty
remains, or the budget cannot fund another grant, in which case
households who would benefit go unserved. The caseload is part of the
solution rather than fixed in advance. The grant is also indivisible,
so a recipient whose gap lies between \(c/2\) and \(c\) receives more
than its gap, even under perfect information. Shifting gaps by \(c/2\)
turns the single-grant case into a selection problem with a maximum
caseload, closely related to the selection problems of
\citeA{chen2023empirical}.

\subsection{Feasible Empirical Bayes Rule and Regret}
\label{sec:ds_eb}

The feasible rule projects the estimated posterior mean poverty gaps
\(\hat{g}(\hat{y},\sigma)\) of Section~\ref{sec:Bayes}, computed under
the NPMLE \(\hat{G}\) of Section~\ref{sec:npmle}, onto the menu, $\hat{\tau}^{\EBlabel}_{\mathcal{C}}(\hat{y},\sigma) = P_{\mathcal{A}_{\mathcal{C}}}\bigl(\hat{g}(\hat{y},\sigma)\bigr)$.
Estimating the posterior mean gaps is the problem of the main text, so
what changes is how estimation error becomes regret. In
Section~\ref{sec:Bayes}, that step is the projection mismatch
inequality of Lemma~\ref{lem:proj_ineq}, which uses the convexity of
\(\mathcal{A}\) to bound the excess loss by the squared error. The
menu is not convex, and the danger comes from households near an
assignment boundary. For example, with a single grant and two
households whose posterior mean gaps are nearly equal, an arbitrarily
small error reverses their order and moves the grant from one to the
other, so the allocation jumps by a fixed distance however small the
error. Such a switch occurs only when the two allocations are nearly
tied under the true posterior mean gaps, which is why its cost stays
small. Lemma~\ref{lem:ds_comparison} makes this precise.

\begin{lemma}
\proptitle{\hyperlink{proof:ds_comparison}{Projection Mismatch Inequality without Convexity}}
\label{lem:ds_comparison}
\hypertarget{lem:ds_comparison}{}
For any \(u,v\in\mathbb{R}^{n}\),
\[
\begin{aligned}
0\le{}&
\|u-P_{\mathcal{A}_{\mathcal{C}}}(v)\|^{2}
-\|u-P_{\mathcal{A}_{\mathcal{C}}}(u)\|^{2}
\le{}&
2\,\|u-v\|\,
\|P_{\mathcal{A}_{\mathcal{C}}}(u)-P_{\mathcal{A}_{\mathcal{C}}}(v)\|.
\end{aligned}
\]
\end{lemma}

When \(u=g_{G}(\hat{y},\sigma)\) and \(v=\hat{g}(\hat{y},\sigma)\),
\(\tau^{\star}:=P_{\mathcal{A}_{\mathcal{C}}}(u)\) and
\(\hat{\tau}:=P_{\mathcal{A}_{\mathcal{C}}}(v)\) denote the oracle and
feasible menu allocations, respectively. The middle expression is then
\(n\) times the posterior expected loss of the feasible allocation in
excess of the oracle allocation. The lemma bounds this excess by the estimation error times
the change in the allocation. An error therefore costs nothing unless
it changes the program, and then costs in proportion to how much it
changes it. The menu and the budget limit that change. A menu pays at
most \(c_{J}\) per household and serves at most
\(s_{B}:=\min\{n,\lfloor B/c_{1}\rfloor\}\) households, so two
allocations in \(\mathcal{A}_{\mathcal{C}}\) differ in at most
\(2s_{B}\) coordinates by at most \(c_{J}\) each, and
\(\|\tau^{\star}-\hat{\tau}\|\le c_{J}\sqrt{2s_{B}}\). Averaging over
the data and combining this cap with the bound on the estimated
posterior mean gaps behind Proposition~\ref{prop:regret} gives the
regret of the feasible rule.

\begin{proposition}
\proptitle{\hyperlink{proof:ds_regret}{Regret of the Feasible Menu Rule}}
\label{prop:ds_regret}
\hypertarget{prop:ds_regret}{}
Let Assumptions~\ref{assump:normality}, \ref{assump:prior},
\ref{assump:approx}, and~\ref{assump:moment} hold, condition on
$\sigma$, and let $\Delta_{n}$, $\bar{\sigma}$, and $p$ be as in
Proposition~\ref{prop:regret}. Write $\hat{\tau}^{\EBlabel}_{\mathcal{C}}$
for the feasible menu rule and $\tau^{\star}_{G,\mathcal{C}}$ for an
oracle Bayes rule over $\mathcal{A}_{\mathcal{C}}$, with Bayes regret
defined as in Section~\ref{sec:bayes_regret_results}. For all $n$ large
enough that $\log n>1/p$,
\[
\operatorname{BR}\bigl(\hat{\tau}^{\EBlabel}_{\mathcal{C}},\tau^{\star}_{G,\mathcal{C}}\mid\sigma\bigr)
\le
2c_{J}\sqrt{\frac{2s_{B}}{n}}\,
\sqrt{2\bar{\sigma}\Delta_{n}+\Delta_{n}^{2}}.
\]
\end{proposition}

The lemma bounds regret by the estimation error itself, while the
statistical ingredient of the main text controls the average squared
error, by \(2\bar{\sigma}\Delta_{n}+\Delta_{n}^{2}\). Taking the
square root of that bound turns it into a bound on the average error,
and this is the square root in the proposition. The bound rises with
\(c_{J}\), which prices each changed assignment, and with \(s_{B}\),
which counts the assignments an error can change. With the menu fixed
and a budget that grows with the population, so that the program
can serve a nonvanishing share of households, the factor
\(2c_{J}\sqrt{2s_{B}/n}\) is bounded, and the regret bound vanishes at
rate \(n^{-p/(4(p+1))}\) up to logarithmic factors, against
\(n^{-p/(2(p+1))}\) in Proposition~\ref{prop:regret}. The regret bound still
vanishes, but more slowly. For \(p=2\), the two rates are \(n^{-1/6}\)
and \(n^{-1/3}\), and as \(p\) grows, they approach \(n^{-1/4}\) and
\(n^{-1/2}\).\footnote{With \(B\) fixed instead, \(s_{B}\) is
eventually constant, and the bound is of order
\(\sqrt{\Delta_{n}/n}\). The faster rate reflects a bounded program
averaged over a growing population rather than an advantage of the
menu, since the lemma holds for nearest points of any set and gives a
bound of the same order when applied to the continuous budget set,
whose diameter is \(\sqrt{2}B\).}

The menu costs accuracy near assignment boundaries but protects it
away from them. Under a menu, an error changes the allocation
only if it
moves the estimated vector far enough to change which allocation is
nearest. If the menu allocation nearest to the true posterior mean
gaps is unique, every other allocation is strictly farther, and
because there are finitely many, the same allocation stays nearest for
every vector close enough to the true one. Estimation errors within
that margin leave the assignment unchanged and cost nothing.

\clearpage

\setcounter{table}{0}
\setcounter{figure}{0}
\renewcommand{\thetable}{\thesection.\arabic{table}}
\renewcommand{\thefigure}{\thesection.\arabic{figure}}
\renewcommand{\theHtable}{\thesection.\arabic{table}}
\renewcommand{\theHfigure}{\thesection.\arabic{figure}}

\section{Additional Figures and Tables}

\subsection{Figures}

\begin{figure}[!ht]
    \centering
    \setstretch{1}
    \setlength{\abovecaptionskip}{0pt}
    \setlength{\belowcaptionskip}{6pt}
    \caption{Perfect-information allocation as a projection}
    \label{fig:projection}
    \input{figures/jmp_v3/figure_projection.tex}

    \vspace{8pt}
    \begin{minipage}{6.25in}
    \footnotesize
    \setlength{\baselineskip}{12pt}
    \emph{Notes.} Under quadratic loss, the gray triangle
    is the feasible set $\mathcal A$. Hollow points are poverty gap vectors; blue
    points are their projections. Orange segments show the remaining poverty gaps.
    For $y$, both households receive transfers, leaving the same gap
    $\lambda^{\PIlabel}(y)$. For $y'$, $0<z-y'_2<\lambda^{\PIlabel}(y')$:
    household~2 is poor but receives nothing.
    \end{minipage}
\end{figure}

\clearpage

\begin{figure}[!ht]
    \centering
    \setlength{\abovecaptionskip}{0pt}
    \setlength{\belowcaptionskip}{6pt}
    \caption{Actual transfer spending saved by EB-OLS at equal poverty gap reduction}
    \label{fig:budget_required}
    \includegraphics[width=6.25in]{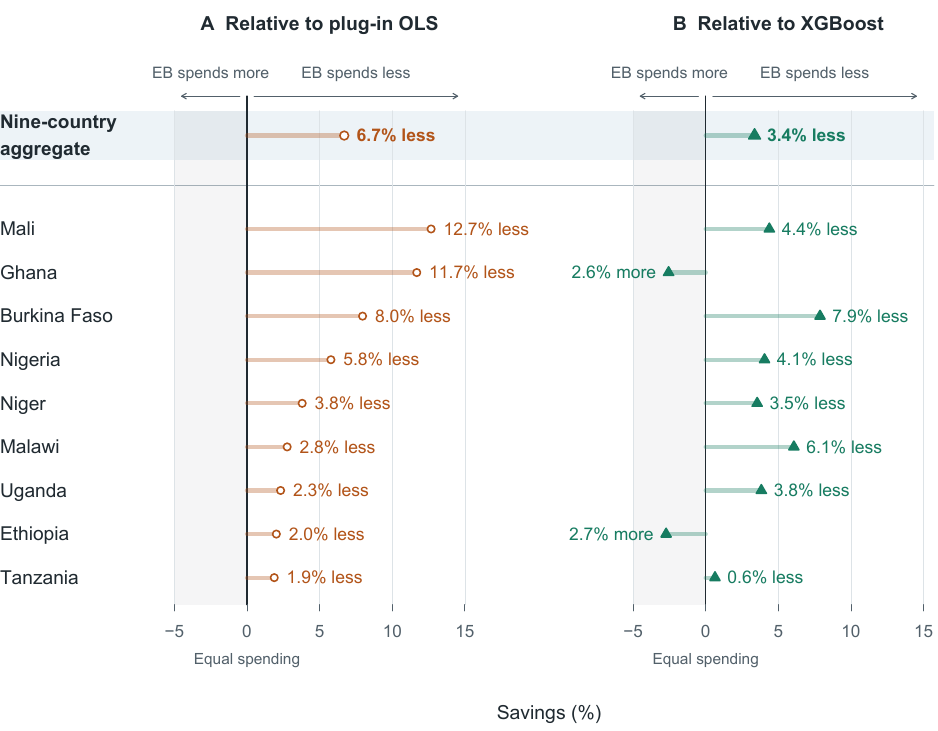}

    \vspace{8pt}
    \begin{minipage}{6.25in}
    \footnotesize
    \setstretch{1}
    \setlength{\baselineskip}{12pt}
    \emph{Notes.} Points report savings in mean actual transfer spending when EB-OLS
    matches each comparator's mean realized poverty gap reduction. Comparators face
    the budget ceiling $b=0.005$. Savings are measured relative to the comparator's
    mean actual spending. Positions show signed savings; labels state how much less
    or more EB-OLS spends than the comparator. Means are computed across simulation
    replications. The nine-country aggregate uses a common EB-OLS budget ceiling to match the
    equal-weighted mean poverty gap reduction across countries, with spending also
    averaged using equal country weights. Countries are ordered by savings
    relative to plug-in OLS, from \mbox{highest to lowest}.
    \end{minipage}
\end{figure}

\clearpage

\begin{figure}[!ht]
    \centering
    \setlength{\abovecaptionskip}{0pt}
    \setlength{\belowcaptionskip}{6pt}
    \caption{Targeting performance under one-sided loss}
    \label{fig:gains_Lplus}
    \includegraphics[width=6.25in]{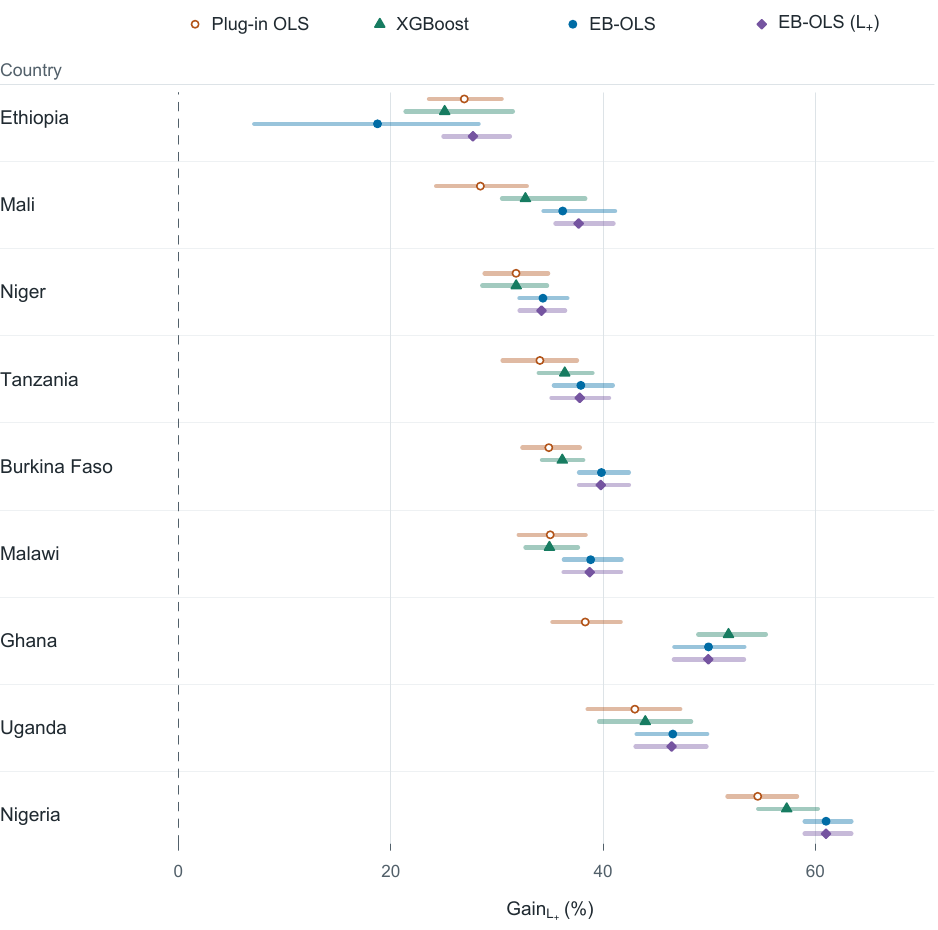}

    \vspace{8pt}
    \begin{minipage}{6.25in}
    \footnotesize
    \setstretch{1}
    \setlength{\baselineskip}{12pt}
    \emph{Notes.} Symbols report country-specific mean gains under one-sided
    loss over \nReps\ simulation replications at the common budget $b=0.005$.
    Gains are reported as percentages of the perfect-information improvement over no transfers.
    A gain of 0\% matches the no-transfer benchmark, and 100\% matches the
    perfect-information benchmark, which is outside the displayed range.
    Pale intervals span the 25th to 75th percentiles across
    replications. Plug-in OLS, XGBoost, and EB-OLS use the allocations
    constructed under quadratic loss and are evaluated here under $L_+$.
    EB-OLS ($L_+$) is reoptimized for the one-sided loss.
    Countries are ordered by mean plug-in OLS
    gain under $L_+$, from lowest to highest.
    \end{minipage}
\end{figure}

\clearpage
\begin{figure}[!ht]
    \centering
    \setlength{\abovecaptionskip}{0pt}
    \setlength{\belowcaptionskip}{6pt}
    \caption{Rankings by headcount reduction depend on the consumption cutoff}
    \label{fig:headcount_reductions}
    \includegraphics[width=6.25in]{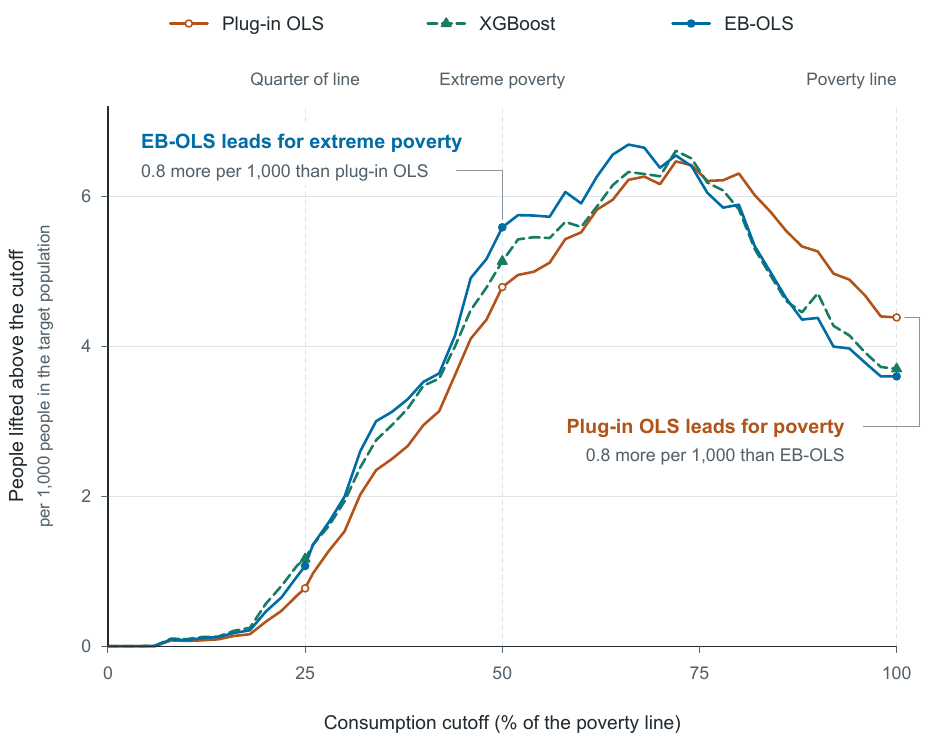}

    \vspace{8pt}
    \begin{minipage}{6.25in}
    \footnotesize
    \setstretch{1}
    \setlength{\baselineskip}{12pt}
    \emph{Notes.} Each curve reports the number of people whose consumption rises
    from at or below a given cutoff to above it, per 1,000 people in the entire
    target population. This equals the reduction in the person-weighted headcount
    relative to no transfers. Cutoffs are expressed relative to each country's
    poverty line, $z$; extreme poverty is defined as consumption at or below $0.5z$.
    All rules retain their quadratic-loss allocations at the common budget ceiling
    $b=0.005$ as the measurement cutoff varies. Results average \nReps\ replications
    within each of nine countries and give countries equal weight. Lines connect
    empirical results at intervals of $0.02z$ and at $0.25z$; symbols mark $0.25z$,
    $0.5z$, and $z$. Callouts report differences in mean headcount reductions.
    \end{minipage}
\end{figure}

\clearpage
\subsection{Tables}
\input{\outdir/tables/tableD1_country_environments.tex}
\FloatBarrier

\clearpage
\input{\outdir/tables/tableD2_robustness_percent.tex}

\clearpage
\input{\outdir/tables/tableD6_program_budgets.tex}

\clearpage
\input{\outdir/tables/tableD3_country_policy_value_percent.tex}

\clearpage
\input{\outdir/tables/tableD4_targeting_mechanisms.tex}

\clearpage
\input{\outdir/tables/tableD5_loss_rule_comparison_percent.tex}

\clearpage

\setcounter{table}{0}
\setcounter{figure}{0}
\renewcommand{\thetable}{\thesection.\arabic{table}}
\renewcommand{\thefigure}{\thesection.\arabic{figure}}
\renewcommand{\theHtable}{\thesection.\arabic{table}}
\renewcommand{\theHfigure}{\thesection.\arabic{figure}}

\section{Technical Results and Proofs for the Online Appendix}
\label{app:Lplus_proofs}
This appendix collects the supporting results and proofs for
Appendices~\ref{app:loss_functions} and~\ref{app:fixed_grants}.

\subsection{Supporting Lemmas for Appendix B}
\begin{lemma}
\proptitle{\hyperlink{proof:Lplus_cap}{Capped optimal allocations under \(L_+\)}}
\label{lem:Lplus_cap}
\hypertarget{lem:Lplus_cap}{}
Under Assumptions~\ref{assump:normality}, \ref{assump:prior},
and~\ref{assump:Lplus_support}, suppose \(0<B<nz\). Imposing
\(\tau_i\le z\) leaves the optimal values of the oracle and fitted
posterior problems with \(\sum_i\tau_i=B\) unchanged, and both problems
admit optimal selections in \([0,z]^n\) that are measurable in the data.
The oracle selection admits a measurable cutoff \(\lambda_{G,+}\ge0\)
satisfying the conditions of
Proposition~\ref{prop:oracle_bayes_rule_Lplus}, and
\(\max_i\tau^\star_{G,+,i}<z\) whenever \(\lambda_{G,+}>0\).
\end{lemma}

\hypertarget{proof:Lplus_cap}{}
\AppendixProofHeading{Proof of Lemma~\ref{lem:Lplus_cap}}
Fix \((\hat y,\sigma)\) and suppress the conditioning arguments. For
\(H\in\{G,\hat G\}\), let
\(Q_H(\tau):=n^{-1}\sum_{i=1}^n\E_H[(z-\mu_i-\tau_i)_+^2\mid\hat y_i,\sigma_i]\),
the average of household losses. This definition is used again in the
proof of Lemma~\ref{lem:Lplus_global_stability}. Because both priors are
supported on nonnegative incomes, each household's posterior loss is
nonincreasing in its transfer and equals zero at every transfer of at
least \(z\).

Take any nonnegative allocation \(\tau\) with total spending \(B\). Set
\(c_i=\min\{\tau_i,z\}\) and \(e=B-\sum_i c_i\ge0\). Capping leaves the
objective unchanged, and
\(\sum_i(z-c_i)=nz-B+e>e\) because \(B<nz\), so the amount \(e\) can be
redistributed among coordinates below the cap. The resulting allocation
spends \(B\), lies in \([0,z]^n\), and has objective no greater than
\(Q_H(\tau)\). Hence imposing the cap does not change the optimal value.

On the fixed compact set \(\{\tau\in[0,z]^n:\sum_i\tau_i=B\}\), the
objective is continuous in \(\tau\) and measurable in the data, since
posterior expectations are ratios of measurable integrals with strictly
positive Gaussian-mixture denominators. Both problems therefore admit
measurable selections of their minimizers \cite{brown1973measurable}.

The selected oracle allocation also minimizes \(Q_G\) over the original
set \(\{\tau\ge0:\sum_i\tau_i=B\}\), a differentiable convex program with
the strictly positive feasible point \((B/n)\mathbf 1\). Apply the KKT
conditions to the summed objective \(nQ_G\), whose budget multiplier is
\(\nu\), and set \(\lambda_{G,+}:=\nu/2\). By the derivative identity in
Lemma~\ref{lem:Lplus_curve_geometry}, the conditions are
\[
g^+_{G,i}(\tau^\star_{G,+,i})\le\lambda_{G,+},
\qquad
g^+_{G,i}(\tau^\star_{G,+,i})=\lambda_{G,+}
\quad\text{if }\tau^\star_{G,+,i}>0.
\]
Since \(B>0\), at least one transfer is positive, so
\(\lambda_{G,+}=\max_i g^+_{G,i}(\tau^\star_{G,+,i})\ge0\), which is
measurable in the data. Finally, \(g^+_{G,i}(z)=0\), so a transfer equal
to \(z\) forces \(\lambda_{G,+}=0\), and a positive cutoff implies
that every oracle transfer is strictly below the cap.
\AppendixQED{lem:Lplus_cap}

\Needspace{13\baselineskip}
\begin{lemma}
\proptitle{\hyperlink{proof:Lplus_curve_geometry}{Geometry of posterior remaining-gap functions}}
\label{lem:Lplus_curve_geometry}
\hypertarget{lem:Lplus_curve_geometry}{}
Let Assumption~\ref{assump:normality} hold, let \(H\) be supported on
\([0,M]\), and fix a household and its signal. On \([0,z]\), the posterior
remaining-gap function \(g^+_{H,i}\) is convex, nonincreasing, and
one-Lipschitz, with \(g^+_{H,i}(z)=0\). The household's posterior expected
loss is convex and continuously differentiable with derivative
\(-2g^+_{H,i}(t)\). Moreover,
\[
g^+_{H,i}(u)\le\frac{z-u}{z-t}\,g^+_{H,i}(t),
\qquad 0\le t\le u\le z,\quad t<z.
\]
\end{lemma}

\noindent
\hypertarget{proof:Lplus_curve_geometry}{}
\AppendixProofHeading{Proof of Lemma~\ref{lem:Lplus_curve_geometry}}
Under the fixed household posterior, write \(Q=z-\mu_i\le z\) and
\(g(t)=\E[(Q-t)_+]\). Each hinge \(t\mapsto(Q-t)_+\) is convex,
nonincreasing, and one-Lipschitz, and vanishes at \(t=z\). Taking
expectations gives these properties for \(g\). By convexity, \(g(u)\) lies
below the line joining \((t,g(t))\) and \((z,0)\), which is the displayed
inequality.

The squared hinge is continuously differentiable with derivative
\(-2(Q-t)_+\), bounded in absolute value by \(2z\) on \([0,z]\).
Differentiation under the expectation therefore gives the loss derivative
\(-2g(t)\), which is continuous because \(g\) is one-Lipschitz and
nondecreasing because \(g\) is nonincreasing, so the loss is convex.
\AppendixQED{lem:Lplus_curve_geometry}

\begin{lemma}
\proptitle{\hyperlink{proof:Lplus_prior_stability}{Posterior remaining-gap stability in the prior}}
\label{lem:Lplus_prior_stability}
\hypertarget{lem:Lplus_prior_stability}{}
Under Assumption~\ref{assump:normality}, there is a deterministic function
\(A:\R\to(0,\infty)\), depending only on \(z\), \(M\), and
\(\sigma_{\min}\), such that
\[
\sup_{t\in[0,z]}
\bigl|g^+_H(t\mid y,\sigma)-g^+_K(t\mid y,\sigma)\bigr|
\le A(y)\,W_2(H,K)
\]
for all probability distributions \(H\) and \(K\) supported on \([0,M]\),
all \(y\in\R\), and all \(\sigma\in[\sigma_{\min},\sigma_{\max}]\).
Moreover, for \(Z\sim N(0,1)\), \(\E[A(\mu+\sigma Z)^2]\le K_A\)
uniformly over \(\mu\in[0,M]\) and \(\sigma\in[\sigma_{\min},\sigma_{\max}]\),
where \(K_A<\infty\) depends only on \(z\), \(M\), \(\sigma_{\min}\), and
\(\sigma_{\max}\).
\end{lemma}

\noindent
\hypertarget{proof:Lplus_prior_stability}{}
\AppendixProofHeading{Proof of Lemma~\ref{lem:Lplus_prior_stability}}
Fix \(y\), \(\sigma\), and \(t\), suppress these arguments, and set
\(h_t(u)=(z-u-t)_+\) and \(w(u)=\exp\{yu/\sigma^2-u^2/(2\sigma^2)\}\). The
Gaussian factor that does not depend on \(u\) cancels from Bayes' formula,
so \(g^+_J=\int h_tw\,dJ/\int w\,dJ\) for \(J\in\{H,K\}\). With
\(c=g^+_K\in[0,z]\), the identity \(\int(h_t-c)w\,dK=0\) gives
\[
g^+_H-g^+_K
=
\frac{\int(h_t-c)w\,d(H-K)}{\int w\,dH}.
\]
On \([0,M]\), \(|\tfrac{d}{du}\log w(u)|=|y-u|/\sigma^2\le
(|y|+M)/\sigma_{\min}^2=:\kappa(y)\), so
\(\sup w/\inf w\le e^{M\kappa(y)}\) and
\(\operatorname{Lip}(w)\le\kappa(y)\sup w\), with the extrema taken over
\([0,M]\). Since \(h_t\) is one-Lipschitz and \(|h_t-c|\le z\),
\(\operatorname{Lip}((h_t-c)w)\le[1+z\kappa(y)]\sup w\). The
Kantorovich--Rubinstein inequality applied to the numerator and
\(\int w\,dH\ge\inf w\) in the denominator give
\(|g^+_H-g^+_K|\le[1+z\kappa(y)]e^{M\kappa(y)}W_1(H,K)=:A(y)W_1(H,K)\).
The bound is uniform in \(t\) and \(\sigma\), and \(W_1\le W_2\), which
proves the first claim.

This choice satisfies \(A(y)\le C_0(1+|y|)e^{C_1|y|}\) for finite
constants depending only on \(z\), \(M\), and \(\sigma_{\min}\). For
\(\mu\in[0,M]\) and \(\sigma\in[\sigma_{\min},\sigma_{\max}]\),
\(|\mu+\sigma Z|\le M+\sigma_{\max}|Z|\), so \(A(\mu+\sigma Z)^2\) is
bounded by \(C(1+|Z|)^2e^{q|Z|}\) with \(C,q<\infty\) depending only on
\(z\), \(M\), \(\sigma_{\min}\), and \(\sigma_{\max}\). This variable has
finite expectation under the standard normal law, which gives \(K_A\).
\AppendixQED{lem:Lplus_prior_stability}

\begin{lemma}
\proptitle{\hyperlink{proof:Lplus_grid}{Grid resolution and likelihood tolerance}}
\label{lem:Lplus_grid}
\hypertarget{lem:Lplus_grid}{}
Let Assumption~\ref{assump:normality} hold and let \(G\) be supported on
\([0,M]\). Let \(\Gamma_n\subset[0,M]\) be a finite grid containing both
endpoints, with mesh \(\delta_n\), and let \(\ell_n\) be the average log
likelihood of Assumption~\ref{assump:Lplus_support}. Suppose \(\hat G\) is
supported on \(\Gamma_n\) and attains a value of \(\ell_n\) within
\(\eta_n\ge0\) of its maximum over distributions supported on \(\Gamma_n\).
Then, for every data realization,
\[
\ell_n(G)-\ell_n(\hat G)\le\frac{\delta_n^2}{8\sigma_{\min}^2}+\eta_n.
\]
\end{lemma}

\noindent
Consequently, choosing \(\delta_n^2+\eta_n\lesssim(\log n)^2/n\) ensures
the approximate-likelihood condition in
Assumption~\ref{assump:Lplus_support} for all sufficiently large \(n\).

\hypertarget{proof:Lplus_grid}{}
\AppendixProofHeading{Proof of Lemma~\ref{lem:Lplus_grid}}
Fix the observed data. On each grid interval \([a,b]\), round \(u\in[a,b]\)
to a random endpoint \(V\) with \(\Pr(V=a)=(b-u)/(b-a)\) and
\(\Pr(V=b)=(u-a)/(b-a)\), so that \(\E[V\mid u]=u\) and
\(\operatorname{Var}(V\mid u)=(u-a)(b-u)\le\delta_n^2/4\). Let \(G^\Gamma\)
be the distribution of \(V\) when \(u\sim G\). It is supported on
\(\Gamma_n\). For observation \(i\), write
\(k_i(v)=\phi_{\sigma_i}(\hat Y_i-v)\). Since \(\log k_i\) is quadratic in
\(v\) and the rounding preserves the mean,
\(\E[\log k_i(V)\mid u]=\log k_i(u)-\operatorname{Var}(V\mid u)/(2\sigma_i^2)\),
and Jensen's inequality gives
\[
\E[k_i(V)\mid u]
\ge\exp\{\E[\log k_i(V)\mid u]\}
\ge k_i(u)\exp\{-\delta_n^2/(8\sigma_i^2)\}.
\]
Integrating over \(u\sim G\), taking logarithms, and averaging over
observations yields
\(\ell_n(G)-\ell_n(G^\Gamma)\le(\delta_n^2/8n)\sum_{i=1}^n\sigma_i^{-2}\le
\delta_n^2/(8\sigma_{\min}^2)\). Since \(G^\Gamma\) is feasible for the
grid problem and \(\hat G\) is within \(\eta_n\) of its maximum,
\(\ell_n(\hat G)\ge\ell_n(G^\Gamma)-\eta_n\). Combining the two
inequalities proves the result.
\AppendixQED{lem:Lplus_grid}

\begin{lemma}
\proptitle{\hyperlink{proof:Lplus_w2}{Support-constrained Wasserstein rate}}
\label{lem:Lplus_w2}
\hypertarget{lem:Lplus_w2}{}
Let Assumptions~\ref{assump:normality}, \ref{assump:prior},
and~\ref{assump:Lplus_support} hold. There are constants \(C_W<\infty\) and
\(n_W<\infty\), depending only on \(M\), \(\sigma_{\min}\),
\(\sigma_{\max}\), and \(C_\ell\), such that for every \(n\ge n_W\),
\[
P_{G,\hat Y}\!\left(W_2^2(G,\hat G)\le\frac{C_W}{\log n}
\,\middle|\,\sigma\right)\ge1-\frac{2}{n^8}.
\]
\end{lemma}

\noindent
\hypertarget{proof:Lplus_w2}{}
\AppendixProofHeading{Proof of Lemma~\ref{lem:Lplus_w2}}
Write \(f_{H,\sigma_i}=\phi_{\sigma_i}*H\) and let
\(\bar h_n^2:=n^{-1}\sum_{i=1}^n h^2(f_{G,\sigma_i},f_{\hat G,\sigma_i})\)
be the average squared Hellinger distance between the true and fitted
signal densities. Theorem 7 of \citeA{soloff2021multivariate}, with
\(d=1\), \(S=[0,M]\), covariance eigenvalues in
\([\sigma_{\min}^2,\sigma_{\max}^2]\), Hellinger localization parameter
proportional to \(\sqrt{\log n}\), and deviation parameter \(\sqrt8\),
applies to any approximate maximizer whose average log likelihood is
within a constant times \((\log n)^2/n\) of \(\ell_n(G)\), which is the
likelihood condition in Assumption~\ref{assump:Lplus_support}. It gives an
event \(\mathcal H_n\) such that
\[
P_{G,\hat Y}(\mathcal H_n^c\mid\sigma)\le\frac{2}{n^8},
\qquad
\bar h_n^2\le C_H\frac{(\log n)^2}{n}
\quad\text{on }\mathcal H_n,
\]
where \(C_H\) depends only on the constants listed in the lemma.

The proof of their Theorem 10 converts Hellinger accuracy into
Wasserstein accuracy by smoothing both priors, applying Nguyen's
transport inequality, and controlling the smoothed density difference by
Fourier inversion and \(\bar h_n\). Its only use of the fitted-support
condition is the moment bound
\(\mathcal R_s\le2^s\{m_s(G)+m_s(\hat G)+2\delta^sm_s(K)\}\) for \(s>2\),
in their notation, where the support of \(\hat G\) is placed in the
observed-data bounding box. Here \(G([0,M])=\hat G([0,M])=1\) gives
\(m_s(G)\le M^s\) and \(m_s(\hat G)\le M^s\) deterministically, so no
additional exceptional event is needed, and equations (34)--(37) of their
supplementary material yield, on \(\mathcal H_n\) for all sufficiently
large \(n\) and \(\bar h_n>0\),
\[
W_2^2(G,\hat G)
\le C\left\{\frac1{-\log\bar h_n}
+(\log n)^{1/2}\bar h_n^{1/12}\right\}.
\]

On \(\mathcal H_n\), \(\bar h_n\le C_H^{1/2}(\log n)/\sqrt n\), so
\(-\log\bar h_n\ge c\log n\) for large \(n\), while the second term is at
most \(Cn^{-1/24}(\log n)^{7/12}=o(1/\log n)\). If \(\bar h_n=0\),
injectivity of Gaussian convolution gives \(\hat G=G\). Enlarging
\(C_W\) and \(n_W\) therefore gives \(W_2^2(G,\hat G)\le C_W/\log n\) on
\(\mathcal H_n\), and the probability bound follows.
\AppendixQED{lem:Lplus_w2}

\begin{lemma}
\proptitle{\hyperlink{proof:Lplus_shadow}{Lower tail of the oracle cutoff}}
\label{lem:Lplus_shadow}
\hypertarget{lem:Lplus_shadow}{}
Let Assumptions~\ref{assump:normality}, \ref{assump:prior},
\ref{assump:Lplus_support}, and~\ref{assump:Lplus_scarcity} hold, and let
\(\lambda_{G,+}\) be the cutoff selected in Lemma~\ref{lem:Lplus_cap}.
There are constants \(\underline\lambda>0\) and \(c_\lambda>0\), depending
only on \(G\), \(z\), \(M\), \(\sigma_{\min}\), \(\sigma_{\max}\), and
\(\bar b\), such that whenever \(0<B/n\le\bar b\),
\[
P_{G,\hat Y}(\lambda_{G,+}<\underline\lambda\mid\sigma)
\le e^{-c_\lambda n}.
\]
\end{lemma}

\noindent
\hypertarget{proof:Lplus_shadow}{}
\AppendixProofHeading{Proof of Lemma~\ref{lem:Lplus_shadow}}
Write \(\mu_0=\inf\operatorname{supp}(G)\), choose \(t\) with
\(\bar b<t<z-\mu_0\), and choose \(R>M\) large enough that
\(p:=\Pr\{|Z|\le(R-M)/\sigma_{\max}\}>\bar b/t\) for \(Z\sim N(0,1)\). The
indicators \(I_i:=\1\{|\hat Y_i|\le R\}\) are independent conditional on
\(\sigma\), and since \(|\hat Y_i|\le M+\sigma_{\max}|Z_i|\), each has
\(\E[I_i\mid\sigma]\ge p\).

Because \(\mu_0\in\operatorname{supp}(G)\) and \(\mu_0<z-t\), the prior
puts positive mass on incomes whose remaining gap after transfer \(t\) is
positive, and the Gaussian likelihood is strictly positive, so
\(g^+_G(t\mid y,s)>0\) for every signal \(y\) and every
\(s\in[\sigma_{\min},\sigma_{\max}]\). This function is continuous in
\((y,s)\), so its minimum \(m\) over \(|y|\le R\) and
\(s\in[\sigma_{\min},\sigma_{\max}]\) is attained and positive. Set
\(\underline\lambda=m\).

On the event \(\{\lambda_{G,+}<\underline\lambda\}\), the conditions of
Proposition~\ref{prop:oracle_bayes_rule_Lplus} give
\(g^+_{G,i}(\tau^\star_{G,+,i}\mid\hat Y_i,\sigma_i)\le\lambda_{G,+}<m\)
for every household. If \(I_i=1\), the household's remaining-gap function
at \(t\) is at least \(m\), so monotonicity forces
\(\tau^\star_{G,+,i}>t\). Since the oracle spends \(B\), the event implies
\(n^{-1}\sum_{i=1}^nI_i<B/(nt)\le\bar b/t\). Hoeffding's inequality for
the independent \([0,1]\)-valued \(I_i\) therefore yields
\[
P_{G,\hat Y}(\lambda_{G,+}<\underline\lambda\mid\sigma)
\le P_{G,\hat Y}\!\left(\frac1n\sum_{i=1}^nI_i\le\frac{\bar b}{t}
\,\middle|\,\sigma\right)
\le\exp\!\left\{-2n\left(p-\frac{\bar b}{t}\right)^2\right\},
\]
and \(c_\lambda=2(p-\bar b/t)^2>0\) proves the result.
\AppendixQED{lem:Lplus_shadow}

\subsection{Proofs of Appendix B Results}
This subsection contains the proofs of the results stated in Appendix~\ref{app:loss_functions}. 

\hypertarget{proof:sandwich_Lplus}{}
\AppendixProofHeading{Proof of Lemma~\ref{lem:sandwich_Lplus}} Fix a household \(i\) and
condition on \((\mu,\hat Y)\). Write \(a_i:=z-\mu_i-\delta_i(\hat Y)\), which
is fixed given \((\mu,\hat Y)\), and note that
\(z-y_i-\delta_i(\hat Y)=a_i-\varepsilon_i\). Since \(x\mapsto x_+^2\) is
convex, conditional Jensen and \(\E[\varepsilon_i\mid\mu,\hat Y]=0\) give
\(\E[(a_i-\varepsilon_i)_+^2\mid \mu,\hat Y] \ \ge\ \big(a_i-\E[\varepsilon_i\mid\mu,\hat Y]\big)_+^2 \ =\ (a_i)_+^2 .\)
For the upper bound, the map \(x\mapsto x_+^2\) is differentiable with
derivative \(2x_+\), which is \(2\)-Lipschitz. The descent inequality
therefore gives, for all \(a,e\in\mathbb R\),
\((a-e)_+^2 \ \le\ a_+^2-2a_+e+e^2 .\)
Setting \(a=a_i\) and \(e=\varepsilon_i\) and taking conditional expectations
eliminates the linear term, so
\(\E[(a_i-\varepsilon_i)_+^2\mid\mu,\hat Y]\le (a_i)_+^2+\E[\varepsilon_i^2\mid\mu,\hat Y]\).
Averaging over \(i\), taking expectations over \(\hat Y\), and applying the
law of iterated expectations yields \eqref{eq:sandwich_Lplus}. This proves the lemma.\AppendixQED{lem:sandwich_Lplus}

\hypertarget{proof:full_info_L_plus}{}
\leavevmode\ProofCorollary{cor:full_info_L_plus}{For brevity, let \(q_i:=(z-y_i)_+\). Fix any feasible \(\tau\in\mathcal A\), and define \(\tilde\tau_i:=\min\{\tau_i,q_i\}\) for each \(i\). Since \(\tilde\tau_i\le \tau_i\) for every \(i\), the vector \(\tilde\tau\) remains feasible. Moreover, replacing \(\tau_i\) by \(\tilde\tau_i\) does not change the term \((z-y_i-\tau_i)_+\): if \(\tau_i\le q_i\), nothing changes, while if \(\tau_i>q_i\), both \((z-y_i-\tau_i)_+\) and \((z-y_i-\tilde\tau_i)_+\) are equal to zero. Hence there exists an optimal solution satisfying \(0\le \tau_i\le q_i\) for all \(i\).

On this restricted region, minimizing \(L_+\) is equivalent to minimizing $\frac{1}{n}\sum_{i=1}^n (q_i-\tau_i)^2$ over \(\mathcal A\). Since \(q_i=(z-y_i)_+\), this differs from the quadratic objective in Proposition~\ref{prop:full_info} only by the additive constant $\frac{1}{n}\sum_{i:y_i\ge z}(z-y_i)^2$, which does not depend on \(\tau\). Therefore \(\tau^{\PIlabel}(y)\) remains optimal under \(L_+\).

If \(B<\sum_{i=1}^n q_i\), every optimum lies in the restricted region. Suppose instead that an optimal \(\tau\) has \(\tau_j>q_j\) for some \(j\). Then \(\sum_{i\ne j}\tau_i\le B-\tau_j<\sum_{i\ne j}q_i\), so some household \(k\ne j\) has \(\tau_k<q_k\). Moving an amount \(\delta\in(0,\min\{\tau_j-q_j,\,q_k-\tau_k\}]\) from \(j\) to \(k\) keeps the allocation feasible, leaves the term for \(j\) at zero, and strictly lowers the term for \(k\), contradicting optimality. On the restricted region, \(L_+\) is a strictly convex quadratic plus a constant and the region is convex, so the minimizer is unique.

If \(B\ge \sum_{i=1}^n q_i\), choosing \(\tau_i=q_i\) for all \(i\) is feasible and yields \(L_+=0\). Hence the minimum value is zero. When \(B=\sum_{i=1}^n q_i\), zero loss requires \(\tau_i\ge q_i\) for all \(i\), and feasibility then forces \(\tau_i=q_i\) exactly, so the minimizer remains unique at the boundary. When \(B>\sum_{i=1}^n q_i\), the set of minimizers is not a singleton, since additional transfers to households already at or above the poverty line do not change the loss.}

\hypertarget{proof:Lplus_simple_dominance}{}
\leavevmode\Proof{prop:Lplus_simple_dominance}{Let $e(x) := \sum_{i=1}^{n}\bigl(\tau_i^{\pluglabel}(x)-z\bigr)_{+}$. For feasibility, fix any \(x\in\mathbb R^n\). By construction, each component of \(\tau^{\cslabel}(x)\) is nonnegative. Moreover, using \(\min\{a,z\}+(a-z)_+=a\) for any \(a\ge 0\), we have
\[
\sum_{i=1}^n \tau_i^{\cslabel}(x)
=
\sum_{i=1}^n \min\{\tau_i^{\pluglabel}(x),z\}+e(x)
=
\sum_{i=1}^n \tau_i^{\pluglabel}(x)
\le B.
\]
Hence \(\tau^{\cslabel}(x)\in\mathcal A\) for every \(x\).

To prove part (i), fix \(\theta\in(-\infty,z]^n\) and \(x\in\mathbb R^n\). If \(\tau_i^{\pluglabel}(x)>z\), then \(\theta_i\le z<\tau_i^{\pluglabel}(x)\) and also \(\tau_i^{\cslabel}(x)\ge z\ge \theta_i\), so both rules generate zero loss in coordinate \(i\). If instead \(\tau_i^{\pluglabel}(x)\le z\), then \(\tau_i^{\cslabel}(x)=\tau_i^{\pluglabel}(x)+e(x)/n\ge \tau_i^{\pluglabel}(x)\), and since \(t\mapsto (\theta_i-t)_+^2\) is nonincreasing, the loss in coordinate \(i\) weakly falls. Therefore,
\(L_+(\tau^{\cslabel}(x),\theta)\le L_+(\tau^{\pluglabel}(x),\theta) \qquad\text{for every }x.\)
Taking expectations yields $R_+(\tau^{\cslabel},\theta)\le R_+(\tau^{\pluglabel},\theta)$.

For part (ii), fix \(\theta\in(-\infty,z]^n\) such that \(\sum_{i=1}^n(\theta_i)_+>B\), and consider any realization \(x\) with \(e(x)>0\). Since
\(\sum_{i=1}^n \tau_i^{\pluglabel}(x)\le B<\sum_{i=1}^n(\theta_i)_+,\)
there must exist some coordinate \(m\) such that $\tau_m^{\pluglabel}(x)<(\theta_m)_+$. Because \(\tau_m^{\pluglabel}(x)\ge 0\), this implies \(\theta_m>0\), hence \((\theta_m)_+=\theta_m\le z\). Therefore coordinate \(m\) is not capped, and
\(\tau_m^{\cslabel}(x)=\tau_m^{\pluglabel}(x)+\frac{e(x)}{n}>\tau_m^{\pluglabel}(x).\)
Since \(t\mapsto(\theta_m-t)_+^2\) is strictly decreasing for \(t<\theta_m\), the loss in coordinate \(m\) falls strictly, while every other coordinate improves weakly by part (i). Hence
\(L_+(\tau^{\cslabel}(x),\theta)<L_+(\tau^{\pluglabel}(x),\theta) \qquad\text{whenever }e(x)>0.\)

It remains to show that \(e(X)>0\) with positive probability. Since \(B>z\), any \(x\) with \(x_1>B\) and \(x_j<0\) for all \(j\ge 2\) satisfies \(x_+=(x_1,0,\ldots,0)\), so its projection onto \(\mathcal A\) is \((B,0,\ldots,0)\). On this event, $e(x)=B-z>0$. Under Assumption~\ref{assump:normality}, the distribution of \(X\) has a density that is strictly positive on \(\mathbb R^n\), so this event has positive probability for every \(\theta\). Therefore \(\Pr_\theta(e(X)>0)>0\).

Combining this with the pointwise strict inequality on \(\{e(X)>0\}\) and weak inequality elsewhere gives $R_+(\tau^{\cslabel},\theta)<R_+(\tau^{\pluglabel},\theta)$. This proves part (ii). The final inadmissibility claim follows immediately from the definition of admissibility.}

\hypertarget{proof:oracle_bayes_rule_Lplus}{}
\leavevmode\Proof{prop:oracle_bayes_rule_Lplus}{
Fix \((\hat y,\sigma)\). Write
\(g^+_{G,i}(t)=g^+_{G,i}(t\mid \hat y_i,\sigma_i)\) and
\(\psi_i(t)=\E_G[(z-\mu_i-t)_+^2\mid \hat y_i,\sigma_i]\) for \(t\ge 0\).

For each realization of \(\mu_i\), the map \(t\mapsto (z-\mu_i-t)_+^2\) is convex and
continuously differentiable on \(\mathbb R_+\), with derivative
\(-2(z-\mu_i-t)_+\). The maintained second-moment condition implies
\(\E_G[(z-\mu_i)_+\mid \hat y_i,\sigma_i]<\infty\), and \(2(z-\mu_i)_+\) is an
integrable envelope for the relevant difference quotients. Dominated convergence
therefore justifies differentiating under the conditional expectation, yielding $\psi_i'(t)=-2g^+_{G,i}(t)$.

The feasible set \(\{\tau\in\mathbb R_+^n:\sum_i\tau_i=B\}\) is nonempty and compact,
and \(\sum_i\psi_i\) is continuous and convex, so an optimum exists. Slater's
condition holds at \(\tau=(B/n)\mathbf 1\), so the KKT conditions are necessary
and sufficient. Since the factor \(1/n\) does not affect the minimizer, work with
\(\sum_i\psi_i(\tau_i)\). There exist \(\nu\in\mathbb R\) and
\(\xi_i\ge 0\) such that, for every \(i\),
\[
\psi_i'(\tau^\star_{G,+,i})+\nu-\xi_i=0,
\qquad
\xi_i\tau^\star_{G,+,i}=0,
\qquad
\tau^\star_{G,+,i}\ge 0,
\qquad
\sum_i\tau^\star_{G,+,i}=B.
\]
Substituting \(\psi_i'(t)=-2g^+_{G,i}(t)\) gives
\(2g^+_{G,i}(\tau^\star_{G,+,i})=\nu-\xi_i\le\nu\). Write
\(\lambda_{G,+}:=\nu/2\), so that
\(g^+_{G,i}(\tau^\star_{G,+,i})\le\lambda_{G,+}\). When
\(\tau^\star_{G,+,i}>0\), complementary slackness gives \(\xi_i=0\), so
\(g^+_{G,i}(\tau^\star_{G,+,i})=\lambda_{G,+}\). If \(\tau^\star_{G,+,i}=0\),
the same display gives \(g^+_{G,i}(0)\le\lambda_{G,+}\). Since \(B>0\) and
\(\sum_i\tau^\star_{G,+,i}=B\), at least one coordinate is active. For any active
coordinate, \(\lambda_{G,+}=g^+_{G,i}(\tau^\star_{G,+,i})\ge 0\). This proves the
KKT characterization.

The cutoff \(\lambda_{G,+}\) is determined by \(\sum_i\tau^\star_{G,+,i}=B\).
}

\bigskip 

\hypertarget{proof:Lplus_global_stability}{}
\AppendixProofHeading{Proof of Lemma~\ref{lem:Lplus_global_stability}}
Fix \((\hat y,\sigma)\), hold the fitted prior \(\hat G=\hat G(\hat y,\sigma)\)
fixed, and let \(Q_H\) be as defined in the proof of
Lemma~\ref{lem:Lplus_cap}, for \(H\in\{G,\hat G\}\). Under the true model,
household independence makes \(Q_G(\tau)\) the posterior expected loss of
\(\tau\) conditional on the full data. Suppress the signal arguments and
abbreviate \(\tau^\star=\tau^\star_{G,+}\),
\(\hat\tau=\hat\tau^{\EBlabel}_+\). The lemma assumes
\(\lambda_{G,+}>0\). By Lemma~\ref{lem:Lplus_cap}, both allocations
belong to \(\{\tau\in[0,z]^n:\sum_i\tau_i=B\}\), every \(\tau_i^\star<z\),
and \(g^+_{G,i}(\tau_i^\star)\le\lambda_{G,+}\) with equality whenever
\(\tau_i^\star>0\).

We first establish quadratic growth away from the oracle allocation. For
any \(\tau\) in the capped budget set, let \(d_i=\tau_i-\tau_i^\star\).
Since \(\sum_id_i=0\), the derivative identity in
Lemma~\ref{lem:Lplus_curve_geometry} gives
\[
n\{Q_G(\tau)-Q_G(\tau^\star)\}
=2\sum_{i=1}^n\int_{\tau_i^\star}^{\tau_i}\{\lambda_{G,+}-g^+_{G,i}(u)\}\,du.
\]
If \(d_i\ge0\), the chord inequality in Lemma~\ref{lem:Lplus_curve_geometry}
gives, for \(0\le v\le d_i\),
\(g^+_{G,i}(\tau_i^\star+v)\le g^+_{G,i}(\tau_i^\star)(z-\tau_i^\star-v)/(z-\tau_i^\star)
\le\lambda_{G,+}(1-v/z)\). If \(d_i<0\), then \(\tau_i^\star>0\), so
\(g^+_{G,i}(\tau_i^\star)=\lambda_{G,+}\), and the chord inequality with
\(t=\tau_i^\star-v\) and \(u=\tau_i^\star\) rearranges to
\(g^+_{G,i}(\tau_i^\star-v)\ge\lambda_{G,+}(z-\tau_i^\star+v)/(z-\tau_i^\star)\ge\lambda_{G,+}(1+v/z)\)
for \(0\le v\le-d_i\). Integrating, with reversed limits when \(d_i<0\),
bounds each coordinate's contribution to the display from below by
\(\lambda_{G,+}d_i^2/z\). Summing yields
\[
Q_G(\tau)-Q_G(\tau^\star)
\ge\frac{\lambda_{G,+}}{z}\,\frac{\|\tau-\tau^\star\|^2}{n}.
\]

We next compare the true and fitted objectives. Let
\(\varepsilon_i:=\sup_{t\in[0,z]}|\hat g_i^+(t)-g^+_{G,i}(t)|\) and
\(q^2:=\|\hat\tau-\tau^\star\|^2/n\), so that
\(r_+^2=n^{-1}\sum_i\varepsilon_i^2\). Convexity of \(Q_G\) gives
\(Q_G(\hat\tau)-Q_G(\tau^\star)\le\langle\nabla Q_G(\hat\tau),\hat\tau-\tau^\star\rangle\),
and optimality of \(\hat\tau\) for \(Q_{\hat G}\) on the same convex set
gives \(\langle\nabla Q_{\hat G}(\hat\tau),\hat\tau-\tau^\star\rangle\le0\).
Subtracting, the derivative identity bounds the \(i\)th coordinate of
\(\nabla Q_G(\hat\tau)-\nabla Q_{\hat G}(\hat\tau)\) by \(2\varepsilon_i/n\)
in absolute value, and the Cauchy--Schwarz inequality gives
\[
Q_G(\hat\tau)-Q_G(\tau^\star)
\le\frac2n\sum_{i=1}^n\varepsilon_i|\hat\tau_i-\tau_i^\star|
\le2r_+q.
\]
Applying the growth bound at \(\tau=\hat\tau\) gives
\((\lambda_{G,+}/z)\,q^2\le2r_+q\). If \(q=0\), the regret is zero.
Otherwise \(q\le2zr_+/\lambda_{G,+}\), and substitution gives
\(Q_G(\hat\tau)-Q_G(\tau^\star)\le4zr_+^2/\lambda_{G,+}\), as claimed.
\AppendixQED{lem:Lplus_global_stability}

\hypertarget{proof:Lplus_curve_error}{}
\AppendixProofHeading{Proof of Lemma~\ref{lem:Lplus_curve_error}}
Fix \(\sigma\). Since both priors are supported on \([0,M]\),
Lemma~\ref{lem:Lplus_prior_stability} applied household by household
gives
\[
r_+^2(\hat Y,\sigma)\le W_2^2(G,\hat G)\,\frac1n\sum_{i=1}^nA(\hat Y_i)^2
\]
for every data realization, including when \(\hat G\) is estimated from
the same signals. Both remaining-gap functions take values in \([0,z]\),
so also \(r_+^2\le z^2\). Averaging the moment bound of
Lemma~\ref{lem:Lplus_prior_stability} over \(\mu_i\sim G\) gives
\(\E[A(\hat Y_i)^2\mid\sigma]\le K_A\) for every household. For
\(n\ge\max\{n_W,2\}\), let
\(\mathcal W_n:=\{W_2^2(G,\hat G)\le C_W/\log n\}\), with \(C_W\) and
\(n_W\) from Lemma~\ref{lem:Lplus_w2}, so that
\(P_{G,\hat Y}(\mathcal W_n^c\mid\sigma)\le2/n^8\). Splitting the
expectation over \(\mathcal W_n\) and its complement,
\[
\E_{G,\hat Y}[r_+^2(\hat Y,\sigma)\mid\sigma]
\le\frac{C_W}{\log n}\,\frac1n\sum_{i=1}^n
\E[A(\hat Y_i)^2\1_{\mathcal W_n}\mid\sigma]
+z^2P_{G,\hat Y}(\mathcal W_n^c\mid\sigma)
\le\frac{C_WK_A}{\log n}+\frac{2z^2}{n^8},
\]
where the second inequality drops the indicator and needs no independence
between the event and the signals. Taking \(C_r=\max\{C_WK_A,2z^2\}\) and
\(n_r=\max\{n_W,2\}\) proves the claim, and the dependence of the
constants follows from Lemmas~\ref{lem:Lplus_prior_stability}
and~\ref{lem:Lplus_w2}.
\AppendixQED{lem:Lplus_curve_error}

\hypertarget{proof:Lplus_regret_rate}{}
\AppendixProofHeading{Proof of Proposition~\ref{prop:Lplus_regret_rate}}
Choose \(n_0\ge\max\{n_r,2\}\) such that \(0<B/n\le\bar b\) for every
\(n\ge n_0\). Since \(\bar b<z-\inf\operatorname{supp}(G)\le z\), also
\(B<nz\). Fix \(n\ge n_0\) and \(\sigma\), and let
\(\Delta(\hat Y):=\E_G[L_+(\hat\tau^{\EBlabel}_+,\mu)-L_+(\tau^\star_{G,+},\mu)\mid\hat Y,\sigma]\).
Oracle optimality gives \(\Delta(\hat Y)\ge0\), nonnegative incomes and
transfers give \(0\le L_+\le z^2\) and hence \(\Delta(\hat Y)\le z^2\),
and iterated expectations give
\(\operatorname{BR}_+(\hat\tau^{\EBlabel}_+,\tau^\star_{G,+}\mid\sigma)=\E[\Delta(\hat Y)\mid\sigma]\).
Let \(\underline\lambda\) and \(c_\lambda\) be supplied by
Lemma~\ref{lem:Lplus_shadow} and set
\(\mathcal S_n=\{\lambda_{G,+}\ge\underline\lambda\}\). On
\(\mathcal S_n\), Lemma~\ref{lem:Lplus_global_stability} gives
\(\Delta(\hat Y)\le(4z/\underline\lambda)\,r_+^2(\hat Y,\sigma)\).
Splitting the expectation over \(\mathcal S_n\) and its complement,
dropping the indicator from the nonnegative curve-error term, and
applying Lemmas~\ref{lem:Lplus_curve_error} and~\ref{lem:Lplus_shadow},
\[
\E[\Delta(\hat Y)\mid\sigma]
\le\frac{4z}{\underline\lambda}\E[r_+^2(\hat Y,\sigma)\mid\sigma]
+z^2P_{G,\hat Y}(\mathcal S_n^c\mid\sigma)
\le\frac{4zC_r}{\underline\lambda}\left(\frac1{\log n}+\frac1{n^8}\right)
+z^2e^{-c_\lambda n}.
\]
Taking \(C=\max\{4zC_r/\underline\lambda,z^2\}\) and \(c=c_\lambda\)
gives the stated bound. For fixed \(G\) and budget sequence, these
constants and \(n_0\) are uniform over noise-level vectors satisfying
Assumption~\ref{assump:normality}.
\AppendixQED{prop:Lplus_regret_rate}

\subsection{Proofs of Appendix C Results}
\label{sec:ds_proofs}

\hypertarget{proof:ds_pi_menu}{}
\AppendixProofHeading{Proof of Proposition~\ref{prop:ds_pi_menu}} The zero
allocation is feasible and $\mathcal{A}_{\mathcal{C}}$ is finite, so an
optimal allocation exists. For part (i), $n$ times the loss of any
allocation is at least $\sum_{i=1}^{n}\min_{c\in\mathcal{C}}(z-y_{i}-c)^{2}$,
with equality exactly when every household receives a menu level nearest
to its poverty gap. If some allocation of this form is feasible, the
optimal allocations are therefore exactly the feasible allocations of
this form. When no poverty gap lies at a midpoint between adjacent menu
levels, each household has a unique nearest level, so exactly one such
allocation exists. For part (ii), suppose an optimal allocation $\tau$ has
$y_{i}<y_{k}$ and $\tau_{i}<\tau_{k}$. Exchanging the two transfers
preserves the multiset of assigned levels and the total cost, hence
feasibility, and changes $n$ times the loss by
\[
(z-y_{i}-\tau_{k})^{2}+(z-y_{k}-\tau_{i})^{2}
-(z-y_{i}-\tau_{i})^{2}-(z-y_{k}-\tau_{k})^{2}
=
-2(y_{k}-y_{i})(\tau_{k}-\tau_{i})<0,
\]
contradicting optimality. Within a group of equal incomes, reordering
the transfers in nonincreasing order changes neither the loss nor the
cost, so an optimal allocation with $\tau_{1}\ge\cdots\ge\tau_{n}$
exists.\AppendixQED{prop:ds_pi_menu}

\hypertarget{proof:ds_oracle_menu}{}
\AppendixProofHeading{Proof of Proposition~\ref{prop:ds_oracle_menu}} By the
posterior loss decomposition in the proof of Lemma~\ref{lem:bayes_rules},
posterior expected loss equals $n^{-1}\|\tau-g_{G}(\hat{y},\sigma)\|^{2}$
plus a term that does not depend on $\tau$, so at each realization the
allocations that minimize posterior expected loss are the menu
allocations nearest to $g_{G}(\hat{y},\sigma)$. Throughout, when several
menu allocations are equally close to a vector,
$P_{\mathcal{A}_{\mathcal{C}}}$ selects the first of them in a fixed
enumeration of $\mathcal{A}_{\mathcal{C}}$. Since $g_{G}(\cdot,\sigma)$
is measurable, the set of realizations at which a given allocation is
nearest is measurable, being defined by finitely many comparisons of
measurable functions, so $P_{\mathcal{A}_{\mathcal{C}}}(g_{G}(\hat{y},\sigma))$
is a measurable rule. Every rule with values in
$\mathcal{A}_{\mathcal{C}}$ has finite Bayes risk, since transfers are
bounded and $G$ has a finite second moment. By iterated expectations, a
rule that minimizes posterior expected loss at almost every realization
is therefore Bayes \cite{berger2013statistical}, and a rule that fails
to do so on a set of realizations of positive probability has strictly
larger Bayes risk.\AppendixQED{prop:ds_oracle_menu}

\hypertarget{proof:ds_one_grant}{}
\AppendixProofHeading{Proof of Corollary~\ref{cor:ds_one_grant}} By
Proposition~\ref{prop:ds_oracle_menu}, it suffices to minimize
$\|\tau-g_{G}(\hat{y},\sigma)\|^{2}$ over $\mathcal{A}_{\mathcal{C}}$ at
each realization. Write $\tau_{i}=cd_{i}$ with $d_{i}\in\{0,1\}$ and
$g_{i}=g_{G,i}(\hat{y}_{i},\sigma_{i})$. Relative to no transfer,
enrolling household $i$ lowers $n$ times posterior expected loss by
$g_{i}^{2}-(g_{i}-c)^{2}=2cg_{i}-c^{2}$, and the budget allows at most
$\lfloor B/c\rfloor$ grants, so the problem is to maximize
$\sum_{i}d_{i}(2cg_{i}-c^{2})$ subject to
$\sum_{i}d_{i}\le\min\{n,\lfloor B/c\rfloor\}$. Since every grant has the
same cost, taking the largest gains first and stopping at capacity or
when the next gain is negative is optimal, which is the stated rule. A
gain of zero, at $g_{i}=c/2$, leaves the objective unchanged, so
enrolling such households while capacity remains is also optimal.
Breaking ranking ties by household index makes the rule measurable, as
in the proof of Proposition~\ref{prop:ds_oracle_menu}.\AppendixQED{cor:ds_one_grant}

\hypertarget{proof:ds_comparison}{}
\AppendixProofHeading{Proof of Lemma~\ref{lem:ds_comparison}} Write
\(p:=P_{\mathcal{A}_{\mathcal{C}}}(u)\) and
\(q:=P_{\mathcal{A}_{\mathcal{C}}}(v)\). Since \(q\) is nearest to \(v\)
in \(\mathcal{A}_{\mathcal{C}}\) and \(p\) belongs to the same set,
\(\|v-q\|^{2}\le\|v-p\|^{2}\). Expanding the squared distances gives
\[
\begin{aligned}
\|u-q\|^{2}-\|u-p\|^{2}
&=
\|v-q\|^{2}-\|v-p\|^{2}
+2\langle u-v,\,p-q\rangle
&\le
2\|u-v\|\,\|p-q\|,
\end{aligned}
\]
by the first inequality and Cauchy--Schwarz. The left side is nonnegative
because \(p\) is nearest to \(u\).\AppendixQED{lem:ds_comparison}

\hypertarget{proof:ds_regret}{}
\AppendixProofHeading{Proof of Proposition~\ref{prop:ds_regret}} Let
$\tau^{\star}=P_{\mathcal{A}_{\mathcal{C}}}(g_{G}(\hat{Y},\sigma))$ and
$\hat{\tau}=P_{\mathcal{A}_{\mathcal{C}}}(\hat{g}(\hat{Y},\sigma))$, both
measurable by the argument of Proposition~\ref{prop:ds_oracle_menu}. By
the posterior loss decomposition in the proof of
Lemma~\ref{lem:bayes_rules}, the difference in posterior
expected losses between $\hat{\tau}$ and $\tau^{\star}$ is
$n^{-1}\{\|g_{G}-\hat{\tau}\|^{2}-\|g_{G}-\tau^{\star}\|^{2}\}$ with
$g_{G}=g_{G}(\hat{Y},\sigma)$, and Bayes regret is its expectation over
$\hat{Y}$. Lemma~\ref{lem:ds_comparison} at $u=g_{G}$ and
$v=\hat{g}(\hat{Y},\sigma)$ bounds the difference at each realization by
$(2/n)\,\|\hat{g}-g_{G}\|\,\|\tau^{\star}-\hat{\tau}\|$, and
$\|\tau^{\star}-\hat{\tau}\|\le c_{J}\sqrt{2s_{B}}$ as shown in the text.
Hence, with all expectations conditional on $\sigma$,
\[
\operatorname{BR}\bigl(\hat{\tau}^{\EBlabel}_{\mathcal{C}},\tau^{\star}_{G,\mathcal{C}}\mid\sigma\bigr)
\le\frac{2c_{J}\sqrt{2s_{B}}}{n}\,\E\|\hat{g}-g_{G}\|
\le2c_{J}\sqrt{\frac{2s_{B}}{n}}\,\sqrt{\frac1n\E\|\hat{g}-g_{G}\|^{2}}
\le2c_{J}\sqrt{\frac{2s_{B}}{n}}\,\sqrt{2\bar{\sigma}\Delta_{n}+\Delta_{n}^{2}},
\]
where the second inequality is Jensen's and the third is the bound on
the estimated posterior mean gaps established in the proof of
Proposition~\ref{prop:regret}, valid for all $n$ with $\log n>1/p$.\AppendixQED{prop:ds_regret}
\renewcommand{\refname}{References for the Online Appendix}
\putbib
\stopcontents[onlineappendix]
\end{bibunit}

\end{document}

%% file: paper/numbers/paper_text_numbers.tex
\newcommand{\nCountries}{9}
\newcommand{\nReps}{500}
\newcommand{\nTrain}{500}
\newcommand{\figureTwoCountry}{Mali}

\newcommand{\povertyRateTarget}{40.1\%}
\newcommand{\budgetShareGapMin}{3.29\%}
\newcommand{\budgetShareGapMax}{5.19\%}

\newcommand{\countriesEBBeatsMLLtwo}{8}
\newcommand{\countriesEBBeatsOLSLplus}{9}
\newcommand{\countriesEBBeatsMLLplus}{8}

\newcommand{\poorReachedRatioEBOLS}{1.77}
\newcommand{\poorReachedEBPct}{12.3\%}
\newcommand{\poorReachedOLSPct}{7.0\%}
\newcommand{\poorReachedMLPct}{11.0\%}

\newcommand{\overshootEBperHundred}{4.69}
\newcommand{\overshootOLSperHundred}{10.62}
\newcommand{\leakageEBperHundred}{19.20}
\newcommand{\leakageOLSperHundred}{19.28}
\newcommand{\unspentEBperHundred}{4.28}
\newcommand{\overshootMLperHundred}{4.85}
\newcommand{\leakageMLperHundred}{21.75}
\newcommand{\unspentMLperHundred}{2.95}
\newcommand{\popTreatedOLS}{3.6\%}
\newcommand{\popTreatedEB}{6.6\%}

\newcommand{\pNinetyTransferOLS}{0.33}
\newcommand{\pNinetyTransferEB}{0.17}

\newcommand{\covExtremePoorEB}{19.6\%}
\newcommand{\covExtremePoorOLS}{12.1\%}

\newcommand{\extrGapClosedShareEB}{2.8\%}
\newcommand{\extrGapClosedShareOLS}{3.0\%}
\newcommand{\extrGapClosedShareML}{2.6\%}
\newcommand{\equivBudgetSavingEBOLSPct}{6.7\%}

\newcommand{\equivBudgetSavingEBXGBPct}{3.4\%}

\newcommand{\nProgramScenarios}{11}
\newcommand{\nProgramScenariosEBBest}{10}

\newcommand{\ethiopiaRtwoOLS}{0.119}
\newcommand{\nigeriaRtwoOLS}{0.481}
\newcommand{\ethiopiaFixedLinePredPoorOLS}{18.6\%}
\newcommand{\olsFixedLinePredPoorMin}{18.6\%}
\newcommand{\olsFixedLinePredPoorMax}{32.4\%}
\newcommand{\olsPoorRecallMin}{27.4\%}
\newcommand{\olsPoorRecallMax}{60.7\%}

%% file: paper/numbers/paper_gain_percent_numbers.tex
\newcommand{\gainEBLtwoPct}{24.5\%}
\newcommand{\gainOLSLtwoPct}{14.6\%}
\newcommand{\gainMLLtwoPct}{20.7\%}
\newcommand{\gainEBLplusPct}{41.5\%}
\newcommand{\gainEBLtwoUnderLplusPct}{40.4\%}
\newcommand{\gainOLSLplusPct}{36.3\%}
\newcommand{\gainMLLplusPct}{38.9\%}
\newcommand{\gainDiffEBOLSMinLtwoPp}{3.3}
\newcommand{\gainDiffEBOLSMaxLtwoPp}{21.8}
\newcommand{\gainEBHighNearLinePct}{19.3\%}
\newcommand{\gainEBLowNearLinePct}{31.0\%}
\newcommand{\ethiopiaGainOLSLtwoPct}{$-$21.3\%}
\newcommand{\ethiopiaGainMLLtwoPct}{$-$7.0\%}
\newcommand{\ethiopiaGainEBLtwoPct}{$-$6.8\%}
\newcommand{\nigeriaGainOLSLtwoPct}{50.0\%}
\newcommand{\nigeriaGainMLLtwoPct}{52.9\%}
\newcommand{\nigeriaGainEBLtwoPct}{58.7\%}
\newcommand{\gainEBOLSWinPp}{12.2}
\newcommand{\gainEBOLSLossPp}{3.9}
\newcommand{\gainEBMLWinPp}{10.0}
\newcommand{\gainEBMLLossPp}{7.8}
\newcommand{\malawiProgramGainAdvantagePp}{3.0}

%% file: paper/tables/table1_policy_value_percent.tex
\begin{table}[!htbp]
\centering
\caption{Targeting outcomes across nine countries}
\label{tab:main_mechanism}
\footnotesize
\renewcommand{\arraystretch}{1.16}
\setlength{\tabcolsep}{3.6pt}
\begin{tabular*}{\textwidth}{@{\extracolsep{\fill}}lcccccccc@{}}
\toprule
& \multicolumn{2}{c}{Gain (\%)} & \multicolumn{2}{c}{Transfer coverage} & \multicolumn{4}{c}{Use of available budget (\%)} \\
\cmidrule(lr){2-3}\cmidrule(lr){4-5}\cmidrule(l){6-9}
Rule & $\mathrm{Gain}_{L_2}$ & $\mathrm{Gain}_{L_+}$ & \shortstack{Population\\treated (\%)} & \shortstack{Poor people\\reached (\%)} & \shortstack{Poverty gaps\\closed} & Overshoot & Leakage & Unspent \\
\midrule
Plug-in OLS & 14.6 & 36.3 & 3.6 & 7.0 & 70.1 & 10.6 & 19.3 & 0.0 \\
XGBoost & 20.7 & 38.9 & 5.9 & 11.0 & 70.4 & 4.9 & 21.8 & 3.0 \\
\addlinespace[2.5pt]
EB-OLS & 24.5 & 40.4 & 6.6 & 12.3 & 71.8 & 4.7 & 19.2 & 4.3 \\
\bottomrule
\end{tabular*}
\vspace{5pt}
\begin{minipage}{\textwidth}
\footnotesize
\textit{Notes:} Equal-weighted averages of country-specific simulation means are reported over 500 draws. Outcomes are evaluated at realized consumption in the held-out target population using person weights. The common budget $b=0.005$ equals 0.5\% of the poverty line per represented person. Gains are reported as percentages of the perfect-information improvement over no transfers. A gain of 0\% matches the no-transfer benchmark, and 100\% matches the perfect-information benchmark. $\mathrm{Gain}_{L_+}$ evaluates the same allocations under $L_+$. Population treated is the share of persons receiving a positive transfer. Poor people reached is the corresponding share among those initially poor. The final four columns divide 100 budget units among poverty gaps closed, overshoot (spending on poor recipients beyond their initial gaps), leakage to initially nonpoor recipients, and unspent resources.
\end{minipage}
\end{table}

%% file: figures/jmp_v3/figure_menu_projection.tex
\begin{tikzpicture}[x=8.5cm,y=8.5cm,
  >={Stealth[length=5.2pt,width=5.2pt]},
  every node/.style={font=\small},
  continuous/.style={-{Stealth[length=6pt,width=6pt]},EBBlue,
      line width=1pt,dashed,shorten >=5.1pt,shorten <=3.8pt},
  menu/.style={-{Stealth[length=6pt,width=6pt]},PlugOrange,
      line width=1pt,shorten >=5.8pt,shorten <=3.8pt},
  menusquare/.style={rectangle,line width=.65pt,minimum size=7.5pt,
      inner sep=0pt},
  squarehalo/.style={rectangle,fill=white,minimum size=10.5pt,
      inner sep=0pt}]

  \draw[line width=.9pt] (.5,0)--(0,.5);
  \draw[->,line width=.75pt] (-.065,0)--(1.14,0) node[right=3pt] {$\tau_1$};
  \draw[->,line width=.75pt] (0,-.065)--(0,.81) node[above=3pt] {$\tau_2$};
  \node[below left=4pt] at (0,0) {$0$};
  \foreach \c/\lab in {.25/{c_1},.5/{c_2=B}} {
    \draw (\c,.009)--(\c,-.009) node[below=4pt] {$\lab$};
    \draw (.009,\c)--(-.009,\c) node[left=4pt] {$\lab$};
  }
  \foreach \x/\y in {0/0,0/.25,0/.5,.25/0,.5/0} {
    \node[squarehalo] at (\x,\y) {};
    \node[menusquare,draw=black!55,fill=black!55] at (\x,\y) {};
  }

  \coordinate (gap) at (.85,.65);
  \coordinate (continuousallocation) at (.35,.15);
  \coordinate (menuallocation) at (.25,.25);

  \draw[continuous] (gap)--(continuousallocation);
  \draw[menu] (gap)--(menuallocation);

  \draw[draw=black!75,fill=white,line width=.9pt] (gap) circle (3pt);
  \node[anchor=west,inner sep=1pt] at (.875,.65) {$z\mathbf 1-y$};

  \fill[white] (continuousallocation) circle (4.8pt);
  \draw[draw=EBBlue,fill=EBBlue,line width=.6pt]
    (continuousallocation) circle (3.2pt);
  \node[squarehalo] at (menuallocation) {};
  \node[menusquare,draw=PlugOrange,fill=PlugOrange] at (menuallocation) {};

  \node[anchor=east,inner sep=1pt,text=PlugOrange] at (.220,.250)
    {$\tau^{\PIlabel}_{\mathcal C}(y)$};
  \draw[EBBlue,line width=.5pt] (.335,.146)--(.265,.137);
  \node[anchor=east,inner sep=1pt,text=EBBlue] at (.257,.137)
    {$\tau^{\PIlabel}(y)$};
\end{tikzpicture}

%% file: figures/jmp_v3/figure_projection.tex
\begin{tikzpicture}[x=10.6cm,y=10.6cm,>={Stealth[length=5.2pt,width=5.2pt]},
  every node/.style={font=\small},
  gapdot/.style={draw=black!75,fill=white,line width=.8pt},
  allocdot/.style={draw=EBBlue,fill=EBBlue},
  proj/.style={-{Stealth[length=6pt,width=6pt]},EBBlue,line width=1pt,dashed,shorten >=2pt,shorten <=2pt},
  cutoffleg/.style={PlugOrange,line width=.85pt,|-|},
  region/.style={font=\small\itshape,text=black!55,align=center}]

  \fill[EBBlue!9] (.5,0)--(1.08,.58)--(1.08,.98)--(.48,.98)--(0,.5)--cycle;
  \fill[PlugOrange!10] (.5,0)--(1.08,.58)--(1.08,-.065)--(.5,-.065)--cycle;
  \fill[black!14] (0,0)--(.5,0)--(0,.5)--cycle;
  \draw[line width=.9pt] (.5,0)--(0,.5);
  \draw[dotted,black!50,line width=.65pt] (.5,0)--(1.08,.58);
  \draw[dotted,black!50,line width=.65pt] (0,.5)--(.48,.98);

  \node[region,rotate=45] at (.69,.875) {Both receive};
  \node[region,rotate=45] at (.16,.82) {Only household 2 receives};
  \node[region] at (.85,.075) {Only household 1\\receives};
  \node at (.14,.105) {$\mathcal A$};

  \draw[->,line width=.75pt] (-.065,0)--(1.14,0) node[right=3pt] {$\tau_1$};
  \draw[->,line width=.75pt] (0,-.065)--(0,1.04) node[above=3pt] {$\tau_2$};
  \draw (.5,.01)--(.5,-.01) node[below=3pt] {$B$};
  \draw (.01,.5)--(-.01,.5) node[left=3pt] {$B$};
  \node[below left=3pt] at (0,0) {$0$};

  \coordinate (u) at (.8,.7);
  \coordinate (p) at (.3,.2);
  \draw[cutoffleg] (.8,.7)--(.3,.7)
     node[midway,above=4pt,text=PlugOrange] {$\lambda^{\PIlabel}(y)$};
  \draw[cutoffleg] (.3,.7)--(.3,.2)
     node[midway,left=4pt,text=PlugOrange] {$\lambda^{\PIlabel}(y)$};
  \draw[proj] (u)--(p);
  \draw[black!60,line width=.55pt] (.325,.225)--(.3,.25)--(.275,.225);
  \draw[gapdot] (u) circle (2.3pt);
  \node[right=5pt] at (u) {$z\mathbf 1-y$};
  \draw[allocdot] (p) circle (2.3pt);
  \node[anchor=north east,text=EBBlue,inner sep=1pt] at (.29,.177)
     {$\tau^{\PIlabel}(y)$};

  \coordinate (v) at (.9,.3);
  \coordinate (q) at (.5,0);
  \draw[cutoffleg] (.9,.3)--(.5,.3)
     node[midway,above=4pt,text=PlugOrange] {$\lambda^{\PIlabel}(y')$};
  \draw[cutoffleg] (.5,.3)--(.5,0)
     node[midway,right=4pt,text=PlugOrange] {$z-y'_2$};
  \draw[proj] (v)--(q);
  \draw[gapdot] (v) circle (2.3pt);
  \node[right=5pt] at (v) {$z\mathbf 1-y'$};
  \draw[allocdot] (q) circle (2.3pt);
  \node[anchor=north west,text=EBBlue,inner sep=1pt] at (.54,-.025)
     {$\tau^{\PIlabel}(y')=(B,0)$};
\end{tikzpicture}

%% file: paper/tables/tableD1_country_environments.tex
\begin{table}[!htbp]
\centering
\caption{Country environments and prediction quality}
\label{tab:country_environments}
\scriptsize
\renewcommand{\arraystretch}{0.96}
\textbf{Panel A: Target populations and poverty}\\[3pt]
\begin{tabular*}{\textwidth}{@{\extracolsep{\fill}}lrccccc}
\toprule
Country & Target $N$ & Poverty rate & Mean gap & Extreme poor & Near line & Budget / poverty gap \\
 & & (\%) & (share of $z$) & (\%) & (\%) & (\%) \\
\midrule
Burkina Faso & 9{,}897 & 39.9 & 0.096 & 2.6 & 16.7 & 5.2 \\
Ethiopia & 4{,}693 & 40.2 & 0.127 & 7.9 & 13.3 & 3.9 \\
Ghana & 4{,}214 & 39.9 & 0.152 & 12.6 & 9.4 & 3.3 \\
Malawi & 3{,}467 & 40.1 & 0.118 & 6.2 & 12.9 & 4.2 \\
Mali & 2{,}740 & 39.9 & 0.130 & 8.4 & 10.0 & 3.8 \\
Niger & 3{,}359 & 40.0 & 0.098 & 3.1 & 16.2 & 5.1 \\
Nigeria & 3{,}814 & 40.0 & 0.132 & 9.1 & 11.0 & 3.8 \\
Tanzania & 4{,}282 & 40.2 & 0.130 & 8.4 & 12.5 & 3.9 \\
Uganda & 2{,}178 & 41.1 & 0.135 & 9.3 & 13.0 & 3.7 \\
\bottomrule
\end{tabular*}
\par\vspace{4pt}
\textbf{Panel B: Overall out-of-sample prediction quality}\\[3pt]
\begin{tabular*}{\textwidth}{@{\extracolsep{\fill}}lcccccc}
\toprule
 & \multicolumn{2}{c}{$R^2$} & \multicolumn{2}{c}{RMSE} & \multicolumn{2}{c}{Weighted rank correlation} \\
\cmidrule(lr){2-3} \cmidrule(lr){4-5} \cmidrule(lr){6-7}
Country & OLS & XGBoost & OLS & XGBoost & OLS & XGBoost \\
\midrule
Burkina Faso & 0.496 & 0.503 & 0.587 & 0.583 & 0.609 & 0.581 \\
Ethiopia & 0.119 & 0.136 & 0.818 & 0.810 & 0.368 & 0.362 \\
Ghana & 0.389 & 0.392 & 1.315 & 1.312 & 0.674 & 0.665 \\
Malawi & 0.401 & 0.373 & 0.673 & 0.689 & 0.566 & 0.535 \\
Mali & 0.365 & 0.352 & 1.001 & 1.012 & 0.563 & 0.555 \\
Niger & 0.414 & 0.410 & 0.488 & 0.489 & 0.551 & 0.544 \\
Nigeria & 0.481 & 0.477 & 0.643 & 0.645 & 0.713 & 0.704 \\
Tanzania & 0.451 & 0.469 & 0.785 & 0.772 & 0.612 & 0.607 \\
Uganda & 0.414 & 0.380 & 0.752 & 0.774 & 0.590 & 0.559 \\
\bottomrule
\end{tabular*}
\par\vspace{4pt}
\textbf{Panel C: Prediction quality in the lower tail}\\[3pt]
\begin{tabular*}{\textwidth}{@{\extracolsep{\fill}}lcccccc}
\toprule
 & \multicolumn{2}{c}{RMSE among poor people} & \multicolumn{2}{c}{Predicted poor} & \multicolumn{2}{c}{Poor correctly classified} \\
\cmidrule(lr){2-3} \cmidrule(lr){4-5} \cmidrule(lr){6-7}
Country & OLS & XGBoost & OLS (\%) & XGBoost (\%) & OLS (\%) & XGBoost (\%) \\
\midrule
Burkina Faso & 0.474 & 0.477 & 30.6 & 23.1 & 51.3 & 38.8 \\
Ethiopia & 0.677 & 0.716 & 18.6 & 8.8 & 27.4 & 13.6 \\
Ghana & 0.988 & 0.991 & 26.7 & 21.7 & 49.9 & 42.0 \\
Malawi & 0.592 & 0.617 & 23.7 & 20.5 & 40.3 & 33.7 \\
Mali & 0.702 & 0.693 & 19.5 & 11.4 & 32.7 & 20.1 \\
Niger & 0.403 & 0.437 & 32.4 & 23.8 & 50.5 & 38.0 \\
Nigeria & 0.505 & 0.488 & 31.7 & 30.6 & 60.7 & 58.5 \\
Tanzania & 0.598 & 0.586 & 31.3 & 26.0 & 51.0 & 43.0 \\
Uganda & 0.569 & 0.592 & 32.0 & 27.4 & 52.9 & 44.5 \\
\bottomrule
\end{tabular*}
\vspace{4pt}
\begin{minipage}{\textwidth}
\scriptsize
\textit{Notes:} Country-specific simulation means are reported over 500 draws. All quantities use person weights in the held-out target population. In Panel A, mean gap is the average pre-transfer poverty gap expressed as a share of $z$. Extreme poor is the share of people with consumption below $0.5z$, and near line is the share with consumption between $0.9z$ and $1.1z$. Budget / poverty gap is the common budget $b=0.005$ as a share of the aggregate pre-transfer poverty gap. In Panel B, the RMSE is expressed relative to $z$, and weighted rank correlation is the person-weighted Spearman correlation between predicted and realized consumption. In Panel C, predicted poor is the share of people whose predicted consumption is below $z$. Poor correctly classified is the share of poor people whose predicted consumption is below $z$.
\end{minipage}
\end{table}

%% file: paper/tables/tableD2_robustness_percent.tex
\begin{table}[!htbp]
\centering
\caption{Robustness of targeting gains}
\label{tab:robustness}
\scriptsize
\renewcommand{\arraystretch}{0.94}
\textbf{Panel A: Common budget levels}\\[3pt]
\begin{tabular*}{\textwidth}{@{\extracolsep{\fill}}lccccc@{}}
\toprule
& \multicolumn{3}{c}{Gain ($L_2$, \%)} & \multicolumn{2}{c}{EB advantage (p.p.)} \\
\cmidrule(lr){2-4}\cmidrule(l){5-6}
\shortstack[l]{Budget per person\\(\% of $z$)} & Plug-in OLS & XGBoost & EB-OLS & over OLS & over XGBoost \\
\midrule
0.125 & 18.9 & 24.0 & 28.6 & 9.7 & 4.6 \\
0.250 & 17.0 & 22.6 & 26.8 & 9.8 & 4.2 \\
0.500 & 14.6 & 20.7 & 24.5 & 9.9 & 3.8 \\
1.000 & 11.0 & 17.7 & 21.2 & 10.2 & 3.5 \\
2.000 & 5.1 & 13.4 & 16.5 & 11.4 & 3.1 \\
\bottomrule
\end{tabular*}
\par\vspace{5pt}
\textbf{Panel B: Alternative design choices at the headline budget}\\[3pt]
\begin{tabular*}{\textwidth}{@{\extracolsep{\fill}}lccccc@{}}
\toprule
& \multicolumn{3}{c}{Gain ($L_2$, \%)} & \multicolumn{2}{c}{\shortstack{Shrinkage-rule\\advantage (p.p.)}} \\
\cmidrule(lr){2-4}\cmidrule(l){5-6}
Design & Plug-in OLS & XGBoost & Shrinkage rule & over OLS & over XGBoost \\
\midrule
Headline specification & 14.6 & 20.7 & 24.5 & 9.9 & 3.8 \\
Standard GLmix support & 14.6 & 20.7 & 23.4 & 8.8 & 2.8 \\
Budget spent in full & 14.6 & 20.6 & 24.3 & 9.7 & 3.7 \\
OLS covariance clustered by enumeration area & 14.5 & 20.9 & 24.5 & 10.0 & 3.6 \\
Person-weighted OLS fit & 9.1 & 20.4 & 21.8 & 12.7 & 1.4 \\
250 training households & $-$0.3 & 14.8 & 17.9 & 18.2 & 3.1 \\
500 training households (common target) & 14.6 & 20.6 & 24.4 & 9.8 & 3.8 \\
1,000 training households & 22.2 & 25.6 & 28.5 & 6.3 & 2.9 \\
Training households retained in target & 16.2 & 25.5 & 26.6 & 10.4 & 1.1 \\
Fixed XGBoost hyperparameters & 14.5 & 13.4 & 24.5 & 10.0 & 11.1 \\
Positive-part James--Stein & 14.5 & 20.9 & 16.1 & 1.6 & $-$4.8 \\
CLOSE-NPMLE & 14.5 & 20.9 & 21.3 & 6.8 & 0.4 \\
\bottomrule
\end{tabular*}
\par\vspace{5pt}
\textbf{Panel C: Documented program budgets}\\[3pt]
\begin{tabular*}{\textwidth}{@{\extracolsep{\fill}}llccccc@{}}
\toprule
& & & \multicolumn{3}{c}{Gain ($L_2$, \%)} & \multicolumn{1}{c}{EB advantage (p.p.)} \\
\cmidrule(lr){4-6}\cmidrule(l){7-7}
Country & Program scenario & \shortstack{Budget per person\\(\% of $z$)} & OLS & XGBoost & EB-OLS & \shortstack{over best\\plug-in} \\
\midrule
Ghana & LEAP 2008 (lower payment) & 0.0048 & 44.1 & 41.6 & 45.1 & 1.0 \\
 & LEAP 2008 (upper payment) & 0.0090 & 42.0 & 41.1 & 44.5 & 2.5 \\
 & LEAP 2010 plan (lower) & 0.0148 & 40.0 & 41.2 & 44.1 & 2.9 \\
 & LEAP 2010 plan (upper) & 0.0277 & 37.2 & 41.9 & 43.7 & 1.8 \\
\addlinespace[1pt]
Malawi & SCTP 2013 (lower bound) & 0.0677 & 29.9 & 21.9 & 28.1 & $-$1.9 \\
 & SCTP 2015 plan & 0.4512 & 19.8 & 15.9 & 22.8 & 3.0 \\
\addlinespace[1pt]
Mali & Jigisemejiri design & 0.2702 & $-$12.2 & 7.7 & 11.3 & 3.6 \\
\addlinespace[1pt]
Niger & PPFS pilot & 0.0097 & 8.6 & 23.6 & 33.9 & 10.4 \\
 & PPFS 2017 plan & 0.2566 & 13.1 & 14.2 & 19.4 & 5.2 \\
\addlinespace[1pt]
Tanzania & PSSN phase costing & 0.2142 & 18.2 & 17.3 & 19.5 & 1.2 \\
 & PSSN realized-scale costing & 0.9425 & 8.5 & 10.7 & 13.7 & 3.0 \\
\bottomrule
\end{tabular*}
\vspace{3pt}
\begin{minipage}{\textwidth}
\scriptsize
\textit{Notes:} Panels A and C and the first three rows of Panel B use 500 draws per country; the remaining rows of Panel B use 100. Gains are reported as percentages of the perfect-information improvement over no transfers. A gain of 0\% matches the no-transfer benchmark, and 100\% matches the perfect-information benchmark. Advantages are differences in gains, measured in percentage points (p.p.). Panel A reports equal-weighted averages of country-specific simulation means at different common budget levels. Budget per person is expressed as a percentage of the poverty line and equals $100b$. Panel B reports the corresponding averages under alternative design choices at $b=0.005$, changing one element of the headline design at a time. ``Shrinkage rule'' refers to EB-OLS except in the James--Stein and CLOSE-NPMLE rows. James--Stein applies the heteroskedastic positive-part estimator of \citeA{xie2012sure} to the OLS-estimated poverty gaps before the same weighted projection. The budget-spent-in-full specification requires every rule to exhaust the available budget. Panel C reports country-specific simulation means at documented program budget levels. Its final column gives the difference between EB-OLS and the better-performing plug-in rule. Table~\ref{tab:program_budgets} documents the program budget conversions.
\end{minipage}
\end{table}

%% file: paper/tables/tableD6_program_budgets.tex
\begin{landscape}
\begin{table}[p]
\centering
\caption{Program budgets and their normalized equivalents}
\label{tab:program_budgets}
\footnotesize
\renewcommand{\arraystretch}{0.96}
\setlength{\tabcolsep}{3.4pt}
\begin{tabular*}{\linewidth}{@{\extracolsep{\fill}}llrrrrrrc@{}}
\toprule
& & \multicolumn{3}{c}{Program evidence} & \multicolumn{2}{c}{Normalization inputs} & & \\
\cmidrule(lr){3-5}\cmidrule(lr){6-7}
Country & Program scenario & Year & \shortstack{Caseload\\(households)} & \shortstack{Annualized\\outlay} & \shortstack{Population\\(millions)} & \shortstack{Annual poverty\\line per person} & \shortstack{Budget per person\\(\% of $z$)} & Source \\
\midrule
Ghana & LEAP 2008 (lower payment) & 2008 & 8{,}200 & GHS 787K & 24.25 & 677 & 0.0048 & \href{https://documents1.worldbank.org/curated/en/275321468036344421/pdf/472230PGD0P113101Official0Use0Only1.pdf}{[1]} \\
 & LEAP 2008 (upper payment) & 2008 & 8{,}200 & GHS 1.48M & 24.25 & 677 & 0.0090 & \href{https://documents1.worldbank.org/curated/en/275321468036344421/pdf/472230PGD0P113101Official0Use0Only1.pdf}{[1]} \\
 & LEAP 2010 plan (lower) & 2010 & 35{,}000 & GHS 3.36M & 25.47 & 894 & 0.0148 & \href{https://documents1.worldbank.org/curated/en/275321468036344421/pdf/472230PGD0P113101Official0Use0Only1.pdf}{[1]} \\
 & LEAP 2010 plan (upper) & 2010 & 35{,}000 & GHS 6.30M & 25.47 & 894 & 0.0277 & \href{https://documents1.worldbank.org/curated/en/275321468036344421/pdf/472230PGD0P113101Official0Use0Only1.pdf}{[1]} \\
\addlinespace[1.5pt]
Malawi & SCTP 2013 (lower bound) & 2013 & 30{,}000 & MWK 972M & 16.16 & 88{,}872 & 0.0677 & \href{https://documents1.worldbank.org/curated/en/929241468056947115/pdf/826170PAD0P1330sclosed01102801300SD.pdf}{[2]} \\
 & SCTP 2015 plan & 2015 & 319{,}000 & MWK 10.3B & 17.09 & 134{,}074 & 0.4512 & \href{https://documents1.worldbank.org/curated/en/929241468056947115/pdf/826170PAD0P1330sclosed01102801300SD.pdf}{[2]} \\
\addlinespace[1.5pt]
Mali & Jigisemejiri design & 2013 & 62{,}000 & XOF 7.44B & 17.46 & 157{,}699 & 0.2702 & \href{https://www.worldbank.org/en/news/press-release/2013/04/30/mali-world-bank-to-support-emergency-social-safety-net-for-over-62000-poor-households}{[3]} \\
\addlinespace[1.5pt]
Niger & PPFS pilot & 2011 & 2{,}281 & XOF 274M & 17.18 & 165{,}015 & 0.0097 & \href{https://documents1.worldbank.org/curated/en/402771474989456640/pdf/WPS7839.pdf}{[4]} \\
 & PPFS 2017 plan & 2017 & 80{,}000 & XOF 9.60B & 21.44 & 174{,}540 & 0.2566 & \href{https://documents1.worldbank.org/curated/en/402771474989456640/pdf/WPS7839.pdf}{[4]} \\
\addlinespace[1.5pt]
Tanzania & PSSN phase costing & 2015 & 250{,}000 & TZS 63.0B & 52.02 & 565{,}388 & 0.2142 & \href{https://documents1.worldbank.org/curated/en/273011479390056768/pdf/110255-WP-P124045-OUO-9-PSSN-IE-Baseline-Report-FINAL-FOR-PUBLISHING.pdf}{[5]} \\
 & PSSN realized-scale costing & 2015 & 1{,}100{,}000 & TZS 277B & 52.02 & 565{,}388 & 0.9425 & \href{https://www.worldbank.org/en/news/press-release/2016/06/16/new-financing-to-increase-incomes-and-opportunities-for-66-million-among-tanzanias-extreme-poor}{[6]} \\
\bottomrule
\end{tabular*}
\vspace{5pt}
\begin{minipage}{\linewidth}
\scriptsize
\textit{Notes:} Budget per person is $100b=100T/(Pz)$, where $T$ is the annualized transfer outlay, $P$ is the program-year population, and $z$ is the annual per-person poverty line in program-year prices. The headline simulation budget $b=0.005$ corresponds to 0.5\% of the poverty line. It exceeds the budget in 10 of the 11 documented scenarios and lies below only the Tanzania PSSN realized-scale costing. The PSSN realized-scale costing applies a monthly transfer of TZS 21,000 to the 1.1 million households reached by December 2015 for a full year, yielding an annualized outlay of TZS 277.2B. This is a costing estimate rather than audited annual expenditure. At the World Development Indicators 2015 period-average exchange rate of 1,991 TZS per US dollar, this is about 139 million dollars a year. The conclusion's dollar illustration applies Tanzania's simulated saving at the headline budget, $b=0.005$, to this outlay; the PSSN costing corresponds to the larger budget $b=0.009425$. Annualized outlays and poverty lines are expressed in the local currency shown in the outlay column. Poverty lines are converted to program-year prices using World Development Indicators consumer price indices, and population figures come from the same source. K, M, and B denote thousands, millions, and billions. The scenarios include historical bounds, pilots, plans, program designs, and constructed costings, so they should not all be interpreted as audited expenditure. Source numbers link to the underlying documents. The replication files record the source locations and conversion steps.
\end{minipage}
\end{table}
\end{landscape}

%% file: paper/tables/tableD3_country_policy_value_percent.tex
\begin{landscape}
\begin{table}[p]
\centering
\caption{Targeting outcomes by country}
\label{tab:country_main_mechanism}
\footnotesize
\renewcommand{\arraystretch}{0.93}
\setlength{\tabcolsep}{4.5pt}
\begin{tabular*}{\linewidth}{@{\extracolsep{\fill}}llcccccccc@{}}
\toprule
& & \multicolumn{2}{c}{Gain (\%)} & \multicolumn{2}{c}{Transfer coverage} & \multicolumn{4}{c}{Use of available budget (\%)} \\
\cmidrule(lr){3-4}\cmidrule(lr){5-6}\cmidrule(l){7-10}
Country & Rule & $\mathrm{Gain}_{L_2}$ & $\mathrm{Gain}_{L_+}$ & \shortstack{Population\\treated (\%)} & \shortstack{Poor people\\reached (\%)} & \shortstack{Poverty gaps\\closed} & Overshoot & Leakage & Unspent \\
\midrule
Burkina Faso & Plug-in OLS & 13.4 & 34.9 & 3.7 & 7.4 & 68.3 & 13.5 & 18.2 & 0.0 \\
 & XGBoost & 14.7 & 36.2 & 7.4 & 13.4 & 68.2 & 6.9 & 24.7 & 0.2 \\
 & EB-OLS & 24.4 & 39.8 & 6.9 & 13.4 & 73.8 & 7.6 & 18.6 & 0.0 \\
\addlinespace[2pt]
Ethiopia & Plug-in OLS & $-$21.3 & 26.9 & 3.5 & 5.6 & 55.4 & 10.9 & 33.7 & 0.0 \\
 & XGBoost & $-$7.0 & 25.1 & 4.9 & 7.9 & 49.2 & 5.7 & 29.0 & 16.2 \\
 & EB-OLS & $-$6.8 & 18.7 & 4.2 & 6.8 & 39.2 & 6.1 & 22.2 & 32.5 \\
\addlinespace[2pt]
Ghana & Plug-in OLS & 22.9 & 38.3 & 2.0 & 4.2 & 72.2 & 12.4 & 15.4 & 0.0 \\
 & XGBoost & 42.3 & 51.8 & 4.2 & 9.1 & 83.5 & 3.6 & 12.9 & 0.0 \\
 & EB-OLS & 39.0 & 49.9 & 4.3 & 8.9 & 81.5 & 3.1 & 15.4 & 0.0 \\
\addlinespace[2pt]
Malawi & Plug-in OLS & 19.2 & 35.0 & 3.7 & 7.0 & 70.2 & 9.9 & 19.9 & 0.0 \\
 & XGBoost & 15.5 & 34.9 & 5.2 & 9.2 & 67.9 & 4.7 & 27.4 & 0.0 \\
 & EB-OLS & 22.5 & 38.8 & 6.5 & 12.1 & 72.1 & 5.0 & 22.9 & 0.0 \\
\addlinespace[2pt]
Mali & Plug-in OLS & $-$12.2 & 28.4 & 3.6 & 6.6 & 60.6 & 14.3 & 25.1 & 0.0 \\
 & XGBoost & 6.3 & 32.7 & 5.8 & 10.6 & 59.7 & 4.8 & 25.3 & 10.1 \\
 & EB-OLS & 9.5 & 36.2 & 6.8 & 12.5 & 65.5 & 4.8 & 24.3 & 5.4 \\
\addlinespace[2pt]
Niger & Plug-in OLS & 10.2 & 31.8 & 5.9 & 10.4 & 64.4 & 9.2 & 26.4 & 0.0 \\
 & XGBoost & 11.6 & 31.8 & 8.4 & 14.3 & 64.5 & 6.1 & 29.3 & 0.1 \\
 & EB-OLS & 15.8 & 34.3 & 11.8 & 19.8 & 66.8 & 4.7 & 28.5 & 0.0 \\
\addlinespace[2pt]
Nigeria & Plug-in OLS & 50.0 & 54.6 & 3.6 & 8.3 & 88.5 & 5.9 & 5.5 & 0.0 \\
 & XGBoost & 52.9 & 57.3 & 5.8 & 13.1 & 90.1 & 2.4 & 7.5 & 0.0 \\
 & EB-OLS & 58.7 & 61.0 & 5.1 & 11.9 & 93.7 & 2.4 & 3.9 & 0.0 \\
\addlinespace[2pt]
Tanzania & Plug-in OLS & 13.2 & 34.0 & 3.3 & 6.1 & 68.4 & 9.4 & 22.2 & 0.0 \\
 & XGBoost & 14.0 & 36.4 & 6.0 & 10.7 & 69.2 & 5.2 & 25.6 & 0.0 \\
 & EB-OLS & 16.6 & 37.9 & 7.9 & 14.0 & 69.3 & 3.7 & 26.5 & 0.6 \\
\addlinespace[2pt]
Uganda & Plug-in OLS & 36.0 & 43.0 & 3.3 & 7.3 & 82.8 & 10.0 & 7.2 & 0.0 \\
 & XGBoost & 35.8 & 44.0 & 5.5 & 10.8 & 81.6 & 4.3 & 14.1 & 0.0 \\
 & EB-OLS & 40.7 & 46.6 & 5.7 & 11.7 & 84.7 & 4.8 & 10.5 & 0.0 \\
\bottomrule
\end{tabular*}
\vspace{5pt}
\begin{minipage}{\linewidth}
\scriptsize
\textit{Notes:} Country-specific simulation means are reported over 500 draws. Outcomes are evaluated at realized consumption in the held-out target population using person weights. The common budget $b=0.005$ equals 0.5\% of the poverty line per represented person. Gains are reported as percentages of the perfect-information improvement over no transfers. A gain of 0\% matches the no-transfer benchmark, and 100\% matches the perfect-information benchmark. $\mathrm{Gain}_{L_+}$ evaluates the same allocations under $L_+$. Population treated is the share of persons receiving a positive transfer. Poor people reached is the corresponding share among those initially poor. The final four columns divide 100 budget units among poverty gaps closed, overshoot (spending on poor recipients beyond their initial gaps), leakage to initially nonpoor recipients, and unspent resources.
\end{minipage}
\end{table}
\end{landscape}

%% file: paper/tables/tableD4_targeting_mechanisms.tex
\begin{table}[p]
\centering
\caption{Recipient composition, transfer amounts, and extreme-poverty outcomes by country}
\label{tab:detailed_mechanisms}
\fontsize{9}{10}\selectfont
\renewcommand{\arraystretch}{0.93}
\setlength{\tabcolsep}{3.2pt}
\begin{tabular*}{\textwidth}{@{\extracolsep{\fill}}llccccc@{}}
\toprule
& & \multicolumn{1}{c}{Recipient composition} & \multicolumn{2}{c}{Transfers among recipients} & \multicolumn{2}{c}{Extreme poverty} \\
\cmidrule(lr){3-3}\cmidrule(lr){4-5}\cmidrule(l){6-7}
Country & Rule & \shortstack{Nonpoor\\recipients (\%)} & \shortstack{Mean\\(share of $z$)} & \shortstack{90th percentile\\(share of $z$)} & \shortstack{People\\reached (\%)} & \shortstack{Gap\\closed (\%)} \\
\midrule
\textit{Cross-country mean} & Plug-in OLS & 21.4 & 0.152 & 0.331 & 12.1 & 3.0 \\
 & XGBoost & 24.5 & 0.087 & 0.185 & 17.2 & 2.6 \\
 & EB-OLS & 22.7 & 0.088 & 0.174 & 19.6 & 2.8 \\
\midrule
Burkina Faso & Plug-in OLS & 19.6 & 0.140 & 0.309 & 12.0 & 3.0 \\
 & XGBoost & 27.2 & 0.070 & 0.155 & 17.1 & 2.4 \\
 & EB-OLS & 21.4 & 0.083 & 0.169 & 19.1 & 2.8 \\
\addlinespace[1.5pt]
Ethiopia & Plug-in OLS & 35.7 & 0.148 & 0.323 & 7.7 & 2.1 \\
 & XGBoost & 35.2 & 0.084 & 0.178 & 10.4 & 1.5 \\
 & EB-OLS & 34.3 & 0.089 & 0.178 & 7.7 & 1.0 \\
\addlinespace[1.5pt]
Ghana & Plug-in OLS & 16.4 & 0.257 & 0.557 & 7.6 & 3.1 \\
 & XGBoost & 14.1 & 0.121 & 0.255 & 15.8 & 3.1 \\
 & EB-OLS & 16.3 & 0.127 & 0.244 & 15.0 & 3.0 \\
\addlinespace[1.5pt]
Malawi & Plug-in OLS & 23.1 & 0.139 & 0.307 & 13.7 & 3.5 \\
 & XGBoost & 28.8 & 0.099 & 0.212 & 16.3 & 2.7 \\
 & EB-OLS & 25.0 & 0.084 & 0.172 & 22.4 & 3.2 \\
\addlinespace[1.5pt]
Mali & Plug-in OLS & 26.6 & 0.150 & 0.320 & 10.6 & 2.2 \\
 & XGBoost & 27.9 & 0.078 & 0.163 & 16.2 & 2.1 \\
 & EB-OLS & 27.1 & 0.076 & 0.137 & 20.1 & 2.5 \\
\addlinespace[1.5pt]
Niger & Plug-in OLS & 29.9 & 0.086 & 0.184 & 16.0 & 3.3 \\
 & XGBoost & 31.8 & 0.061 & 0.129 & 18.3 & 2.4 \\
 & EB-OLS & 32.1 & 0.050 & 0.098 & 25.6 & 2.8 \\
\addlinespace[1.5pt]
Nigeria & Plug-in OLS & 7.0 & 0.140 & 0.306 & 17.5 & 4.4 \\
 & XGBoost & 9.3 & 0.088 & 0.188 & 25.0 & 4.1 \\
 & EB-OLS & 6.0 & 0.103 & 0.210 & 24.4 & 4.5 \\
\addlinespace[1.5pt]
Tanzania & Plug-in OLS & 25.0 & 0.156 & 0.341 & 9.4 & 2.4 \\
 & XGBoost & 27.8 & 0.086 & 0.183 & 15.4 & 2.2 \\
 & EB-OLS & 28.2 & 0.077 & 0.154 & 20.5 & 2.3 \\
\addlinespace[1.5pt]
Uganda & Plug-in OLS & 9.6 & 0.154 & 0.332 & 14.3 & 3.6 \\
 & XGBoost & 18.5 & 0.095 & 0.204 & 19.7 & 3.2 \\
 & EB-OLS & 13.8 & 0.104 & 0.206 & 21.6 & 3.4 \\
\bottomrule
\end{tabular*}
\vspace{5pt}
\begin{minipage}{\textwidth}
\scriptsize
\textit{Notes:} Country-specific simulation means are reported over 500 draws. The cross-country mean gives equal weight to each country. Outcomes are evaluated at realized consumption in the held-out target population using person weights. The common budget is $b=0.005$. Nonpoor recipients is the share of recipients who were initially nonpoor. Mean and 90th-percentile transfers are calculated among recipients and expressed as shares of $z$. Replications in which a rule makes no transfer are excluded from these conditional transfer-amount statistics. Extreme poor denotes people with pre-transfer consumption below $0.5z$. The final two columns report the share of extremely poor people receiving a positive transfer and the share of their aggregate pre-transfer poverty gap closed.
\end{minipage}
\end{table}

%% file: paper/tables/tableD5_loss_rule_comparison_percent.tex
\begin{table}[!htbp]
\centering
\caption{Empirical Bayes allocations optimized for alternative losses}
\label{tab:eb_L2_vs_Lplus}
\footnotesize
\renewcommand{\arraystretch}{1.05}
\textbf{Panel A: Gain by country and evaluation loss}\\[4pt]
\begin{tabular*}{\textwidth}{@{\extracolsep{\fill}}lcccc@{}}
\toprule
& \multicolumn{2}{c}{$\mathrm{Gain}_{L_2}$ (\%)} & \multicolumn{2}{c}{$\mathrm{Gain}_{L_+}$ (\%)} \\
\cmidrule(lr){2-3}\cmidrule(l){4-5}
Country & $\hat\tau^{\mathrm{EB}}$ & $\hat\tau^{\mathrm{EB}}_+$ & $\hat\tau^{\mathrm{EB}}$ & $\hat\tau^{\mathrm{EB}}_+$ \\
\midrule
\textit{Cross-country mean} & 24.5 & 23.6 & 40.4 & 41.5 \\
\midrule
Burkina Faso & 24.4 & 24.1 & 39.8 & 39.8 \\
Ethiopia & $-$6.8 & $-$12.0 & 18.7 & 27.7 \\
Ghana & 39.0 & 39.0 & 49.9 & 49.9 \\
Malawi & 22.5 & 22.4 & 38.8 & 38.7 \\
Mali & 9.5 & 8.2 & 36.2 & 37.7 \\
Niger & 15.8 & 15.4 & 34.3 & 34.2 \\
Nigeria & 58.7 & 58.7 & 61.0 & 61.0 \\
Tanzania & 16.6 & 16.1 & 37.9 & 37.8 \\
Uganda & 40.7 & 40.5 & 46.6 & 46.5 \\
\bottomrule
\end{tabular*}
\vspace{9pt}
\textbf{Panel B: Cross-country transfer coverage and use of available budget}\\[4pt]
\begin{tabular*}{\textwidth}{@{\extracolsep{\fill}}lcccccc@{}}
\toprule
& \multicolumn{2}{c}{Transfer coverage} & \multicolumn{4}{c}{Use of available budget (\%)} \\
\cmidrule(lr){2-3}\cmidrule(l){4-7}
Allocation rule & \shortstack{Population\\treated (\%)} & \shortstack{Poor people\\reached (\%)} & \shortstack{Poverty gaps\\closed} & Overshoot & Leakage & Unspent \\
\midrule
$\hat\tau^{\mathrm{EB}}$ & 6.6 & 12.3 & 71.8 & 4.7 & 19.2 & 4.3 \\
$\hat\tau^{\mathrm{EB}}_+$ & 7.6 & 13.9 & 74.1 & 5.0 & 20.8 & 0.0 \\
\bottomrule
\end{tabular*}
\vspace{5pt}
\begin{minipage}{\textwidth}
\footnotesize
\textit{Notes:} Country-specific simulation means are reported over 500 draws. Outcomes are evaluated at realized consumption in the held-out target population using person weights. Both allocation rules use the same OLS signals and fitted restricted GLmix prior at $b=0.005$. The rule $\hat\tau^{\mathrm{EB}}$ minimizes the posterior expected quadratic loss. The rule $\hat\tau^{\mathrm{EB}}_+$ minimizes the posterior expected one-sided loss. Panel A evaluates each allocation under both losses. The cross-country mean gives equal weight to each country. Gains are reported as percentages of the perfect-information improvement over no transfers. A gain of 0\% matches the no-transfer benchmark, and 100\% matches the perfect-information benchmark. Panel B reports equal-weighted averages of country-specific simulation means. Population treated is the share of persons receiving a positive transfer. Poor people reached is the corresponding share among those initially poor. The final four columns divide 100 budget units among poverty gaps closed, overshoot, leakage, and unspent resources.
\end{minipage}
\end{table}